\documentclass[pre,superscriptaddress,showpacs,floatfix,onecolumn]{revtex4-2}
\usepackage[utf8]{inputenc}
\usepackage[T1]{fontenc}
\usepackage[english]{babel}    
\usepackage{graphicx}
\usepackage{enumitem}
\usepackage{amssymb}
\usepackage{hyperref}
\usepackage{amsthm}
\usepackage{appendix}
\usepackage{tensor}
\usepackage{physics}
\usepackage{bbold}
\usepackage{amsmath,amsfonts,amssymb}
\usepackage{placeins}
\usepackage{xcolor}

\hypersetup{
    colorlinks,
    citecolor=blue,
    filecolor=black,
    linkcolor=blue,
    urlcolor=black
}

\newcommand{\smeq}{ \! = \!}

\newcommand{\tria}{{\mathcal{I}}_{\alpha}} 
\newcommand{\ria}{r_{I,\alpha}}
\newcommand{\ngamin}{{n^\gamma_{\alpha ,\min}}}
\newcommand{\namin}{{n_{\alpha ,\min}}}

\definecolor{ao}{rgb}{0.0, 0.5, 0.0}   

\usepackage[normalem]{ulem}
\usepackage{cancel}

\numberwithin{equation}{section}

\begin{document}

\title{Mean Field Game Approach to Non-Pharmaceutical Interventions in a Social Structure model of Epidemics}



\author{Louis Bremaud}
\affiliation{Universit\'e Paris-Saclay, CNRS, LPTMS, 91405 Orsay, France}

\author{Olivier Giraud}
\affiliation{Universit\'e Paris-Saclay, CNRS, LPTMS, 91405 Orsay, France}
\affiliation{MajuLab, CNRS-UCA-SU-NUS-NTU International Joint Research Unit, Singapore}
\affiliation{Centre for Quantum Technologies, National University of Singapore, Singapore}

\author{Denis Ullmo}
\affiliation{Universit\'e Paris-Saclay, CNRS, LPTMS, 91405 Orsay, France}

\begin{abstract}
The design of coherent and efficient policies to address infectious diseases and their consequences requires to model not only epidemics dynamics, but also individual behaviors, as the latter has a strong influence on the former. In our work, we provide a theoretical model for this problem, taking into account the social structure of a population. This model is based on a Mean Field Game version of a SIR compartmental model, in which individuals are grouped by their age class and interact together in different settings. This social heterogeneity allows to reproduce realistic situations while remaining usable in practice. In our game theoretical approach, individuals can choose to limit their contacts by making a trade-off between the risks incurred by infection and the cost of being confined. The aggregation of all these individual choices and optimizations forms a Nash equilibrium through a system of coupled equations that we derive and solve numerically. The global cost born by the population within this scenario is then compared to its societal optimum counterpart (i.e.\ the cost associated with the optimal set of strategies from the point of view of the society as a whole),  and we investigate how the gap between these two costs can be partially bridged within  a {\em constrained Nash equilibrium} for which a governmental institution would, under specific conditions, impose ``partial lockdowns'' such as the ones that were imposed during the Covid-19 pandemic. Finally we consider the consequences of the finiteness of the population size $N$,  or of a time $T$ at which an external event (e.g.\ a vaccine) would end the epidemic, and  show that the variation of these parameters could lead to {\em first order phase transitions} in the choice of optimal strategies. In this paper, all the strategies considered to mitigate epidemics correspond to non-pharmaceutical interventions (NPI), and we provide here a theoretical framework within which guidelines for public policies depending on the characteristics of an epidemic and on the cost of restrictions on the society could be assessed.
 \end{abstract}

\date{\today}

\maketitle

\setcounter{tocdepth}{1}
\makeatletter
\def\l@subsubsection#1#2{}
\makeatother
\tableofcontents
 
\section{Introduction}

\begin{figure}[htb]
    \centering
    \includegraphics[scale=0.6]{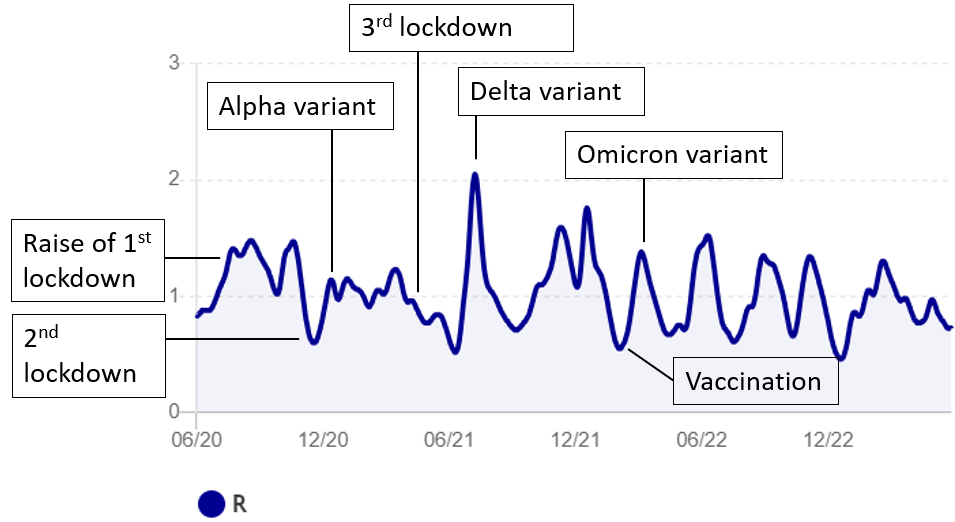}
    \caption{Evolution of $R_\text{eff}$ in France during the Covid-19 pandemic between June 2020 and June 2023.  $R_{\text{eff}}$ corresponds to the effective reproduction number of the virus, that is, the average number of people to which the virus is transmitted by a sick individual. If $R_{\text{eff}} > 1$, the epidemic grows, and it decreases if $R_{\text{eff}} < 1$. We see that there are very significant variations of $R_{\text{eff}}$ which range from $0.6$ to $2$.  We marked on the figure some peaks and valleys that have clearly identified origins (Data from ``Santé Publique France'', author: Guillaume Rozier [\url{https://covidtracker.fr}].) 
    \label{fig:R_eff_France}}
\end{figure}

As  our history with Covid-19 has made rather explicit, modeling as precisely as possible  the dynamics of epidemics is crucial if one wishes to design public policies able to mitigate effectively their negative impact.   One major difficulty encountered toward this goal is that, most often, the parameters one would naturally choose to build such models have significant, and sometimes very fast, variations.  This is illustrated for instance by the graph plotted in Fig.~\ref{fig:R_eff_France}, which shows the time dependence of $R_\text{eff}$, the average number of people to which the virus is transmitted by a sick individual, for the Covid-19 pandemic in France. 

The figure reveals that there are huge variations of $R_{\text{eff}}$ over time. Some of them can easily be associated with known events (lockdown, new variant, etc) but some other remain unexplained.  Indeed, $R_\text{eff}$ is impacted by many phenomena, such as natural immunity, vaccination, but also by behavioral changes that have important consequences on the spreading of the disease. While data such as immunity or vaccination rate are taken into account in even the most basic models, this is not the case for the evolution of social interactions.

However, these modifications of social behavior, either under governmental influence or because people change their individual habits, significantly affect  epidemics dynamics. These individual or collective strategies against the virus sometimes prevented a health disaster  \cite{Impact_NPI_Imperial} by significantly decreasing the total number of infected people and the time at which the peak occurs \cite{Impact_NPI_Imperial, Impact_saniraty_system}. As a counterpart, they had  significant worldwide negative impact, for the  economy \cite{world_economic_impact}, or in terms of health (as medical acts had to be postponed), time, money, social interactions, psychological pressure \cite{social_impact_covid} (domestic violence, depression), etc, which in turn could increase the stress on the sanitary system \cite{Impact_saniraty_system}.  In such a context, any policy or any individual decision must consider the trade-off between the cost of reducing social interaction and the cost of the epidemic; see for instance  \cite{Optimising_cost_effectiveness, morton_wickwire_1974, optimal_isolation_policies}, where realistic impacts and constraints on the quarantine and isolation strategies have been considered, and  \cite{threshold_epidemics_2012, Individual_behavior_physrev} where the  individual behavioral response to isolation policy has been investigated. This individual response is of course greatly influenced by cultural habits together with social, economic, religious needs of the population. 

In models currently used to describe the propagation of epidemics, social interactions are often described by constant parameters, or at best by time-dependent parameters  which are {\em extrinsic}, in the sense that their time evolution is not predicted by the model itself, but ideally obtained from epidemic data  \cite{Wynantsm1328, Impact_NPI_Imperial}. However,  given  the amplitude and time scale of these variations, and in spite of the large amounts of data used, exploiting these data involves a lot of guesswork and lead to  predictions \cite{Covid_France_science, guan2020modeling} which could be inaccurate, especially on long time scales.

To overcome these difficulties, one needs to introduce models for which the {\em extrinsic} parameters have no time dependence (at least on the time scale of the epidemic), and which can therefore be fitted in a reliable way on field data.  On the other hand, all time-dependent  parameters, and in particular the ones modeling social interactions, should  be {\em intrinsic}, in the sense that their dynamics should be predicted by the model. This naturally calls for a game theoretical approach (for a review, see \cite{huang2022game}). Here we will follow an approach known as mean field game theory. 

Introduced by Lasry and Lions almost two decades ago \cite{Lions_MFG, Lions_optimal_control, Lions_stationnaire} and independently by Huang, Malhamé and Caines \cite{Huang_Malhame_MFG}, Mean Field Games (MFG) focus on the derivation of a Nash equilibrium within a population containing a large number of individuals.  Readers may refer to \cite{Dynamic_Game_theory_Caines,continous_time_2013,Probabilistic_theory_MFG_Carmona} for a complete mathematical description, and to \cite{Ullmo_Quadratic_MFG, MFG_Ullmo_Shrodinger} for an introduction aimed at physicists.  Applications of MFG include finance \cite{Gueant_MFG_applications}, economics \cite{MFG_Bertrand_Cournot}, crowd modeling \cite{Pedestrians_Butano}, and opinion dynamics \cite{Opinion_dynamics_MFG} among many others.  


The introduction of MFG models to describe epidemics dynamics has been first used a decade ago by Reluga {\it et al.}~\cite{reluga2010game} about social distancing. Mean Field Games have been then used to describe vaccination rates, which appears to be an extrinsic parameter with a dynamics mainly influenced by individuals choices. Pioneers on this matter are Laguzet {\it et al.}~\cite{Turinici_vaccination} (see also \cite{Turinici_vaccination, hubert2018nash, salvarani2018optimal}). Recently, a similar approach has been proposed by Elie  {\it et al.}~in \cite{Turinici_contact_rate_SIR_simple} to study the impact of individual decisions regarding distancing and isolation, that is, to study human impact on the dynamics of the epidemic (see \cite{olmez2022modeling, olmez2022does} for a mathematical perspective). An extensive review of recent progresses in this new field can be found in \cite{roy2023recent}. 

The significant advances made in \cite{Turinici_contact_rate_SIR_simple} deserve to be push further, on more complex models, to be relevant from a practical, public policy point of view. The goal of this paper is to show that MFG models can implement a high degree of complexity into the differentiated response of individuals towards an epidemic. In particular, these can include a description of the social structure of the society in which the epidemic develops. Furthermore we shall see that with our MFG approach, questions of direct practical importance, such as defining the best government strategy in terms of the timing of lockdowns, can be addressed. This second part of our work is refereed, in the literature, as the non-pharmaceutical interventions (NPI) strategies to mitigate epidemics. Preliminary results can be found in \cite{Bremaud_Ullmo_PhysRevE}.

The manuscript is organized as follows. In Section \ref{section:model}, we introduce the SIR model with a social structure on which we base our discussion. In Section \ref{section:MFG}, we implement the MFG paradigm on this model, that is, we present the individual optimization scheme and its consequences at the society scale,  and find the corresponding Nash equilibrium. 
In Section \ref{section:epidemics dynamics} we discuss the resulting epidemics dynamics.  After introducing the form of the cost function and our choice of parameters in Section~\ref{section:cost_function}, we compute the corresponding Nash equilibrium in Section  \ref{section:Nash}.  We then consider in Section \ref{section:Nash_constraints} a modified Nash equilibrium associated with constraints (partial lockdowns) imposed by a centralized authority, then in Section~\ref{section:societal_optimum} the societal optimum, which corresponds to the situation  obtained when a global planner control perfectly  the behaviour of each agent in order to minimize the total costs borne by the society, and in Section \ref{section:Cost_comparison}, we compare the different scenarios. 
In Section \ref{section:strategies} we then consider the possibility that the total size $N$ of the population, or that the final time $T$ of the epidemic dynamics (associated with an external event such as the expected occurrence of a vaccine) could be finite.  We show that as a function of these parameters, one observe {\em first order phase transitions} where the optimal strategies exhibit discontinuous changes, and discuss the specific character of the different phases. Finally concluding remarks are assembled in Section \ref{section:conclusion}.  Some mathematical and numerical details, as well as  a more general exploration of the parameter space of our model, are gathered in the Appendix.

\section{A social structure based modeling of the epidemics dynamics}
\label{section:model}
\subsection{The SIR model}

Since the early twentieth century, many models have been proposed to model epidemic dynamics, one of the simplest being the SIR (Susceptible-Infected-Recovered) compartment model \cite{SIR_Mckendrick} and its variations \cite{mathematics_infectious_diseases}. Recently, this model has been refined to take into account the structure of  social contacts \cite{Inferring_social_structure, mistry2020inferring}, as well as spatial or geographic aspects of the dynamics \cite{spatiotemporal_dynamics_epidemics, Nature_social_networks}.

The SIR model is defined as follows. Individuals can be in three possible states $x=s,i$ or $r$, with $s$=``susceptible'', $i$=``infected'' and $r$=``recovered''.
Starting from some initial configuration at $t=0$, one then assumes that the evolution of the system is Markovian. Between times $t$ and $t+dt$, individuals can switch from one state to another with a certain probability, which depends on their contact rate with the rest of the population and of the status of people they meet. In a population composed of $N$ individuals, the probability for an individual $k$ to have contact with another individual $l$  during the interval $[t,t+dt[$ is $\frac{1}{N} \chi_k(t) dt$, with $\chi_k(t)$ a (possibly time dependent) given parameter corresponding to the total contact rate of the individual $k$. We make the assumption that all individuals can be met by $k$ with equal probability (in other words, the population considered from the point of view of $k$ is homogeneous). If individual $l$ is infected and $k$ susceptible, then there is a probability $q$ that the disease be transmitted from $l$ to $k$ upon contact. Lastly, infected individuals have a probability $\xi dt$ to recover from their illness during the interval $[t,t+dt[$, after which they are immune to the disease. 

Noting $S(t)$, $I(t)$ and $R(t)$, respectively, the relative proportion of susceptible, infected, and recovered individual at time $t$ (thus $S(t)+I(t)+R(t)=1$), and $\langle S(t) \rangle$, $\langle I(t) \rangle$ and $\langle R(t) \rangle$ the corresponding average over realizations of the Markov process, the evolution of the epidemic is governed by the system of equations \cite{SIR_Mckendrick}
\begin{equation}
\label{eq:SIR}
\begin{aligned}
\langle \dot{S}\rangle &= - q \chi(t) \langle S(t) \rangle \langle I(t)\rangle \\
\langle \dot{I}\rangle &= q \chi(t) \langle S(t)\rangle \langle I(t)\rangle -  \xi \langle I(t)\rangle  \\
\langle \dot{R}\rangle &=   \xi \langle I(t) \rangle\,.
\end{aligned}
\end{equation}
This system of equations is almost a century old \cite{SIR_Mckendrick}; we derive it for completeness in appendix~\ref{app:SIReqs} to prepare for the slightly more involved situation that we are going to consider in this paper.  Let us highlight here the two main underlying hypotheses of the derivation of \eqref{eq:SIR}:  i) the total contact rate of individual $k$, $\chi_k(t)$, only depends on the state $x_k(t)$ of the individual $k$, which means that $\chi_k(t) = \chi_{x_k(t)}(t)$; additionally, as only susceptible individuals can be contaminated, we denote this contact rate by $\chi(t)$; ii) $N$ is large enough to consider the states of two randomly chosen individuals $k$ and $l$ as independent. We shall keep both these hypotheses to derive dynamical equations for our model introduced in Section \ref{SIRsocial}; while hypothesis ii) is rather harmless in practice where $N$ is large, hypothesis i) is an important assumption which can be discussed in practice. 

\begin{figure}[ht]
    \centering
    \includegraphics[scale=0.5]{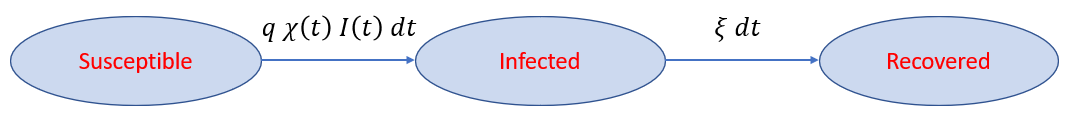}
    \caption{Illustration of the Markov process for the classic SIR model with the transition rates to move from one state to another between time $t$ and $t+dt$. An individual susceptible at $t$ has a probability $q \chi(t) I(t) dt$ to become infected (we omit the bracket notation on $I$). If this individual is already infected at $t$, she will have a constant probability $\xi dt$ to recover from the disease.}
    \label{fig:markovprocess}
\end{figure}

Figure \ref{fig:markovprocess} summarizes the process that drives an individual from state $s$ to $i$ to $r$. The system of equations \eqref{eq:SIR} only involves average quantities $\langle S\rangle, \langle I\rangle$ and $\langle R\rangle$, which are  determined as solutions of the system. Furthermore it is characterized by two extrinsic parameters, the recovery rate $\xi$ and the product of the contact rate $\chi(t)$ by the probability $q$ of transmitting the disease, which must be obtained from observation data \cite{guan2020modeling}. For virus epidemics like Covid-19, with a very fast dynamics, this is a challenging task. Major efforts have been invested by the epidemiologist community to extract these parameters, or their counterpart in more complex models, from the actual data observed on the field.  While $\xi$ is mainly fixed by biological considerations, and considered constant in time in the present model, the contact rate  $\chi(t)$ on the other hand depends a lot on the agent's behavior, that is, how social they are (or are allowed to be); that behavior may vary strongly with time, and in a way that may depend on the dynamics of the epidemic itself.  A consequence of this retroaction is that it is essentially impossible to fit the time dependence of $\chi(t)$ on past data. In models used to advise public policies, this time dependence is thus either simply ignored, or involves a lot of guesswork \cite{Covid_France_science}, leading to predictions that can be trusted only for a rather short amount of time \cite{guan2020modeling}(see nevertheless \cite{Mobility_data_covid,Impact_NPI_Imperial}).

What we discussed above is the simplest version of the SIR model. A number of variations can be found in the literature, that aim to gain in precision. The most common ones are the SIRD model (D for deceased \cite{SIRD}), SIRV (V for vaccination \cite{SIRV}), MSIR (M for maternally derived immunity \cite{mathematics_infectious_diseases}), SIRC (C for carrier but asymptomatic \cite{SIRC}), or SEIR models (E for exposed class \cite{SEIR}) to name a few - see \cite{mathematics_infectious_diseases} for a more detailed literature on the subject of compartmental models. However, there are two essential limitations of these models: they assume that the population is entirely uniform, and they take parameters such as the contact rates as extrinsic. 

Let us expand slightly on these two issues. The first limitation is that these models assume a homogeneous population: all individuals are expected to act in the same way, have the same contact rate with all other individuals (in a given compartment), and behave similarly with respect to the epidemic. Of course this is not true, and social heterogeneity has an important impact on epidemics modeling. As an example, epidemics inside schools have a different and faster dynamics than can be expected from the SIR model, because children have a lot of contacts with each other and they live together during a long part of the day. To address this issue, SIR models with a structure of social contacts were proposed in \cite{Inferring_social_structure} and \cite{mistry2020inferring} to get a more detailed description of the society at a mesoscopic scale. We will address that limitation by introducing a refined model in Section \ref{SIRsocial}. The second limitation of SIR models, already discussed in the introduction, is that the contact rates are extrinsic parameters, fixed at the beginning of the dynamical process. A more realistic approach is to consider that people change their behavior as the epidemic unfolds, so that contact rates should be updated according to the dynamics of the epidemic. We shall circumvent this issue by taking a MFG approach to our model with a social structure in Sec.~\ref{section:MFG}, where contact rates will become intrinsic parameters, co-evolving with the epidemic.

\subsection{The SIR model with social structure}
\label{SIRsocial}

\subsubsection{Description of the model}
\begin{figure}[!ht]
    \centering
    \includegraphics[scale=0.5]{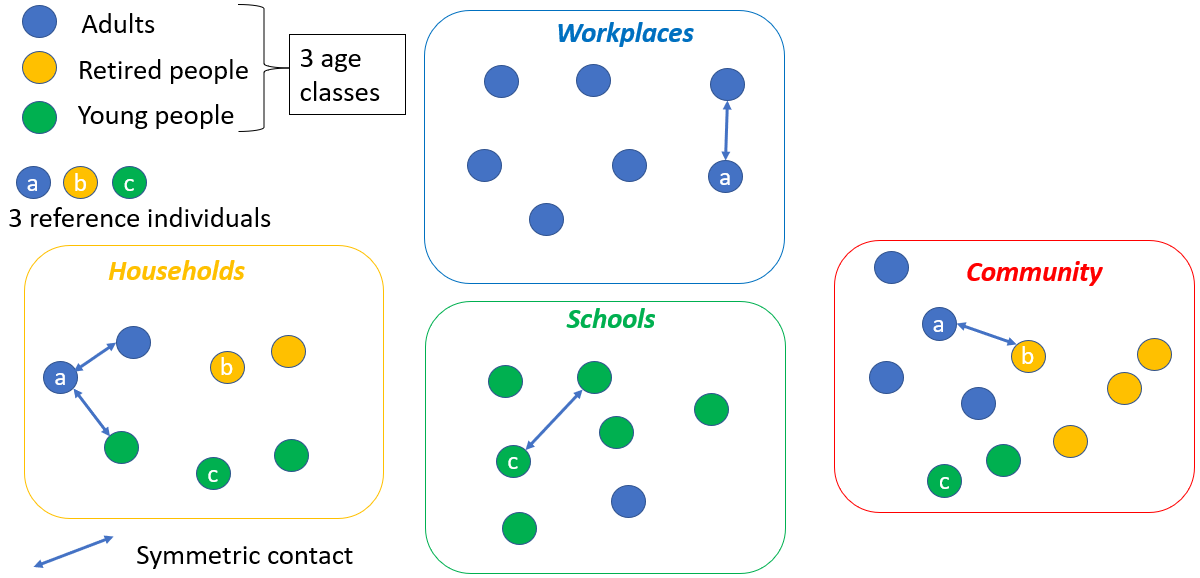}
    \caption{Graphical illustration of the social structure we implemented. A reference individual ($a$, $b$ and $c$ for each age class) will have (symmetric) contacts in each setting, with different type of individuals (more adults at workplaces, more children at school, etc). The precise structure of interactions is given by the matrix of contacts $M_{\alpha \beta}$ introduced in the following section.}
    \label{fig:social_structure}
\end{figure}

We now introduce a SIR model with a social structure, in the spirit of \cite{Inferring_social_structure}. In this model, rather than taking society as monolithic, we consider a refined description of social contacts. Namely, we introduce three age classes : young, adult and retired, and we assume that individuals have contacts with one another in four different settings:   schools,  households,  community and  workplaces; of course a larger number of age classes and settings could easily be implemented. Interactions between individuals may differ between different age classes and  in different settings. As a result, the dynamics of the epidemic will be different in each subcategory. The structure of the population is illustrated in Fig.~\ref{fig:social_structure}. We assume the total size of the population, $N$, large. 

\begin{figure}[!ht]
    \centering
    \includegraphics[scale=0.6]{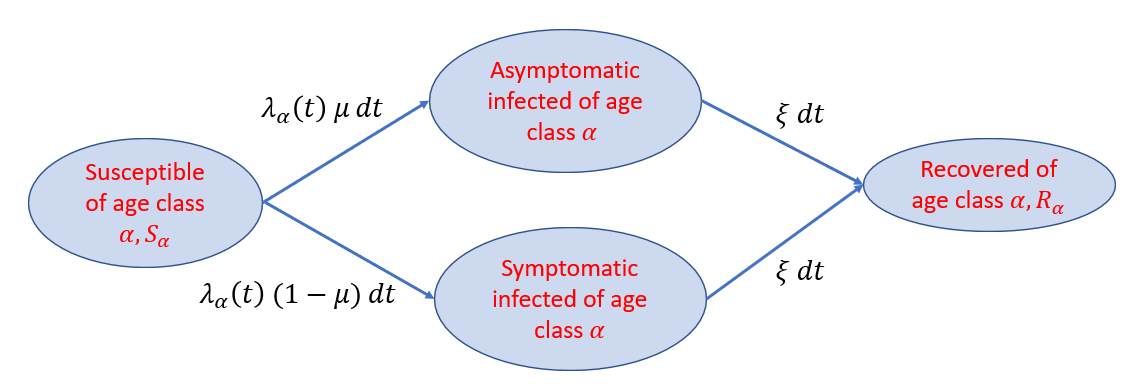}
    \caption{Graphical illustration of the particular SIR model we use. An individual infected at time $t$ has a probability $\mu$ to be asymptomatic and $1-\mu$ to be symptomatic. The force of infection $\lambda_\alpha$ is derived in the following section and drives the probability of infection $\lambda_\alpha dt$. Then, all individuals have a constant recovery rate $\xi$ to recover from the disease.}
    \label{fig:markov_process_SIIR}
\end{figure}

Secondly, we have in mind a game-theory framework in which each individual has its own specific behavior. In the present context, this behavior is described by an individually chosen contact rate which may depend on the epidemic situation, age, health status, own aversion to risk, and so on. In order to account for these preferences, each individual associates a cost to the constraints of being deprived of social contacts, being ill, etc, and at any given time chooses the contact rate that minimizes the expected total future cost. In our approach we concentrate on susceptible individuals, which are assumed to minimize a cost function $C$ (defined in the next section). The behavior of infected individuals is fixed as a hypothesis of the model, and can vary from a completely egoistic approach, where they stop doing any effort, to a very altruistic one where they completely isolate from the rest of population. To make things more concrete, we choose this latter option, but assume that a fraction $\mu$ of the population is asymptomatic (they do not know if they are infected or not) and hence  behave as susceptible, while the other fraction $1-\mu$ is symptomatic and stay home to protect others. This additional status (symptomatic or asymptomatic) is random in the population and is fixed at the beginning of the epidemic. Therefore, the epidemic is only spread by individuals who are both asymptomatic and infected. They represent a fraction $\mu I(t)$ of the population. We summarize our model in Fig.~\ref{fig:markov_process_SIIR}.

\subsubsection{Definition of the parameters}

In our model, following \cite{Inferring_social_structure}, interactions between individuals depend on two factors: the setting $\gamma\in\{$school,  household,  community, workplace$\}$ in which they meet, and their age class $\alpha\in\{$young, adult, retired$\}$. We denote by $N_\alpha$ the number of individuals in class $\alpha$. We first consider the simple case of a single setting where interactions only depend on age class, which will be labeled by the Greek letters $\alpha$ or $\beta$; extension to the case of multiple settings is then straightforward.

\begin{figure}[!ht]
    \centering
    \includegraphics[scale=0.5]{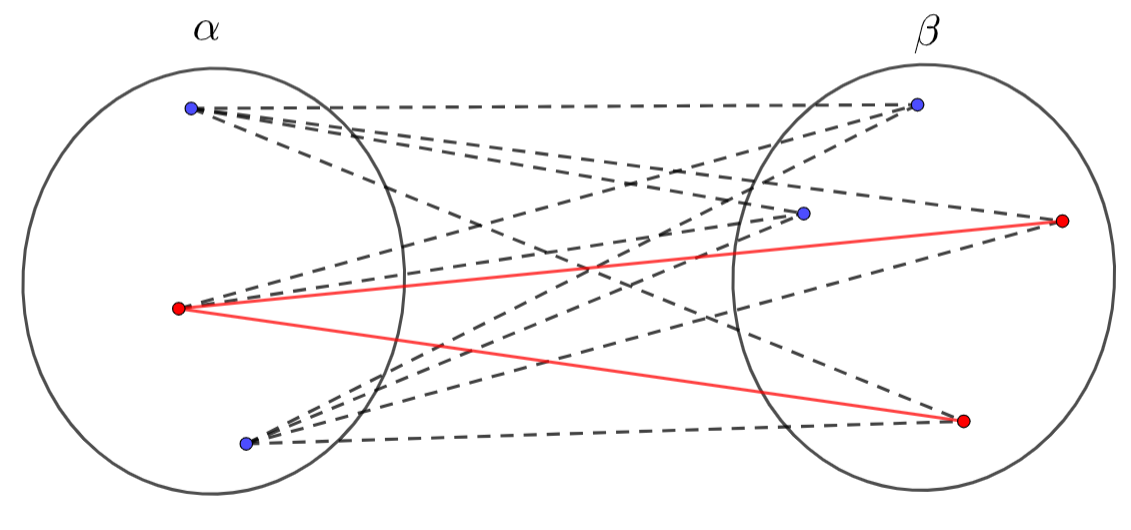}
    \caption{Graphical illustration of the interactions in our model. Two age classes $\alpha$ and $\beta$ are represented, here with $N_\alpha = 3$ individuals of age class $\alpha$ and $N_\beta = 4$ of class $\beta$. Each vertex is either ``active'' (in red) if the corresponding individual is willing to have contact with the other class, or ``inactive'' (in blue). The $N_\alpha N_\beta$ possible contacts are represented in dashed black lines, and effective contacts between pairs of active individuals are red solid lines. Here we have $w_{\alpha\beta}N_\alpha=1$ active individual of age class $\alpha$ and $w_{\beta\alpha}N_\beta=2$ active individuals of age class $\beta$, which gives $w_{\alpha \beta} = \frac{1}{3}$ and $w_{\beta \alpha} =  \frac{1}{2}$. The probability for a randomly chosen pair to be in contact is $W_{\alpha \beta} = w_{\alpha \beta} w_{\beta \alpha} = \frac{1}{6}$. The average number of contacts with $\beta$ for an individual $a\in\alpha$ is $M_{\alpha \beta}=W_{\alpha \beta}N_\beta=\frac{2}{3}$. Similarly, $M_{\beta \alpha} = \frac{1}{2}$. The total number of contacts between the two classes, corresponding to the number of red links in the graph, is given by the sum rule $N_\alpha M_{\alpha \beta} = N_\beta M_{\beta \alpha} = 2$.}
    \label{fig:willingness_illustration}
\end{figure}

For two given age classes $\alpha$ and $\beta$ we define $W_{\alpha \beta}dt$ as the probability for a pair of individuals $a\in\alpha, b\in\beta$ drawn at random to be in contact during a time interval $dt$. This means that among all possible $N_\alpha N_\beta$ pairs, only $W_{\alpha \beta}N_\alpha N_\beta dt$ encounters occur during $dt$. This is illustrated by the graph of Fig.~\ref{fig:willingness_illustration}; it is similar to Erdös-Renyi graphs, where each potential edge is realised with some probability. In the present case, all potential edges between vertices from one class to the other are realized with some probability that depends on the two classes they connect. A given individual $a\in \alpha$ encounters on average a number $M_{\alpha\beta}dt=W_{\alpha\beta}N_\beta dt$ of individuals of class $\beta$ during $dt$.

A natural assumption, in the spirit of compartmental models, is that behaviour of individuals toward different age classes is differentiated, but that a given age class is considered homogeneous from the point of view of an individual. That is, an individual $a\in \alpha$ can decide whether she chooses to encounter members of class $\beta$ or not, but does not decide which individuals she may encounter in that class. In other words, any individual $a\in \alpha$  willing to meet someone from class $\beta$ will possibly meet all individuals from class $\beta$ who themselves are willing to meet individuals from class $\alpha$. At each time,  an individual $a\in \alpha$ can decide whether she is open or close to interactions with class $\beta$. Let us denote by $w_{\alpha\beta}\in [0,1]$ the fraction of  individuals $a\in \alpha$ open to meet people from class $\beta$. The willingness $w_{\alpha\beta}$ thus indicates the probability of an individual $a$ taken at random in $\alpha$ to be open to contacts with class $\beta$. There are $w_{\alpha\beta}N_\alpha$ individuals $a\in \alpha$ willing to meet people with class $\beta$, and $w_{\beta\alpha}N_\beta$ individuals $b\in \beta$ willing to meet people from class $\alpha$. A contact becomes effective (i.e.~occurs with probability $dt$ in the interval $[t,t+dt[$) only if both individuals are willing, and therefore among all $N_\alpha N_\beta$ possible links between $\alpha$ and $\beta$, only $w_{\alpha\beta}N_\alpha\times w_{\beta\alpha}N_\beta dt$ are realized during $dt$.  As mentioned above, the number of pairs effectively realized can also be expressed as $W_{\alpha \beta}N_\alpha N_\beta dt$, hence $W_{\alpha\beta}=w_{\alpha\beta}w_{\beta\alpha}$ (and $W_{\alpha\beta} $ is actually symmetric, as it should be).

We make the assumption that the contact willingness of an individual $a\in\alpha$ with class $\beta$ is $w_{\alpha \beta}(0)$ at the onset of the epidemic. In a game-theoretical setting, agents can adapt their behaviour individually at each time step, so as to optimize a certain foreseeable cost. Individual $a$ can control her willingness as $w_{a \beta}(t) = n_a(t) w_{\alpha\beta}(0)$, that is, her initial willingness is modulated by a time-dependent coefficient $n_a(t)$ which measures the effort made by $a$ to limit her contacts with others. 
For simplicity we assumed that this effort is independent of $\beta$, but a $\beta$ dependence can easily be implemented to this model and only slightly changes the equations. We additionally assume that $n_a(t) \in [\namin,1]$, with $\namin$ the maximum effort that can be expected from an agent in class $\alpha$; the upper bound 1 corresponds to the natural assumption that the epidemic situation can only reduce the initial willingness. In the SIR context we shall assume that all individuals of a given age class behave in the same way, so that all $n_a(t)$ are taken equal to some $n_{\alpha}(t)$; however, in the MFG setting of Section \ref{section:MFG} we will need to make a distinction between the individual behavior $n_a(t)$ and the generic behavior $n_{\alpha}(t)$. In this section we use the notation $n_a(t)$, however we turn to the notation $n_{\alpha}(t)$ whenever the distinction is irrelevant.

The time evolution equations will also depend on three global parameters $q$, $\mu$ and $\xi$, taken to be independent of individuals and time. Firstly, as in the simple SIR model, contact does not necessarily imply contamination, and we denote by $q$ the probability of transmission of the virus per effective contact between a susceptible and an infected individual. Secondly, as mentioned above, our choice is that contamination occurs only via the fraction $\mu$ of the population which is asymptomatic, that is, which behaves as susceptible whatever their epidemic status; in our study we will take a small $\mu$ ($\mu \ll 1$). Thirdly, the recovery rate $\xi$ is supposed to be constant for all individuals. Finally, in the multiple-setting scenario, the quantities $M_{\alpha \beta}$, $W_{\alpha \beta}$, $w_{\alpha \beta}$ and $n_{\alpha}$ may differ from one setting to another, and we will index them by the superscript $\gamma$. In Table \ref{table:parameters_def} we summarize the set of parameters we used in our model.

\subsubsection{Time evolution equations}

\begin{table}
\begin{center}
\begin{tabular}{|| c | c | c || } 
\hline
Variable & Mathematical definition & Physical definition \\ 
\hline
$\;$ & $\;$ &
\\
$S_\alpha, I_\alpha, R_\alpha$ & Equations \eqref{eq:SIR-ss} & Proportion of susceptible, infected and recovered individuals ($\alpha$) \\
$\;$ & $\;$ &
\\
$N_\alpha$ & - & Number of individuals of age class $\alpha$ \\
$\;$ & $\;$ &
\\
$K_\alpha$ & - & Proportion of individuals of age class $\alpha$ \\
$\;$ & $\;$ &
\\
$q$ & - & Probability of transmission per contact \\
$\;$ & $\;$ & \\
$\xi$ &- &  Recovery rate \\ 
$\;$ & $\;$ & \\
$M^\gamma_{\alpha \beta}$ & $M^\gamma_{\alpha \beta} = W^\gamma_{\alpha \beta} N_{\beta}$ & Average frequency of contacts ($M^{\gamma(0)}_{\alpha \beta}$ in absence of epidemic)\\
$\;$ & $\;$ & \\
$W^\gamma_{\alpha \beta}$ & $W^\gamma_{\alpha \beta} = w^{\gamma }_{\alpha \beta} w^{\gamma}_{\beta \alpha}$ & Symmetric willingness of contacts between two age classes $\alpha$ and $\beta$  \\
$\;$ & $\;$ &
\\
$w^{\gamma }_{a \beta}$(t) & $w^{\gamma }_{a \beta}(t) = w^{\gamma (0)}_{\alpha \beta} n^{\gamma}_{a}(t)$ & Willingness of an individual $a$ to have contact with someone of class $\beta$ \\
$\;$ & $\;$ &
\\
$n^{\gamma}_{\alpha}(t)$ & $n^{\gamma}_{\alpha}(t) \in [\ngamin,1] $ & Coefficient modeling the reduction of contact willingness \\
$\;$ & $\;$ &
\\
$x_a(t)$ & $x_a(t) = s, i, r$ & State of a individual $a$ at $t$ (susceptible, infected or recovered) \\
$\;$ & $\;$ &
\\
$\mu$ & - & Proportion of asymptomatic individuals in the population \\
$\;$ & $\;$ &
\\
\hline
\end{tabular}
\end{center}
\caption{Parameters used in our model and their definitions.}
\label{table:parameters_def}
\end{table}

We now derive the time evolution equations of the epidemic quantities for this model. 
The fraction of susceptible (resp.~infected, recovered) individuals in class $\alpha$ is $S_\alpha$ (resp.~$I_\alpha, R_\alpha$), with $S_\alpha+I_\alpha+R_\alpha=1$. In order to establish the mean-field equations, we single out a reference individual $a\in\alpha$ who is susceptible at time $t$ and has status $x_a(t)=s,i$ or $r$ at subsequent times.
This reference individual has her  own time-dependent strategy $n_a(t)$ and willingness $w_{a\beta}(t)=n_a(t)w^{(0)}_{\alpha\beta}$. Let $b\in \beta$ be an individual of class $\beta$, whose willingness to meet class $\alpha$ is
$w_{b\alpha}(t)=n_b(t)w^{(0)}_{\beta\alpha}$. In order for $a$ to be contaminated by $b$ during $[t,t+dt[$, $b$ must be infected {\em and} asymptomatic, and $a$ and $b$ must meet; contamination then occurs with probability $q$. The probability that $a$ become infected by $b$ during $[t,t+dt[$ is therefore
\begin{equation}
 P_{ab}\Bigl(x_a(t+dt) \smeq i \, | \, x_a(t) \smeq s\Bigr) = q  n_a(t) n_b(t) \frac{M^{(0)}_{\alpha \beta}}{N_{\beta}}     \mu\delta_{x_b(t),i}  dt \; \; ,
\end{equation}
where we used the fact that $w^{(0)}_{\alpha\beta}w^{(0)}_{\beta\alpha}=W^{(0)}_{\alpha \beta}=M^{(0)}_{\alpha \beta}/N_\beta$ (see Table \ref{table:parameters_def}). Taking the sum over all $b\in\beta$ we get (at first order in $dt$) the probability $P_{a\beta} \equiv \sum_{b=1}^{N_\beta} P_{ab} $ that $a$ become infected by someone in $\beta$ during $[t,t+dt[$,
\begin{equation}  \label{eq:average_proba}
 P_{a\beta}\Bigl(x_a(t+dt) \smeq i \, | \, x_a(t) \smeq s\Bigr)  
 =   q\,  n_a(t)\frac{M^{(0)}_{\alpha \beta}}{N_{\beta}} \sum_{b=1}^{N_\beta} n_b(t) \mu \delta_{x_b(t),i} dt  \, .
\end{equation}
We then follow the same reasoning as in the SIR case (see Eq.~\eqref{eq:average_S}). 
Averaging over all individuals $a\in\alpha$ and over realizations of the Markov process, and summing over age classes $\beta$, we obtain 
\begin{equation}
\label{eq:SIR_ss_average}
   \langle S_\alpha(t+dt) \rangle - \langle S_\alpha(t) \rangle = - \Big\langle  \frac{1}{N_\alpha} \sum_{a=1}^{N_\alpha} \delta_{x_a(t),s} \sum_\beta P_{a\beta}(x_a(t+dt) = i \ | \ x_a(t) = s )\Big\rangle \; .
\end{equation}
Using \eqref{eq:average_proba}, and applying the same argument of independence between events as in \eqref{eq:average_S_2}, we get
\begin{equation}
\label{eq:SIR_ss_average2}
    \frac{d \langle S_\alpha(t) \rangle}{dt} = - q \sum_\beta    M^{(0)}_{\alpha \beta} \Big\langle \frac{1}{N_\alpha} \sum_{a=1}^{N_\alpha} n_a(t) \delta_{x_a(t),s} \Big\rangle \Big\langle \frac{1}{N_{\beta}} \sum_{b = 1}^{N_\beta}n_b(t) \mu \delta_{x_b(t),i}  \Big\rangle \, .
 \end{equation}
The mean-field approach then consists in subsuming all individual behaviors $n_a(t)$ of susceptible individuals $a$ of a given class $\alpha$ under a single strategy $n_{\alpha}(t)$; this means that in each age class we neglect the individual variations of $n_a(t)$ around $n_\alpha(t)$. Note that the above equations are valid even without that hypothesis, and we will make use of this in the next section.
One could think of distinguishing the strategies of susceptible agents and that of infected ones; but our hypothesis that only asymptomatic individuals are responsible for contaminations implies that both susceptible and asymptomatic infected agents will behave in the same way, given by the function $n_\alpha(t)$ or $n_\beta(t)$. Equation \eqref{eq:SIR_ss_average2} then becomes
\begin{equation}
\label{eq:SIR_ss_average3}
  \frac{d \langle S_\alpha(t) \rangle}{dt}    =  - q \sum_\beta   n_{\alpha} (t)  n_{\beta}(t) M^{(0)}_{\alpha \beta} \Big\langle S_\alpha(t) \Big\rangle \Big\langle \mu I_\beta(t) \Big\rangle \, ,
\end{equation}
that is, 
\begin{equation}
\label{eq:lambda1}
   \frac{d \langle S_\alpha(t) \rangle}{dt} = - \lambda_\alpha(t)\langle S_\alpha(t) \rangle \,,\qquad \lambda_\alpha(t) \equiv  \mu  q n_{\alpha}(t)\sum_{\beta} n_{\beta}(t) M^{(0)}_{\alpha \beta} \ \langle I_{\beta}(t) \rangle  \,.
\end{equation}
This equation generalizes in a straightforward way when we include different settings $\gamma$ in the model. In that case we have
\begin{equation}
\label{eq:lambda2}
\lambda_\alpha(t) \equiv  \mu q \sum_{\beta=1}^{n_\textrm{cl}} \sum^{n_\textrm{set}}_{\gamma = 1} n^{\gamma}_{\alpha}(t) 
n^{\gamma}_{\beta}(t) M^{\gamma(0)}_{\alpha \beta} \langle I_{\beta}(t) \rangle  \; \; .
\end{equation}
Equation \eqref{eq:lambda1} is the analog of the SIR Eq.~\eqref{eq:chi_mean_field} but in the case of a population with social structure. The two other equations analogous to the system \eqref{eq:SIR} are derived in the same way. Since all epidemic quantities that will be useful in the following sections are quantities averaged over realizations, from now on we shall omit the brackets $\langle \  \rangle$. The system of coupled differential equations for the SIR model with social structure finally reads
\begin{equation}
\label{eq:SIR-ss}
\begin{aligned}
\dot S_\alpha & = -  \lambda_\alpha (t) S_\alpha(t) \; \\
\dot I_\alpha & = \lambda_\alpha (t)   S_\alpha(t) -  \xi I_\alpha(t) \; \\
\dot R_\alpha & =   \xi \ I_\alpha(t)\,.
\end{aligned}
\end{equation}
These equations are the main equations of our SIR model with a social structure. For any given interaction strategies $n^{\gamma}_{\alpha}(.)$ for each age class $\alpha$ and each setting $\gamma$, one can solve \eqref{eq:SIR-ss} and obtain the relative proportion of susceptible, infected and recovered in each class. However, for rational agents interaction strategies should depend on the evolution of the epidemic. To address this interplay, we need the machinery of mean-field games, which we now address.

\section{Mean-field game approach : individual optimization}
\label{section:MFG}

In our model, an individual $a$ can choose at each time the value of her own control parameter $n^{\gamma}_a (t)$, which reflects her desire to have contact with someone in each setting $\gamma$. Individual $a$ can be in one of the three states $s_\alpha, i_\alpha, r_\alpha$, depending of her age class $\alpha$ and on whether she is susceptible, infected or recovered. We denote by $x_{a}(t)$ the state of $a$ at time $t$. We do not make a distinction between susceptible and asymptomatic individuals as far as the calculation of the cost function is concerned, since agents know their infected status only when they are infected and symptomatic. 

In practice, each agent will adjust her control parameter $n^{\gamma}_a(t)$ to minimize her foreseeable cost over a certain time interval. In the mean-field setting, the agent will see individuals in a given age class as indistinguishable; therefore the time-dependent cost function only depends on the stochastic number of agents in the different possible states. Furthermore, since the stochastic realizations  are peaked around the mean because of the large size of the population, an individual realizing the optimization will consider the average (over stochastic realizations) proportions of agents in each state, i.e.~the quantities $S_\alpha(.), I_\alpha(.),R_\alpha(.)$, which in turn depend on the control parameters $n^{\gamma}_\beta(t)$ via Eq.~\eqref{eq:SIR-ss}. In this section, we derive the optimization made by the agents, following in the spirit the work of Turinici \textit{et al.}~in \cite{Turinici_contact_rate_SIR_simple}.

\subsection{Low asymptomaticity \texorpdfstring{$\mu \ll 1$}{mu << 1}}

\label{subsection:littlemu}
Let us first consider the case $\mu \ll 1$. In that case, almost all infected individuals are symptomatic, and thus individuals with no symptoms can estimate their future cost neglecting the probability that they might be infected. Note however that contamination still occurs via the few infected asymptomatic individuals.

Consider a fixed individual $a\in\alpha$. Individual $a$ makes the assumption that all individuals in each age class $\beta$ will follow the same strategy $n^{\gamma}_\beta(t)$. If $a$ has no symptoms at time $t$, he estimates the cost that the epidemic will incur as the sum of two terms~: one which is due to the cost of efforts to avoid infection, and one due to the cost of infection if it happens. This cost depends on the strategies that $a$ will follow in each of the settings $\gamma$. If $a$ becomes infected at some time $\tau >t$, the total cost paid between $t$ and the end of the optimization process at $T$ is 
\begin{equation}
\label{eq:cost_agent}
C_a\left(n^\gamma_a(\cdot),\{n^\gamma_\beta (.) \},t, \tau \right) \equiv \tria(I(\tau)) \mathbb{1}_{\tau < T}+\int_t^{\min(\tau,T)}\!\!\!\!\!\!\!\! f_\alpha \left (n^\gamma_a(s) \right) ds   \; \; .
\end{equation}
This cost is a function of the strategies $n^\gamma_a(\cdot)$ of $a$ in each setting and at each time between $t$ and $\min(\tau,T)$; it also depends on all the strategies $\{n^\gamma_\beta (.) \}$ for all age classes $\beta$ (including $\alpha$) and settings $\gamma$ in the same time interval. 
The first term in Eq.~\eqref{eq:cost_agent} is the total cost of infection $\tria(I(\tau))$ paid by the agent after she is infected. We assume that this cost of infection depends on the age class and on the (average) proportion $I(\tau) \equiv \sum_\alpha K_\alpha I_\alpha(\tau)$ of infected in the population (reflecting the pressure on the sanitary system). The cost depends on all the strategies $\{n^\gamma_\beta (.) \}$ via $I(\tau)$. 
In the second term,  $f_\alpha \left(n^\gamma_a(s) \right)$ measures the cost (both psychological and financial) associated with the limitation of social contacts; this cost can be different according to the age class of the individual, and depends on the behaviour of the individual only. At each time $s$ between $t$ and $\tau$ (the time of infection) or $T$ (if the agent is never infected) the agent will pay a cost $f_\alpha \left(n^\gamma_a(s)\right) ds$;  for $s>\tau$ we have $f_\alpha=0$, as the individual is either infected (in which case the social cost is included in the term  $\tria$) or recovered (as there is no possible new infection in our model).

From the perspective of agent $a$ at time $t$, and since the epidemic propagation is a stochastic process, the time of infection $\tau$ is a random variable that changes from one realization of the epidemic to the other, with some probability distribution $P_a(\tau)$; note this probability also depends on $t$ since the agent has acquired information about whether or not she has been infected in the interval $[0,t]$. The cost in Eq.~\eqref{eq:cost_agent} is thus also a stochastic variable, and at each time $t$, a rational agent should choose her future strategies in each setting $n^\gamma_a(s), s>t$, as the ones that minimize the {\em average} value of $C_a$ over random realizations,
\begin{equation}
\label{eq:avcost}
    C_a\left(n^\gamma_a(\cdot),\{n^\gamma_\beta (.) \},t \right)  \equiv \int_t^\infty d\tau\ P_a(\tau) \ C_a\left(n^\gamma_a(\cdot),\{n^\gamma_\beta (.) \},t, \tau \right)\,,
\end{equation}
where formally we understand $\tau>T$ as an absence of infection (so that we can normalize $\int_t^\infty P_a(\tau) d\tau =1$, and $C_a\left(n^\gamma_a(\cdot),\{n^\gamma_\beta (.) \},t, \tau > T \right) = \int_t^{T}\!\! f_\alpha \left (n^\gamma_a(s) \right) ds$).
We now need to evaluate the probability $P_a(\tau)$ for an agent $a$ who is assumed to follow a specific strategy $n^\gamma_a(\cdot)$. Let $\phi_a(\tau)$ be the corresponding cumulative probability, that is, the probability for $a$ to be infected before time $\tau$. The probability that the infection time for $a$ is between $\tau$ and $\tau+d\tau$ is 
\begin{equation}
\label{eq:infected_at_t}
\phi_a'(\tau) d\tau = P_a(\tau) d\tau  = P\Bigl(x_a(\tau+d\tau) = i_\alpha | x_a(\tau)= s_\alpha\Bigr) \times P(x_a(\tau)  = s_\alpha)\,,
\end{equation}  
where the probability that $a$ is susceptible at time $\tau$ is $P(x_a(\tau)  = s_\alpha)=1-\phi_a(\tau)$. The probability for $a$ to become infected between $\tau$ and $\tau+d\tau$ by someone of class age $\beta$  in the setting $\gamma$ is then given by averaging \eqref{eq:average_proba}, which reads 
\begin{equation}
 P\Bigl(x_a(\tau+d\tau) \smeq i_\alpha \, | \,  x_a(\tau) \smeq s_\alpha \Bigr) = \lambda_a(\tau) d\tau \; ,
\end{equation} 
with $\lambda_a$ given by 
\begin{equation}
\label{eq:lambda_a}
\lambda_a(t) \equiv  \mu q \sum_{\beta=1}^{n_\textrm{cl}} \sum^{n_\textrm{set}}_{\gamma = 1} n^{\gamma}_{a}(t) 
n^{\gamma}_{\beta}(t) M^{\gamma(0)}_{\alpha \beta} I_{\beta}(t)  \; \; ,
\end{equation}
which is the force of infection seen by individual $a$, with her behavior $n^{\gamma}_{a}$ replacing the collective one $n^{\gamma}_{\alpha}$ which appears in Eq.~\eqref{eq:lambda2}. Equation \eqref{eq:infected_at_t} thus leads to $\phi_a'(\tau)=\lambda_a(\tau) (1 - \phi_a(\tau))$, which together with $\phi_a(t)=0$ gives
\begin{equation}
\phi_a(\tau) = 1  - \exp\left(-\int_t^\tau \lambda_a(s) ds\right) \; .
\end{equation}
The average cost \eqref{eq:avcost} then reads
   \begin{equation}
   \begin{aligned}
   C_a\left(n^\gamma_a(\cdot),\{n^\gamma_\beta (.) \},t \right)   &= \int_t^{T} d\tau\ P_a(\tau)\tria(I(\tau))   +\int_t^\infty d\tau\ P_a(\tau) \int_t^{\min(\tau,T)} ds\ f_\alpha \left (n^\gamma_a(s) \right ) \; \\
    &= \int_t^{T} ds\ P_a(s)\tria(I(s)) +\int_t^T  ds\ f_\alpha \left (n^\gamma_a(s) \right ) \int_s^{\infty} d\tau\ P_a(\tau)\; .
   \end{aligned}
\end{equation}
We then use the fact that $\phi_a'(\tau)=P_a(\tau)=\lambda_a(\tau) (1 - \phi_a(\tau))$
to get
  \begin{equation}
\label{eq:final_cost}
   C_a\left(n^\gamma_a(\cdot),\{n^\gamma_\beta (.) \},t \right) = \int_t^{T}   \left[\lambda_a(s) \
      \tria(I(s))+ f_\alpha \left (n^\gamma_a(s)\right)\right](1 - \phi_a(s)) ds\,.
\end{equation}
In the following, we will often use  $C_a\left(n^\gamma_a, t \right)$ for simplicity, but the cost still depends implicitly on all the $n^\gamma_\beta(\cdot)$.

\subsection{Arbitrary asymptomaticity}
\label{subsection:general_mu}
In the general case $\mu \in [0,1]$, the equations change only slightly. As before, only asymptomatic infected individuals participate to the propagation of the disease. Asymptomatic individuals ignore their status, and if infected feel no harm; as a consequence, they will not change their behavior upon contamination at time $\tau$ (thus the integral in \eqref{eq:cost_agent} will extend up to $T$), nor bear the health costs (thus the second term in \eqref{eq:cost_agent} will be zero for them). The cost for asymptomatic individuals thus reads
\begin{equation}
\label{eq:cost_agentasymp}
C_a\left(n^\gamma_a(\cdot),\{n^\gamma_\beta (.) \},t, \tau \right) \equiv \int_t^{T}\!\! f_\alpha \left (n^\gamma_a(s) \right) ds .
\end{equation}
Since the agent ignores whether she is asymptomatic or not, the average cost she anticipates is with probability $(1-\mu)$ the estimated cost \eqref{eq:final_cost} and with probability $\mu$ the cost \eqref{eq:cost_agentasymp} (which is independent of $\tau$), therefore
\begin{equation}
\begin{aligned}
\label{eq:costmu}
 C^\mu_a \left(n^\gamma_a (\cdot),t\right) &= (1-\mu)\int_t^{T}   \left( f_\alpha \left (n^\gamma_a(s)\right )+\lambda_a(s) \
      \tria(I(s))\right)(1 - \phi_a(s)) ds  + \mu \int_t^{T} f_\alpha (n^\gamma_a(s) )  ds \\
& = \int_t^T  \left[ (1-\mu) \lambda_a(s) \tria(I(s)) (1 - \phi_a(s)) + f_\alpha (n^\gamma_a(s)) (1 - (1-\mu) \phi_a(s)) \right] ds  \, .
\end{aligned}
\end{equation}
The term $(1-\mu) \phi_a(s)$ can be interpreted as the probability for an individual of age class $\alpha$ to be infected and symptomatic before $s$, since the two events ``have been infected before $s$'' and ``be symptomatic'' are independent.  In the limit of $\mu \ll 1$, we recover the cost derived before in \eqref{eq:final_cost}; note that to allow an epidemic growth in this limit we assume that $\mu q$  and thus $\lambda_a$ are of the same order in $\mu$ as $\xi$ (the recovery rate), that is, of order $0$ in $\mu$. 

Considering a finite $\mu$ makes notations slightly heavier without  changing qualitatively the dynamics of the epidemics.  Therefore in the following, we will be dealing only with the case $\mu \ll 1$.

\subsection{Hamilton-Jacobi-Bellman equations}
The expected cost at time $t$ for agent $a$ is a function of her own strategy $n_a$ and of the epidemic functions $S(.),I(.),R(.)$. The next step is to solve the optimization problem, that is, find the optimal strategy $n^*_a$ for a given epidemic $S(.),I(.),R(.)$. Following a standard approach in this context \cite{continous_time_2013}, we introduce the {\em value function}

\begin{equation}
\label{eq:value function}
    U_a(t) = 
    \left\{
\begin{aligned}
&\underset{n^\gamma_a(\cdot)}{\min}\, C_a\left(n^\gamma_a(\cdot),t\right) \,,&\quad a \quad \textrm{susceptible at } t\\
&0,&\quad a \quad \textrm{infected at } t.\\
\end{aligned}
\right.
\end{equation} 
This corresponds to the minimal cost that an agent has to pay between $t$ and the end of the game (averaged over random realizations of the game, and assuming that all other players follow some given strategies $n^\gamma_\beta$). Note that in \eqref{eq:cost_agent} we assumed that  the total cost of infection is paid right after infection, so that individuals do not incur any additional cost at later times. The Markov process of the game is described by the following equations, illustrated in Fig.~\ref{fig:markov_process_SIIR}:
\begin{equation}
\label{eq:Markov} 
\left\{
\begin{aligned}
P_a(x_{a}(t+dt) &= i_\alpha | x_{a}(t) = s_\alpha) = \lambda_a(t) dt \\
P_a(x_{a}(t+dt) &= s_\alpha | x_{a}(t) = s_\alpha) = 1 - \lambda_a(t) dt \\
P_a(x_{a}(t+dt) &= r_\alpha | x_{a}(t) = i_\alpha) =  \xi\ dt\,. \\
\end{aligned}
\right.
\end{equation} 
We use a standard Bellman argument to find the evolution of $U_a$: the lowest possible cost at time $t$ is given by adding two quantities: the lowest possible cost at time $t+dt$, and the cost incurred in the interval $[t,t+dt[$ associated with the optimal strategy at $t$. Assuming a status $x_{a}(t) = s_\alpha$ at time $t$, this can be expressed as
\begin{equation}
\label{eq:Bellman}
U_a(t) = \\
\underset{n^\gamma_a(t)}{\min}\ \mathbb{E}_{x_{a}(t+dt)} \left [ U_a(t+dt) + c_a(t) \right ] \; ,
\end{equation}
with $c_a(t)$ the cost paid in the interval $[t,t+dt[$. At time $t+dt$, the agent either is still susceptible, or becomes infected. If $x_{a}(t+dt) = s_\alpha$ then the only cost at $t$ is $c_a(t) = f_\alpha(n^\gamma_a(t)) dt$, whereas if $x_{a}(t+dt) = i_\alpha$ then $a$ has to bear the costs due to infection, and thus $c_a(t) = \tria(I(t))$. Following \eqref{eq:value function}, if $a$ is susceptible at $t+dt$ then the quantity $U_a(t+dt)$ involves the average cost $C_a(n^\gamma_a(\cdot),t+dt)$, which is an average over all random realizations of the epidemic at times $t'>t+dt$; if $a$ is infected at $t+dt$ then $U_a(t+dt)=0$. The expectation value in \eqref{eq:Bellman} is therefore taken over random realizations of the status $x_{a}(t+dt)$.

Writing explicitly the expectation in \eqref{eq:Bellman} and using the probabilities given by \eqref{eq:Markov} we get
\begin{equation}
\label{eq:HJB_discret}
U_a(t) = \underset{n^\gamma_a(t) }{\min}  \left[  \tria(I(t)) \lambda_a(t) dt  + \\ (1 - \lambda_a(t) dt) \left (U_a(t+dt) +  f_\alpha(n^\gamma_a(t) ) dt \right ) \right]  \; .
\end{equation}
At first order in $dt$, this gives the Hamilton-Jacobi-Bellman (HJB) equation of our Mean Field Game
\begin{equation} \label{eq:HJB}
- \frac{dU_a(t)}{dt} = \\ \underset{n^\gamma_a(t)}{\min} \left [ \lambda_a(t) \left ( \tria (I(t)) - U_a(t) \right ) + f_\alpha (n^\gamma_a(t)) \right ]   \; ,
\end{equation}
and the optimal strategy $n^{\gamma*}_\alpha(t)$ at time $t$ is given by  
\begin{equation} \label{eq:nstar}
n^{\gamma*}_a(t) = \\ \underset{n^\gamma_a(t)}{\text{argmin}} \left [ \lambda_a(t) \left ( \tria (I(t)) - U_a(t) \right ) + f_\alpha (n^\gamma_a(t)) \right ]    \; ,
\end{equation}
where the optimization is now performed for a given, {\em fixed}, time.
By taking a particular form for $f_\alpha$, one can compute $n^{\gamma*}_a(t)$ by setting to zero the derivative of the right hand side with respect to $n$. Thus, for a given epidemic, we can obtain the optimal individual behavior backward in time by solving HJB Eq.~\eqref{eq:HJB}. More details will be given in Sec.~\ref{section:Nash}.

\section{Epidemics dynamics}
\label{section:epidemics dynamics}
\subsection{Cost function and choice of the parameters}
\label{section:cost_function}
With the Mean Field Game  introduced above,  the dynamics of time-dependent quantities such as $\lambda_a(t)$ is now an output of the model, and the ``extrinsic'' model's parameters can be thought of as constant in time.  They can thus in principle be extracted from field data.  We shall, however, not attempt this ambitious goal, and rather, in this section, illustrate the kind of dynamics that emerges from our model on a typical example. A more thorough exploration of the model's parameter space will be performed in appendix~\ref{sec:exploration}.

For the cost of infection we take
\begin{equation}
\begin{aligned}
\label{eq:rI}
& \tria(I(s)) = \ria + g_\alpha(I(s)) = \kappa_\alpha r_I + \kappa_\alpha \left[r_I 
    \left ( \exp\left[\nu_{\rm sat} \frac{I(t)-I_{\rm sat}}{I_{\rm sat}}\right] - 1 \right) \right]   \\
&=  \kappa_\alpha r_I 
    \exp\left[\nu_{\rm sat} \frac{I(t)-I_{\rm sat}}{I_{\rm sat}}\right]    \; .
\end{aligned}
\end{equation}
This includes a base cost $\ria$ and an additional cost $g_{\alpha}(I(s))$ which models the saturation of the sanitary system, and more generally the dependence of the cost of infection on $I(s)$. Here we use a constant cost of infection $\ria$,  within an age class $\alpha$, but we could consider more complex versions; for instance the cost of infection could depends on the duration of the stay in the infected compartment. We take $\ria = \kappa_\alpha r_I$, with a factor $\kappa_\alpha$ which increases with the age class $\alpha$. The term $g_\alpha(I(s))$ models the possible saturation of health services; we take an exponential increase of the strain on human and material resources as the saturation threshold $I_{\rm sat}$ is approached, with a slope $\nu_{\rm sat}$ corresponding to the impact of saturation on the cost; the  ``$-1$'' term in $g_\alpha(I(s))$ is here to recover the usual cost $\kappa_\alpha r_I$ when one take $\nu_{\rm sat} = 0$. This extra term $g$ is often referred in the literature as the limited capacity of intensive care units (ICU).   

Turning now to $f$, the cost of modifying social contacts, we choose to follow the same form as Turinici and al. in \cite{Turinici_contact_rate_SIR_simple}, namely 
\begin{equation}
\label{eq:falpha}
    f_\alpha(n^\gamma_a(t)) = \sum_{\gamma} \left [\left (\frac{1}{n^\gamma_a(t)} \right )^{\mu_\gamma} - 1 \right ] \; ,
\end{equation}
where $\mu_\gamma $ models the degree of ``attachment''  to the setting $\gamma$: for example it is usually easier to reduce contacts at work than inside families. Moreover, $f$ is decreasing with a positive second derivative, meaning that the more one decreases her social contacts, the higher the price to pay. 

The set of values chosen in this section for our parameters is summarized in Table \ref{table:M}. We take a society with $25 \%$ of young individuals, $25 \%$ of retired individuals and $50 \%$ of adults. The time scale chosen is the week, which means that a matrix element $M^{\gamma (0)}_{\alpha \beta}$ corresponds to the number of contacts per week for an individual of age class $\alpha$ with someone of age class $\beta$ in a setting $\gamma$, in the absence of any epidemic. This choice of a week as a time step is also why we have a high recovery rate ($\xi = 1.2$) compared to the literature. 

The parameter $T$ denotes the time at which agents end their optimization process. This corresponds for instance to the time where herd immunity is reached, or it can depend on other circumstances such as the expected production of a vaccine, the seasonality of the virus, among others. In Sec.~\ref{section:epidemic_dynamics_results}, our simulations are performed on a duration of $T=40$ weeks to focus on scenarios where collective immunity is  reached and to avoid short end-time effects. Scenarios for which, due to short end-time,  collective immunity is not reached at the end of the optimization period will be studied more specifically in Sec.~\ref{section: strategies_comparison}. Since the main wave of the epidemic appears in the first $10$ weeks, we often present the results on a duration of $15$ weeks. 

\begin{table}
    \centering
    \begin{tabular}{ c c c c c c c c }
     \hline 
     $M^S$ & $M^W$ & $M^C$ & $M^H$ & $\kappa_\alpha$ & $K_\alpha$ \\ 
     \hline
    $\begin{pmatrix}
    100 & 0 & 0 \\
    0 & 0 & 0 \\
    0 & 0 & 0
    \end{pmatrix}$  & 
    $\begin{pmatrix}
    0 & 0 & 0 \\
    0 & 75 & 0 \\
    0 & 0 & 0
    \end{pmatrix}$ & $\begin{pmatrix}
    12.5 & 25 & 12.5 \\
    12.5 & 25 & 12.5 \\
    12.5 & 25 & 12.5
    \end{pmatrix}$ & $\begin{pmatrix}
    15 & 25 & 10 \\
    12.5 & 32.5 & 5 \\
    10 & 10 & 30
    \end{pmatrix}$  & $(1,10,100)$ & $(0.25, 0.5, 0.25)$  \\ \hline 
    $I_\alpha(0)$ & $n^\gamma_{\min}$ & $(I_{\rm sat},\nu_{\rm sat})$  & ($\xi, q, \sigma, \mu$) & ($I_l$, $I_d$) & $\mu_{\gamma}$  \\ \hline
    $(0.01, 0.01, 0.01)$ & $\left(\frac{1}{3},\frac{1}{5},\frac{1}{5},\frac{1}{2}\right)$  & (0.01,0.01) & (1.2, 0.2, 0.35, 0.1) & (0.12, $4.10^{-4}$) & $(2,2,1,3)$ \\ \hline 
    \end{tabular}
    \caption{Table of parameters used in our simulations. The matrix entries $M^{\gamma (0)}_{\alpha \beta}$ correspond to the average frequency of contacts (per week) between an individual of age class $\alpha$ and someone of age class $\beta$ in the setting $\gamma$. $\kappa_\alpha$ is the coefficient appearing in $\tria$. $K_\alpha$ is the proportion of the population in each age class. $n^\gamma_{\min}$ is the minimum contact willingness in each setting $\gamma$,  while $\mu_\gamma$ weights the cost of contact reduction in each setting. $I_\alpha(0)$ are the initial proportion of infected for each age class. $\xi$ is the recovery rate (per week), $q$ the transmission rate per contact. $I_l, I_d$ are the thresholds for the best lockdown and $\sigma$ its intensity level. $\mu$ correspond to the proportion of asymptomatic individuals in the population. Finally, $\alpha = 1, 2, 3$ for age class of young, adults and retired individuals respectively while $\gamma = 1, 2, 3, 4$ for respectively schools, workplaces, community and households. Unless explicit mention, these parameters are used in all our numerical simulations. With our choice of parameters, we get $\tilde{R}_0 = 2.9$ with the method described in \cite{Inferring_social_structure,diekmann1990definition} to compute $\tilde{R}_0$ at the beginning of epidemics in heterogeneous populations. This value is coherent with the literature for viruses like Covid-19 ones \cite{guan2020modeling} and allow us to extract relevant insights from our simulations.}
    \label{table:M}  
\end{table}

\subsection{Epidemics dynamics}
\label{section:epidemic_dynamics_results}

The characteristic features of the Nash equilibrium are better revealed if one compares the corresponding epidemic dynamics with other scenarios.   In addition to the \emph{(unconstrained) Nash equilibrium} described until now, we shall therefore in the next subsections consider also  the case of a \emph{``constrained'' Nash equilibrium}, where individuals have to deal with global constraints imposed by an authority (these constraints can be either naive or optimally chosen), as well as the \emph{the societal optimum}, which is the idealistic case where everybody strives to optimize the global cost.
In each of these cases,  the epidemic dynamics is driven by the system Eqs.~\eqref{eq:SIR_kolmogorov}, but with different $\{n_\alpha^\gamma(.)\}$, and thus different forces of infection $\{\lambda_\alpha\}$. The results for each of the above strategies will be presented and discussed in the following subsections; technical details about the numerical implementation are given in appendix~\ref{app:numerics}. 

\begin{figure}[t!]
    \centering
    \includegraphics[scale=0.4]{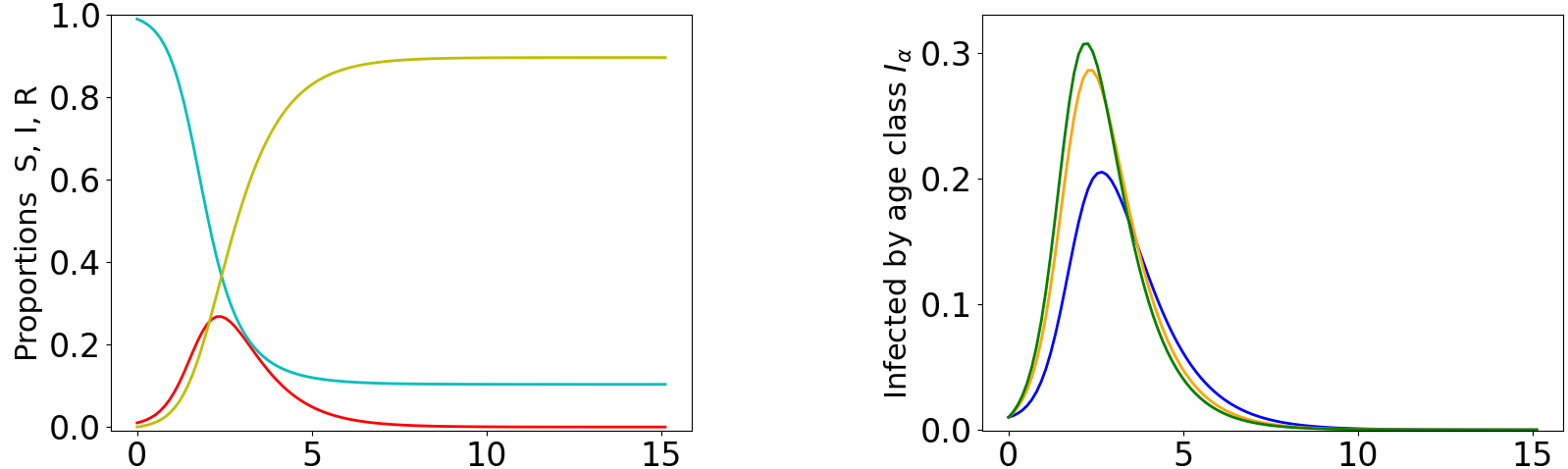}
    \includegraphics[scale=0.4]{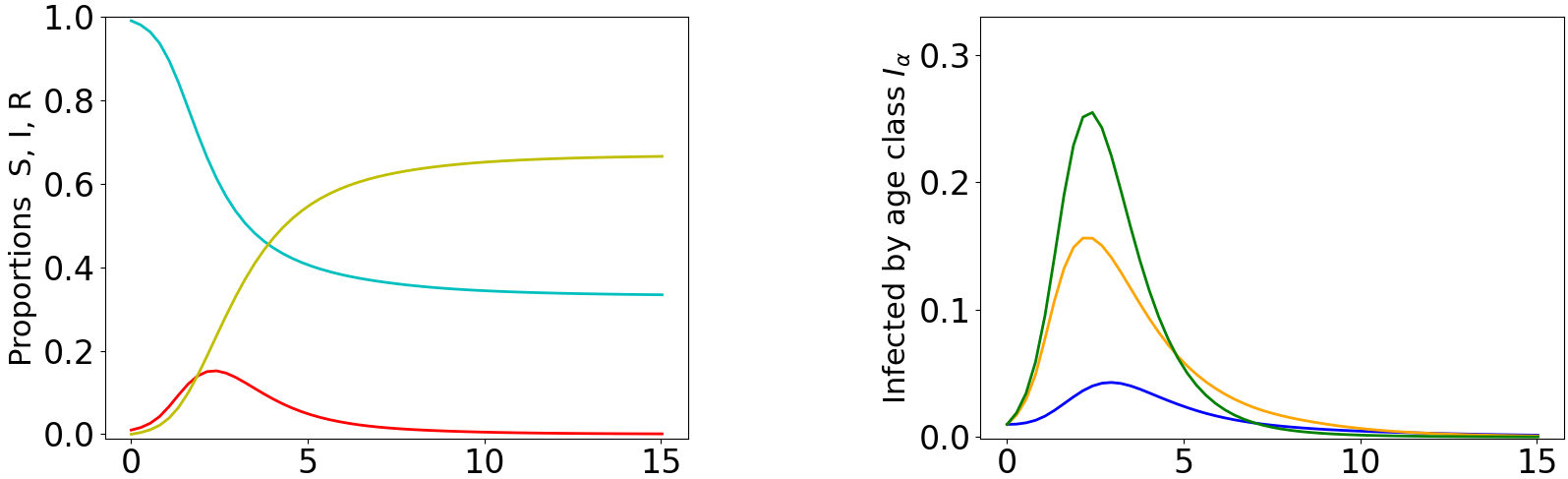}
    \includegraphics[scale=0.4]{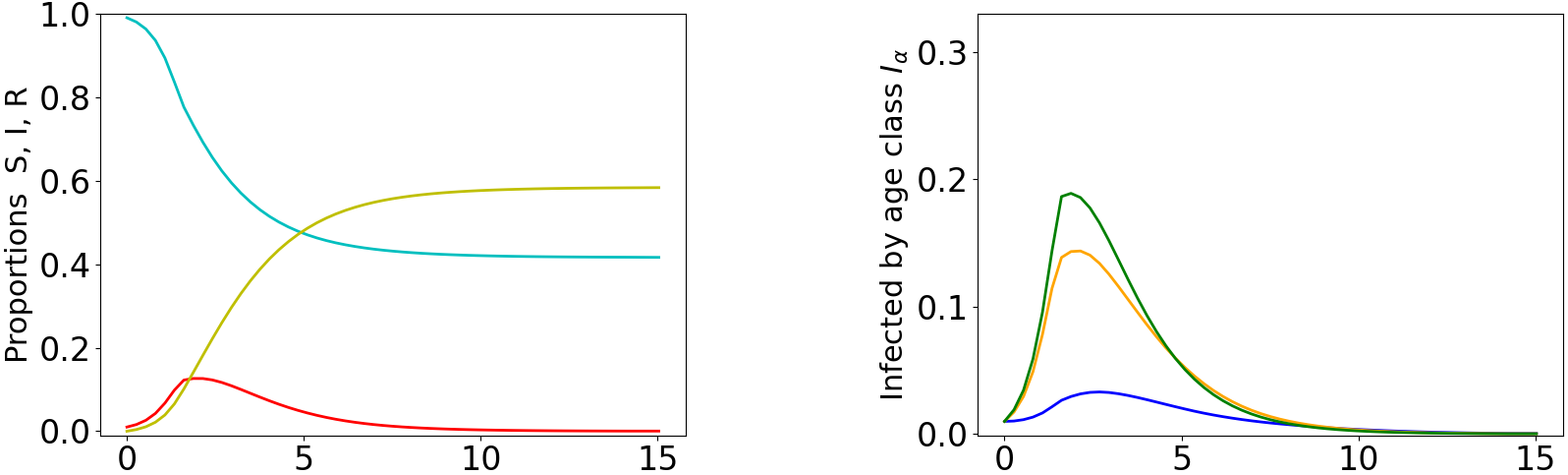}
    \includegraphics[scale=0.4]{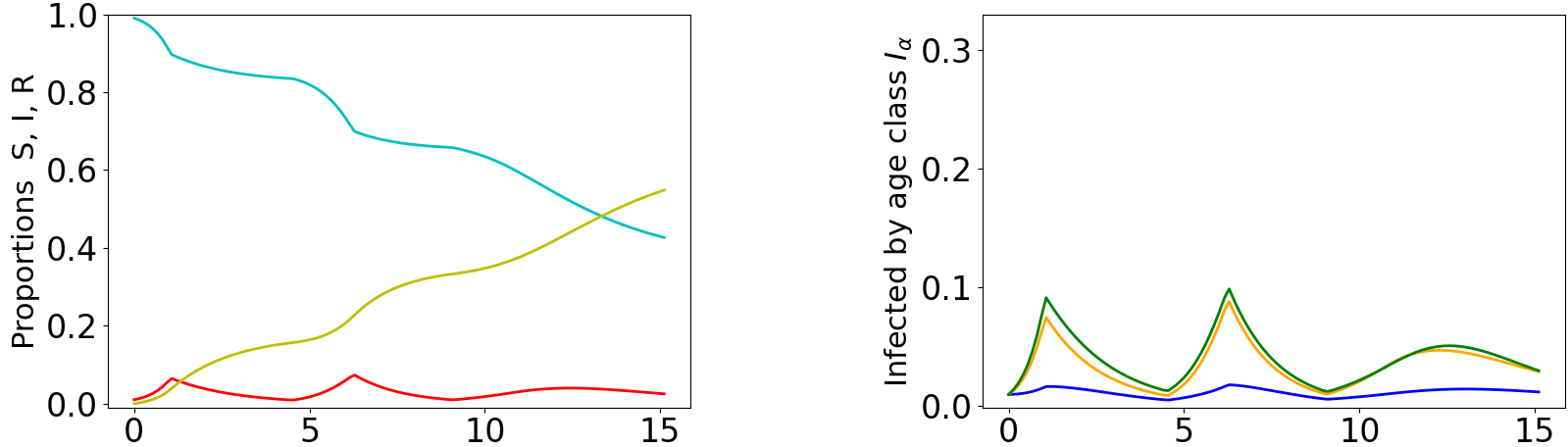}
    \includegraphics[scale=0.4]{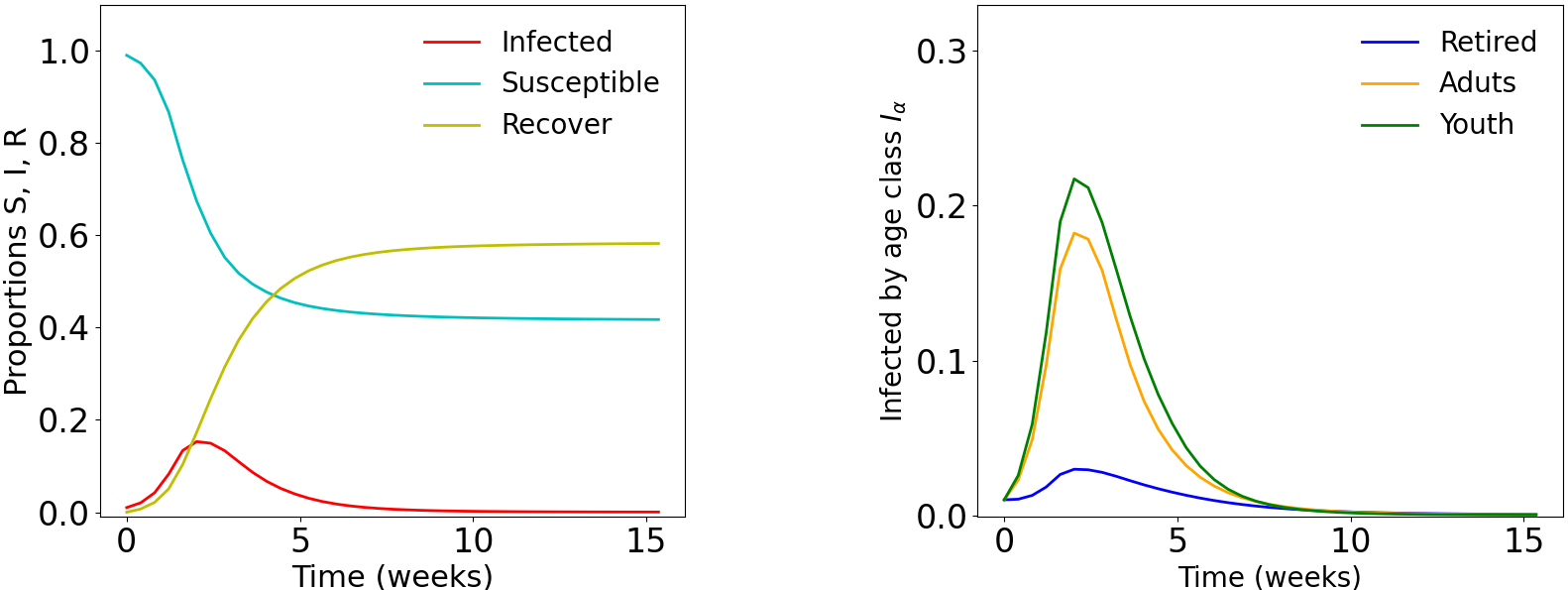}
    \caption{Time evolution of the epidemic quantities with $r_I = 1$ and parameters of Table \ref{table:M}. From top to bottom : Business as usual (no efforts), (unconstrained) Nash equilibrium, Nash equilibrium under optimal constraints, Nash equilibrium with naive constraints, societal optimum. Left: time evolution of the proportion of susceptible $S$ (cyan), infected $I$ (red) and recovered $R$ (yellow) in the population. Right: time evolution of the proportion of infected in each age class $I_\alpha$, retired people are in blue, adults in orange and youth in green.}
    \label{fig:epidemic_simulations}
\end{figure}

For all these scenarios, as well as for the one corresponding to \emph{business as usual}  (for which no modification of the contact parameter is done), we summarize in Fig.~\ref{fig:epidemic_simulations} the results for the dynamics of $S$, $I$ and $R$, for the set of parameters defined in Table \ref{table:M}. 
There are notable similarities between the different ``optimized'' scenarios (Nash, constrained Nash and societal optimum) that distinguish them from the  business as usual one.  For instance, the number of susceptible individuals at the end of the epidemic  is   $S_\infty \simeq 0.4$ in all cases but for the business as usual scenario, where it is significantly below (first row).  This is due to the fact that in all circumstances one needs to reach herd immunity to escape from the disease, and the fact that $S_\infty$  is much below this required value is a clear indication of the business as usual sub-optimal character. In the same way, for all optimized scenarios there is a significant difference between the height of the infection wave for the different age class, as retired individuals and adults are more impacted by the disease than the youths, and therefore protect themselves.  In the business as usual scenario the difference is much less significant, and only due to the relative proportion of contacts in each age class.  On the other hand, the constrained Nash equilibrium with ``naive'' constraints differs from all the others because of the existence of two epidemic waves, which can be understood as originating from an excessive limitation of contacts that prevents the society from reaching herd immunity. Other differences, which are mainly quantitative, also exist between these different scenarios, and will be discussed in more details in Section~\ref{sec:comparison}. We now turn to the detailed description of each strategy.

\FloatBarrier

\subsection{The (unconstrained) Nash equilibrium}
\label{section:Nash}
Let us first consider the (unconstrained) Nash equilibrium. We have seen that it is described by two sets of differential equations. The first one is the rate equation of the epidemic, Eq.~\eqref{eq:SIR-ss} (also known as the Kolmogorov equation in this context), which is forward in time, that is, starting from initial conditions $S_\alpha(0), I_\alpha(0), R_\alpha(0)$, populations at later time $t$ in age class $\alpha$ are obtained by solving
\begin{equation}
\label{eq:SIR_kolmogorov}
\begin{aligned}
\dot{S}_\alpha & = - \lambda_\alpha(t) S_\alpha(t) \\
\dot{I}_\alpha & = \lambda_\alpha(t) S_\alpha(t) -  \xi I_\alpha(t) \; , \\
\dot{R}_\alpha & =   \xi I_\alpha(t)
\end{aligned}
\end{equation}
with $\lambda_\alpha(t)$ given by Eq.~\eqref{eq:lambda2}.
The second set of equations corresponds to the Hamilton-Jacobi-Bellman equation Eq.~\eqref{eq:HJB}, with one reference individual $a$ for each age class $\alpha$, 
\begin{equation} \label{eq:HJB2}
- \frac{dU_a(t)}{dt} = \underset{n^\gamma_a(t)}{\min} \left [ \lambda_a(t) \left ( \tria (I(t)) - U_a(t) \right) + f_\alpha (n^\gamma_a(t)) \right ]   \;.
\end{equation}
As only the terminal condition on $U$ is fixed, namely, $U_a(T) = 0$, Eq.~\eqref{eq:HJB2} is backward in time.
%
%
At equilibrium, all individuals will follow their own optimal strategy; but as all agents in a given age class are equivalent, this optimal strategy should be the same for all agents $a$ of age class $\alpha$. Thus we have the additional self-consistency condition 
\begin{equation} \label{eq:Nash}
 n^{\gamma *}_a (t)= n^\gamma_\alpha (t) \; .
\end{equation} 
This equation imposes that if all other agents follow the strategy solution of the self-consistent system Eqs.\eqref{eq:SIR_kolmogorov}-\eqref{eq:HJB2}-\eqref{eq:Nash}, deviating from that strategy implies a higher cost.  The solution of the MFG equation thus corresponds to a Nash equilibrium.
The two equations \eqref{eq:SIR_kolmogorov} and \eqref{eq:HJB2}, together with the self consistency condition \eqref{eq:Nash}, form a system of equations coupling all epidemic rates $S(.)$, $I(.), R(.)$ and all age-class strategies $n^\gamma_\alpha$ via the individual optimal strategies $n^{\gamma *}_a$. Indeed, the epidemic rates in \eqref{eq:SIR_kolmogorov} depend on $\lambda_\alpha(t)$ given in \eqref{eq:lambda2}, which depend on the global strategies ${n^\gamma_\beta}$. In turn, the optimal strategy ${n^{\gamma *}_a}$ for a reference individual $a$ is a solution of HJB equation~\eqref{eq:HJB2}; with the precise form of the costs $\tria(I(s))$ and $f_\alpha(n^\gamma_a(t))$ chosen in Section \ref{section:cost_function}, it can be computed explicitly and reads 
\begin{equation}
\label{eq:nstar_explicit}
    n^{\gamma *}_a(t) = \left( \frac{\mu q}{\mu_\gamma} \left[ \tria(I(t))-U_a(t) \right] \sum_{\beta=1}^{n_{\rm{cl}}} n^\gamma_\beta(t) M^{\gamma(0)}_{\alpha \beta} I_\beta(t) \right) ^{-\frac{1}{\mu_\gamma+1}}\; ,
\end{equation}
 which depends explicitly on the global strategies ${n^\gamma_\beta}$ and on the epidemic rate $I(.)$. 
One obtains in this way an initial-terminal value problem (ITVP), which can be solved numerically in different ways; we present some of them briefly in appendix~\ref{app:numerics-Nash}. 

\begin{figure}[t!]
    \centering
    \includegraphics[scale=0.4]{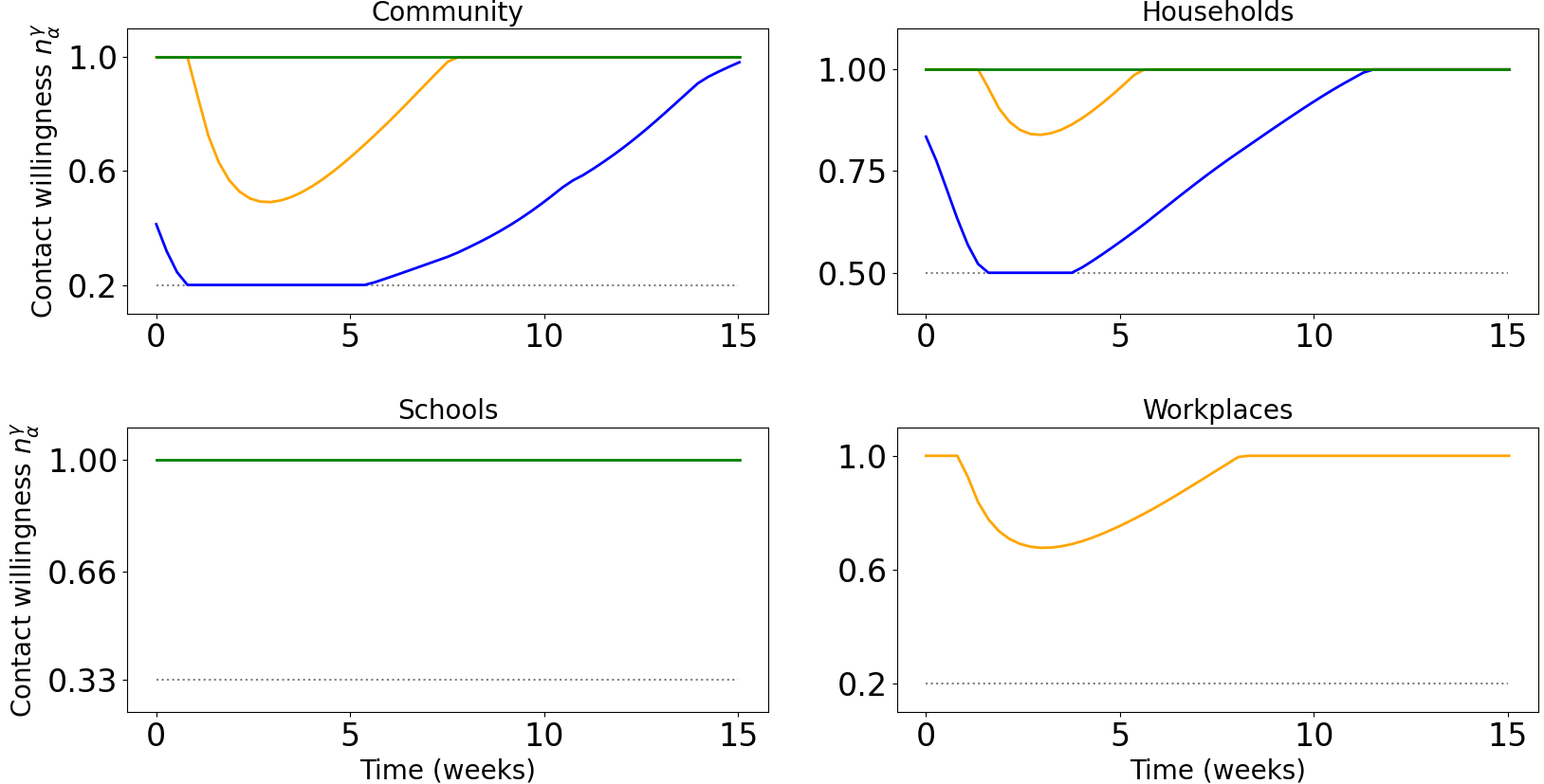}
    \caption{Time evolution of the contact willingness $n^\gamma_\alpha(t)$ with $r_I = 1$ at the Nash equilibrium. We plot $n^\gamma_\alpha(t)$ for each type of individual according to their age class (retired people in blue, adults in orange and youth in green) in community (upper left), households (upper right), schools (lower left, for the young) and workplaces (lower right, for the adults). The dotted gray horizontal lines correspond to the minimum contact willingness allowed (maximum effort).}
    \label{fig:contact_nash}
\end{figure}

The solutions of the MFG system \eqref{eq:SIR_kolmogorov}-\eqref{eq:Nash} are displayed in the second row of Fig.~\ref{fig:epidemic_simulations} for the set of epidemics quantities $S_\alpha(.), I_\alpha(.), R_\alpha(.)$, and in Fig.~\ref{fig:contact_nash} for the set of optimal strategies $n^\gamma_\alpha(.)$.  For our choice of parameters, young individuals do not modify at all their behaviour, when retired people reach maximal effort for significant amount of time in both community and household settings, and adults do some efforts, but without ever reaching the maximum one. 

\FloatBarrier

\subsection{The Nash equilibrium under constraints}
\label{section:Nash_constraints}

In the Nash equilibrium considered above, each agent optimises for herself, and the resulting Nash equilibrium can lead to a  global cost for the society, 
\begin{equation}
\label{eq:global-cost}
 C_{\rm glob} \left(\{n_\beta\} \right) \; \equiv \; \sum_{\alpha} K_\alpha C_\alpha \left(n_a \smeq n_\alpha, \{n_\beta\} \right) \; ,
\end{equation}
which is sub-optimal.  In Eq.~\eqref{eq:global-cost}, $\{ n_\beta \}$ is the set of strategies followed by  each age class, $n_a = n_{\alpha}$ means that any given individual $a$ of class $\alpha$ follows the strategy $n_\alpha$ assigned to age class $\alpha$, and  the cost for each age class is weighted by the proportion $K_\alpha$ of individuals in that class. 
A question that naturally arises from a public policy point of view is to know whether one could improve the global wellbeing of the population by 
driving the position of the Nash equilibrium through constraints on the population. This is, in some sense, what has been attempted in many countries during Covid-19 pandemic. The restrictions taken then, however, involved a lot of guesswork, both about the precise decisions to take, and about their potential effects on society (individuals behavioral response, impact on economic, health, etc).

Here we present a possible quantitative approach to study such restriction policies, which aim at  reducing the societal cost by constraining the behavior of individuals. Again, we remain here at the level of a ``proof of concept'', as practical implementations of our formalism would  require determining realistic  forms of the cost functions  and of the constraints, which is clearly beyond the scope of our work.

With the free (i.e. unconstrained) Nash equilibrium, individuals choose their contact willingness $n^\gamma_\alpha(t)$ in the range  $[n^\gamma_{\alpha, \min},1]$, where the maximum $1$ correspond to the situation without epidemic. We now add a constraint similar to a partial lockdown, by setting this maximum to $n^\gamma_{\alpha, l} < 1$ when some epidemic level is reached. In that way, everyone is required to make a minimal amount of efforts to preserve the sanitary system and reduce the societal cost \eqref{eq:global-cost}. This ``lockdown'' is implemented when the proportion of infected $I(t)$ reaches a certain threshold $I_l$, and, as the proportion of infected decreases we assume the lockdown is lifted when $I(t)$ goes below a value $I_d < I_l$ (which is assumed lower than $I_l$ to avoid unrealistic oscillations around $I_l$). The lockdown has thus a hysteresis form, and is implemented in the  following way (with $L$ a Boolean variable which is 1 if the lockdown is active and 0 otherwise):
\begin{equation}
\label{eq:lockdown}
\begin{cases}
\displaystyle{\textrm{if} \; I(t) < I_d : n^\gamma_\alpha(t) \in [n^\gamma_{\alpha, \min}, 1] \quad \; \; \& \quad L \mapsto 0 \qquad  \textrm{  no constraints }}\\
\displaystyle{\textrm{if} \;  I(t) > I_l : n^\gamma_\alpha(t) \in [n^\gamma_{\alpha, \min}, n^\gamma_{\alpha, l}] \quad \& \quad L\mapsto 1\qquad \textrm{  active constraints}}\\
\displaystyle{\textrm{if} \; I_d < I(t) < I_l \textrm{  and  } L=0 : n^\gamma_\alpha(t) \in [n^\gamma_{\alpha, \min}, 1] \; \; \textrm{  no constraints}}  \\
\displaystyle{\textrm{if} \; I_d < I(t) < I_l \textrm{  and  } L=1 : n^\gamma_\alpha(t) \in [n^\gamma_{\alpha, \min}, n^\gamma_{\alpha, l}] } \; \; \textrm{  active constraints.} 
\end{cases} \qquad 
\end{equation}
In Eq.~\eqref{eq:lockdown}, we choose ${n^\gamma_{\alpha, l} = \sigma n^\gamma_{\alpha, \min} + (1- \sigma)}$,  with $\sigma \in [0,1]$ a variable measuring the intensity of the lockdown: $\sigma = 0$ corresponds to the free situation without any constraint, while $\sigma = 1$ corresponds to a strict lockdown with no freedom, as $n^\gamma_\alpha(t)$ is fixed to $n^\gamma_{\alpha, \min}$. Therefore, the lockdown is described by a set of three variables ($\sigma, I_l, I_d$): the intensity $\sigma$, the first threshold $I_l$, and the second threshold $I_d$. 
The numerical implementation of this set of equations is briefly discussed in appendix~\ref{app:numerics-constr}.

\begin{figure}[t!]
    \centering
    \includegraphics[scale=0.4]{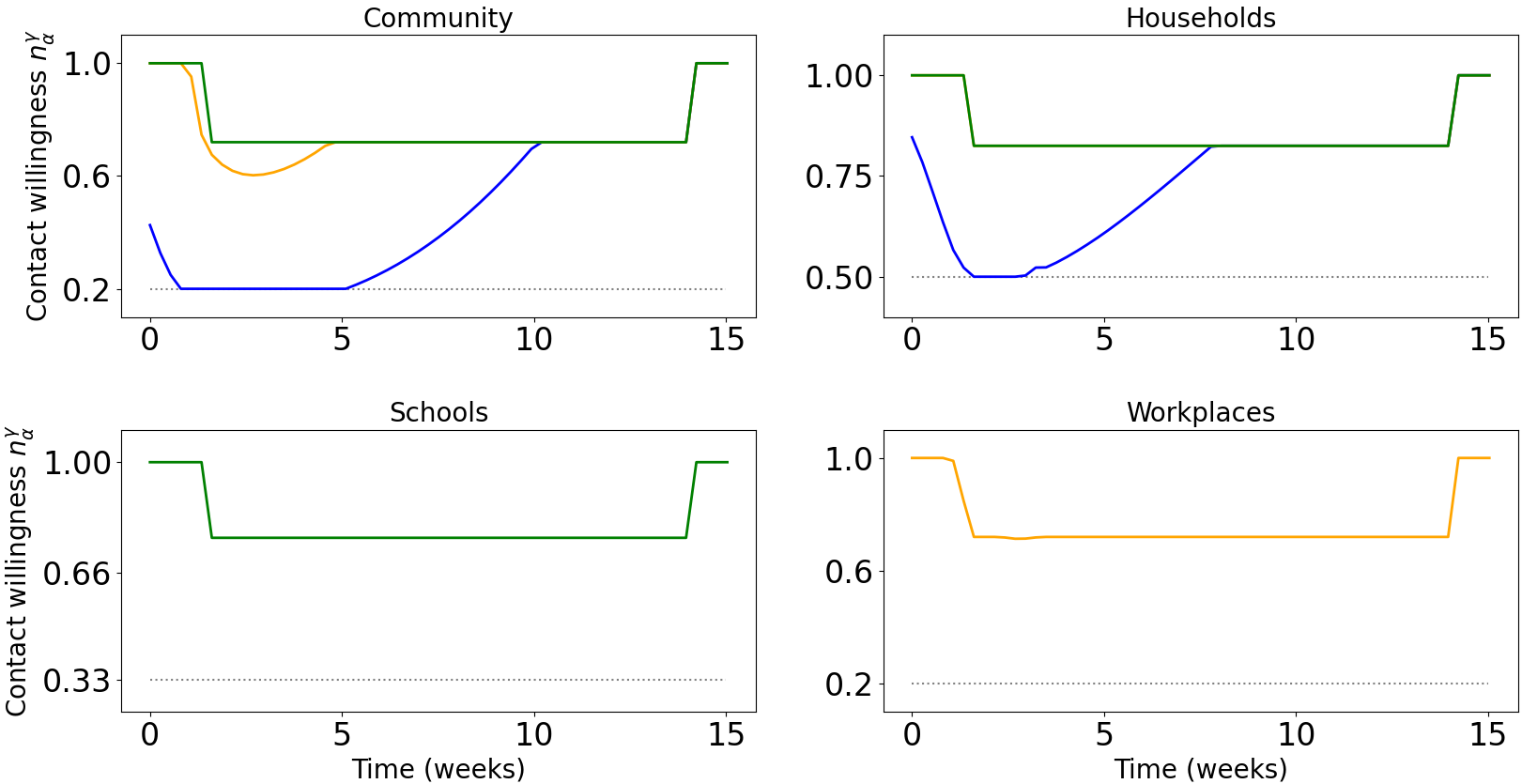}
    \caption{Time evolution of the contact willingness $n^\gamma_\alpha(t)$ with $r_I = 1$ for the Nash equilibrium under optimal constraints $(\sigma \smeq 0.35, I_l \smeq 0.12, I_d \smeq 4. 10^{-4})$. We plot $n^\gamma_\alpha(t)$ for each type of individual according to their age class (retired people in blue, adults in orange and youth in green) in community (upper left), households (upper right), schools (lower left, for the young) and workplaces (lower right, for the adults). The dotted gray horizontal lines correspond to the minimum contact willingness allowed.}
    \label{fig:contact_constraints}
\end{figure}

In Fig.~\ref{fig:epidemic_simulations} (third row) we show the evolution of the epidemic quantities for the choice of parameters $(\sigma \smeq 0.35, I_l \smeq 0.12, I_d \smeq 4. 10^{-4})$. As will be discussed in Sec.~\ref{section:Cost_comparison}, this choice corresponds to an optimal value in the sense that these parameters minimise the global cost Eq.~\eqref{eq:global-cost} among all possible constraints in the parameter space ($\sigma, I_l, I_d$). In Fig.~\ref{fig:contact_constraints} we display the corresponding strategies chosen by individuals under these constraints.  The constraints are enforced after 2 or 3 weeks into the epidemic, and are raised after almost 14 weeks (over 40 for the total epidemic time) when the proportion of infected is low and there is no risk of any epidemic rebound.  The values of the  constraints appear as straight lines followed by youth individuals, whose behavior is not dictated by their own ``egoistic'' optimisation but by the fact they are forced to respect the lockdown as soon as it is imposed. Retired people on the other hand choose most of the time to limit their contact even more than required by the constraints; adults most of the time just follow the lockdown, but sometimes limit their contacts further.

\begin{figure}[t!]
    \centering
    \includegraphics[scale=0.4]{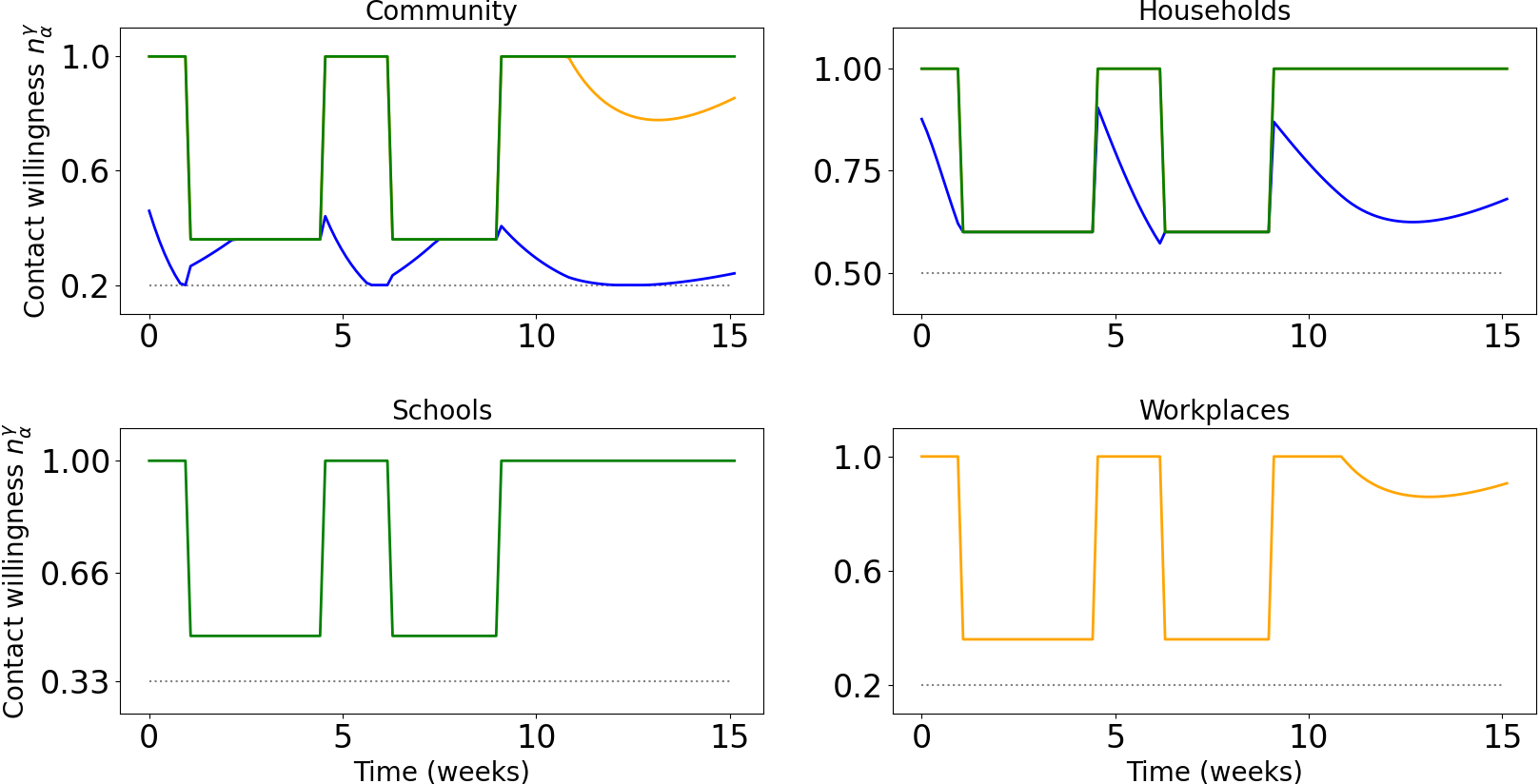}
    \caption{Time evolution of the contact willingness $n^\gamma_\alpha(t)$ with $r_I = 1$ for the Nash equilibrium under naive constraints ($\sigma \smeq  0.8, I_l \smeq 0.06, I_d \smeq 0.01$). We plot $n^\gamma_\alpha(t)$ for each type of individual according to their age class (retired people in blue, adults in orange and youth in green) in community (upper left), households (upper right), schools (lower left, for the young) and workplaces (lower right, for the adults). The dotted gray horizontal lines correspond to the minimum contact willingness allowed.}
    \label{fig:contacts_naive_lockdown}
\end{figure}

As we shall discuss in Sec.~\ref{sec:comparison} this optimal lockdown, despite the fact that it depends on only three parameters, can improve on the free Nash equilibrium, in the sense that the societal cost Eq.~\ref{eq:global-cost} is lower.   However, public policies executives have to be careful about their choice as it can generate situations which are clearly worse than the free Nash equilibrium. We illustrate this situation in Figs.~\ref{fig:epidemic_simulations} (fourth row) and \ref{fig:contacts_naive_lockdown} with parameters $(\sigma \smeq  0.8, I_l \smeq 0.06, I_d \smeq 0.01)$: in that case one imposes a very strong but short lockdown. Since we consider here a long end-time configuration with $T=40$ weeks, for which collective immunity is required to end the epidemic, this leads to epidemic rebounds and increases significantly the epidemic cost. Indeed, all drastic efforts that are made while the epidemic is low, and before collective immunity is obtained, are essentially useless, and just add to the global cost endured by the population. In what follows we shall thus distinguish Nash under optimal constraints (NOC) and Nash under ``naive'' (uncarefully chosen) constraints (NNC).

\FloatBarrier

\subsection{The societal optimum}
\label{section:societal_optimum}

In the previous two scenarios, each agent performs a personal, possibly constrained, but essentially egoistic, optimization.  To set the scale of what is the cost associated with these egoistic approaches, it may be useful to compare them with the ``societal optimum'' that could be imposed by a ``benevolent global planner'', i.e.~a well-meaning government with full empowerment. This amounts to finding the minima of the global cost Eq.~\eqref{eq:global-cost}. There is already a rich literature on topics related to societal optimization (see for example \cite{optimal_isolation_policies,morton_wickwire_1974,Turinici_contact_rate_SIR_simple,tchuenche2011optimal,bertsekas2012dynamic,abakuks1973optimal,abakuks1974optimal,kantner2020beyond,kruse2020optimal,khouzani2011optimal}) on various types of models, as this problem is reduced to a single global optimization.  
%
%
The difference between this minimization and the Nash equilibrium discussed above is referred to as  ``the cost of anarchy'': while there is no cooperation between individuals in the Nash equilibrium, the societal optimum case corresponds to ``the best'' (from a societal cost point of view) that one can obtain for $C_{\rm glob}$ among all possible strategies. 

The numerical construction of this societal optimum is briefly discussed in appendix~\ref{app:numerics-optim}. In Fig.~\ref{fig:epidemic_simulations} (fifth row) we show the epidemic quantities associated with the societal optimum. This situation is optimal from a society point of view if we look for the global cost only, that is, the addition of all individual costs. However, the total number of infected individuals is not the lowest possible, as infection within the youths does not carry the same cost as within the retired agents. The total amount of infected at the end of the epidemic is still relatively high, because in our framework, one has to reach collective immunity to definitely escape from the disease. Also, the epidemic peak is still at a rather high level, as it is efficient to allow an epidemic spread while keeping the epidemic under control to reach quickly herd immunity. However, the precise distribution of infected proportion in each age class is different from the free Nash equilibrium.

\begin{figure}[t!]
    \centering
    \includegraphics[scale=0.4]{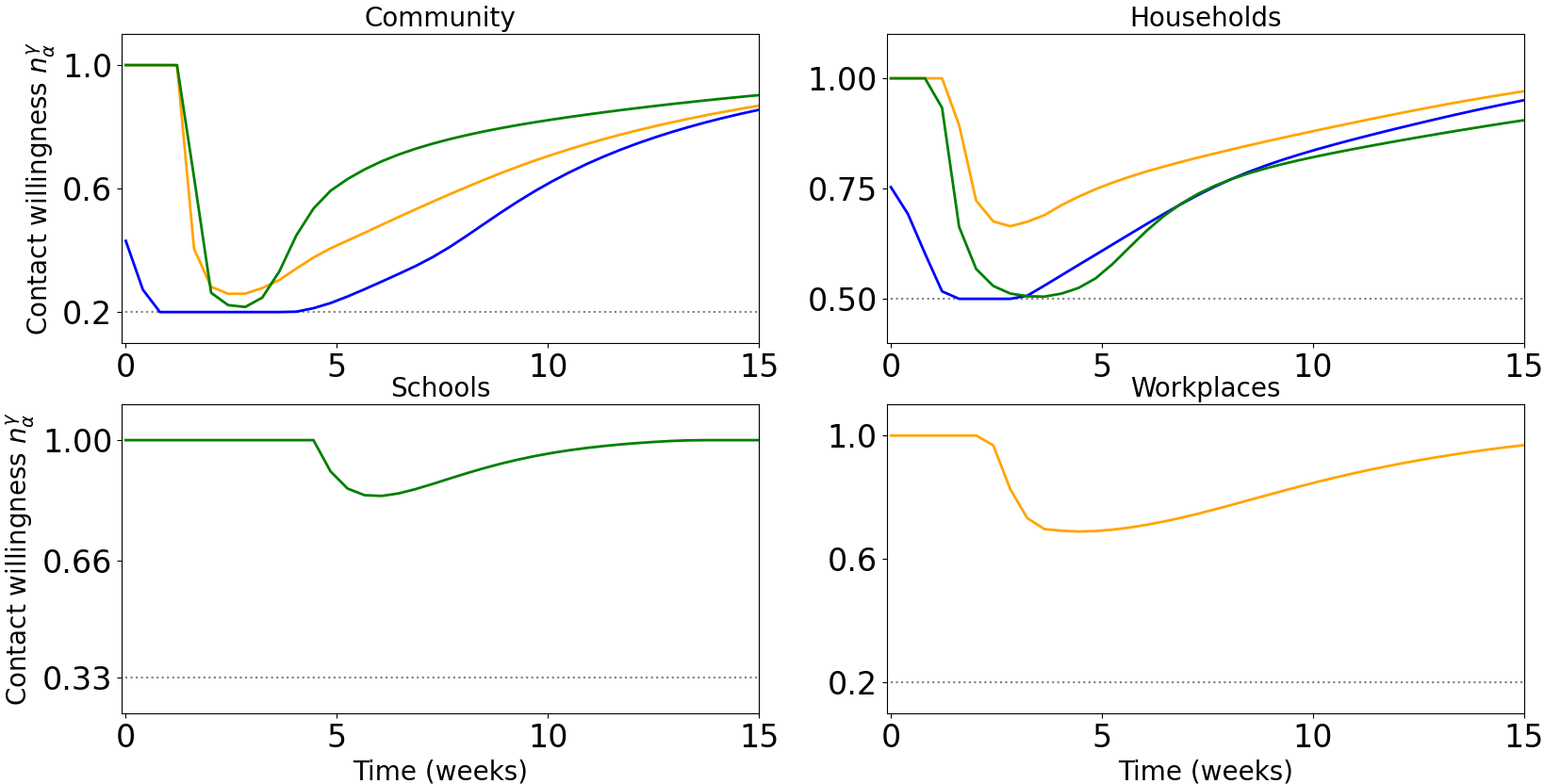}
    \caption{Time evolution of the contact willingness $n^\gamma_\alpha(t)$ with $r_I = 1$ for the societal optimum. We plot $n^\gamma_\alpha(t)$ for each type of individual according to their age class (retired people in blue, adults in orange and youth in green) in community (upper left), households (upper right), schools (lower left, for the young) and workplaces (lower right, for the adults). The dotted gray horizontal lines correspond to the minimum contact willingness allowed.}
    \label{fig:contact_nash_OS}
\end{figure}

In Fig.~\ref{fig:contact_nash_OS} we show the corresponding optimal contact willingnesses. They do not correspond to individual optimum;
rather, there is a cooperation between individuals in different age classes to get an epidemic which will make lower damage with a reasonable amount of efforts. In the community setting and in households, we observe that all individuals make significant efforts during the epidemic peak to avoid a global infection peak that would saturate the sanitary system: they do it in particular in those two settings to avoid a too strong diffusion to retired people. On the other hand, efforts are done with less intensity in schools and workplaces. Once  the epidemic peak is reached, we see that the epidemic continues to spread, in particular in young and adults classes, so that collective immunity can be reached and in this way protect retired people. Thus, the efforts in schools and workplaces are here to smooth sufficiently the epidemic, avoid any rebound, and get a relative collective immunity as fast as possible, making it possible to lift  the efforts in communities and households.

\subsection{Comparison between the different scenarios }
\label{sec:comparison}

\subsubsection{Comparison of global costs}
In order to compare quantitatively the scenarios presented above, we normalize the costs with respect to the total cost of the societal optimum, which we set equal to 100.

\begin{figure}[t!]
    \centering
    \includegraphics[scale=0.6]{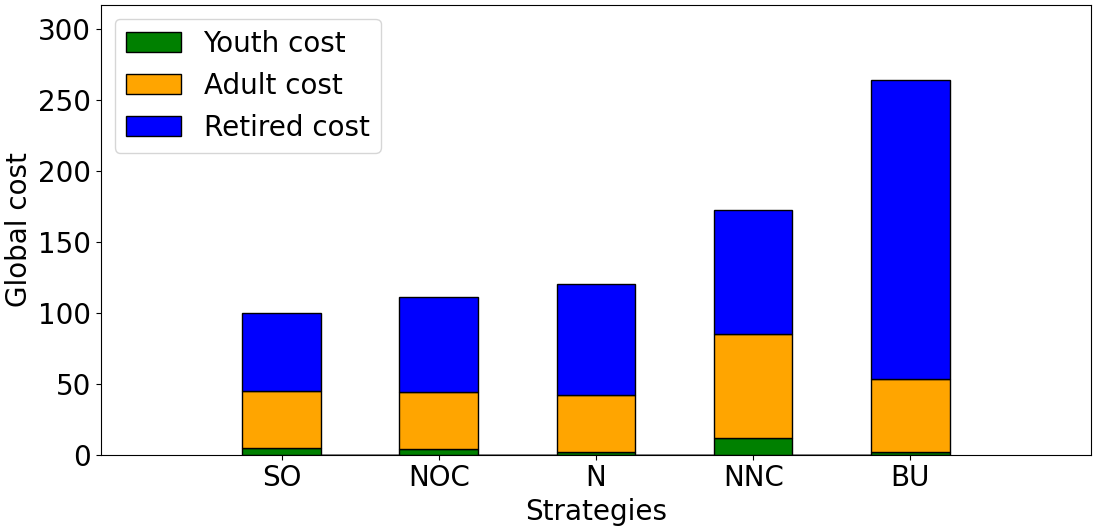}
    \caption{Comparison of costs for the different strategies studied: SO (Societal Optimum), NOC (Nash under Optimal Constraints), N (free Nash equilibrium), NNC (Nash under naive constraints), BU (Business as Usual). The costs are represented on a base of $100$ for SO; the color bars represent the total cost of each age class. Thus, the level of each bar comes from the cost per individual multiplied by the proportion $K_\alpha$ of his age class.}
    \label{fig:cost_comparison}
\end{figure}

In Fig.~\ref{fig:cost_comparison} we show, for the choice of parameters given in Table~\ref{table:M}, the global costs obtained with the different kinds of strategies considered above. As expected, the societal optimum (SO) is the best strategy at society level, followed quite closely by the Nash equilibrium under optimal constraints (NOC), which itself is better than the free Nash equilibrium (N). 
As the imposition of societal-optimal strategies implies a lack of freedom for the individual, as well as a coordination cost which may be significant and which is not included in Eq.~\ref{eq:global-cost}, we argue that the constrained Nash equilibrium presumably forms in practice a good compromise between effectiveness and practicability. One should bear in mind, however, that with a naive choice for the constraints, such as for the NNC strategy of Fig.~\ref{fig:cost_comparison}, one could easily obtain a result worse than for the free Nash equilibrium. 

The color bars in Fig.~\ref{fig:cost_comparison} illustrate the relative importance of each age class in the total cost paid by the society.  This shows that, to reach a global optimum, the key point is to reduce as much as possible the cost for retired people whose contribution is large.  This contribution is actually larger than that of adults, despite the latter representing twice as many people as retired individuals in our population choice. Note that, from the point of view of adults or young people, the free Nash equilibrium is the best strategy, as they do not have to make efforts for others. We can also notice that making a wrong choice for the constraints will not lead to the same ``extra cost'' for everyone. Indeed, for the NNC scenario, the cost for retired people is still relatively low  because the epidemic is maintained at a low level, but the cost of social restrictions becomes very high for adults and young individuals. This has to be contrasted with the business as usual scenario where the extra cost is borne almost exclusively by retired people.

\label{section:Cost_comparison}

\subsubsection{Comparison of contact willingness for the two best strategies}

\begin{figure}[t!]
    \centering
    \includegraphics[scale=0.4]{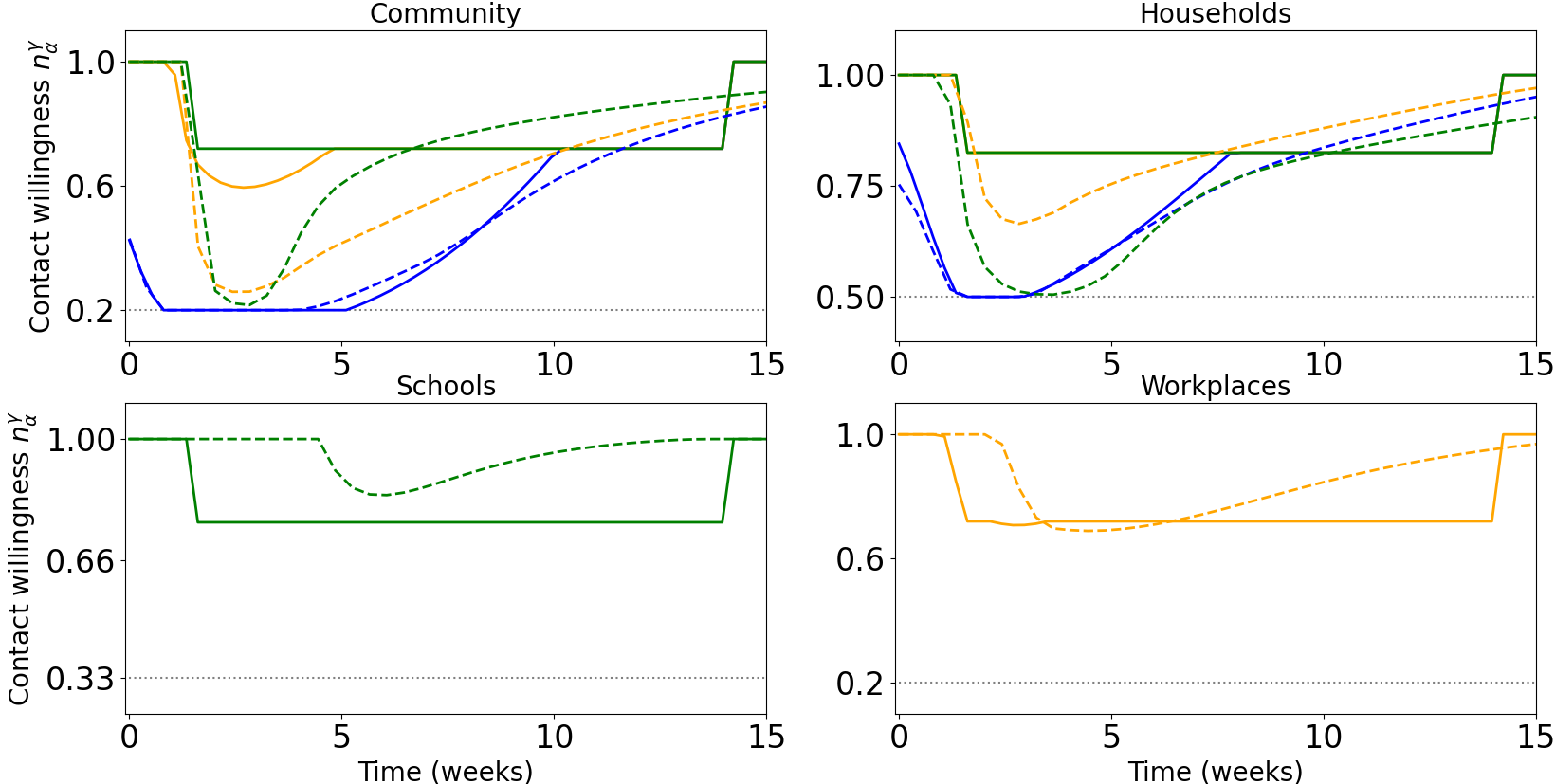}
    \caption{Comparison of contact willingness for the Societal Optimum (dashed line) and the Nash equilibrium under optimal constraints (solid line). We plot $n^\gamma_\alpha(t)$ for each type of individual according to their age class (retired people in blue, adults in orange and youth in green) in community (upper left), households (upper right), schools (lower left, for the young) and workplaces (lower right, for the adults). The dotted gray horizontal lines correspond to the minimum contact willingness allowed.}
    \label{fig:Contact_OS_vs_Constraints_V1}
\end{figure}

In Fig.~\ref{fig:Contact_OS_vs_Constraints_V1}, we show the comparison between the contact willingness obtained with the societal optimum (dashed line) and the Nash equilibrium under optimal constraints (solid line). We see that for the Nash equilibrium under constraints we get constraints which start at almost  the same time as the ones of the societal optimum (after typically $2$ weeks); but since it is a Nash equilibrium, these constraints are raised after a long time, around $14$ weeks, so that even without individual efforts from adults and youth the epidemic is kept under control. At a global level, these constraints are not too strong compared to the ones of the societal optimum, but since they are less localized, both spatially (in the good settings) and temporally (during the epidemic peak with a progressive release afterwards), they are less effective to protect retired people who suffer from a higher epidemic with a larger total number of infected people at the end of the epidemic. 

These two strategies, the societal optimum and the Nash equilibrium under constraints, suggest interesting guidelines for public health executives to mitigate an epidemic through collective immunity. First, quite naturally, sufficiently strong constraints should be imposed at the epidemic peak to avoid saturation of the sanitary system; and the constraints need to protect people at risk, which implies to limit contact both among these people as well as between the rest of the society and these individuals. On the other hand, in a perhaps less intuitive way, constraints on people who are not at risk should be relatively light. 
Indeed, the epidemic needs to spread on the population, in a controlled way, to reach as fast as possible the collective immunity. After the epidemic peak, one can lift progressively the constraints, until the collective immunity is reached. At this point, the epidemic will be back at a low level and will stay low while the constraints can be completely lifted. The precise characteristics of the constraints, such as their intensity or their timing, will depend on the characteristics of the population and of the disease under consideration. 
However, strategies that induce epidemic rebound, like the Nash scenario with naive constraints described above, are quite ineffective in such a context, because the time span between the peaks does not help reaching collective immunity and is very costly in terms of  constraints on the society.

\subsubsection{Comparison of global cost for the Nash equilibrium under different constraints}

\begin{figure}[t!]
    \centering
    \includegraphics[scale=0.5]{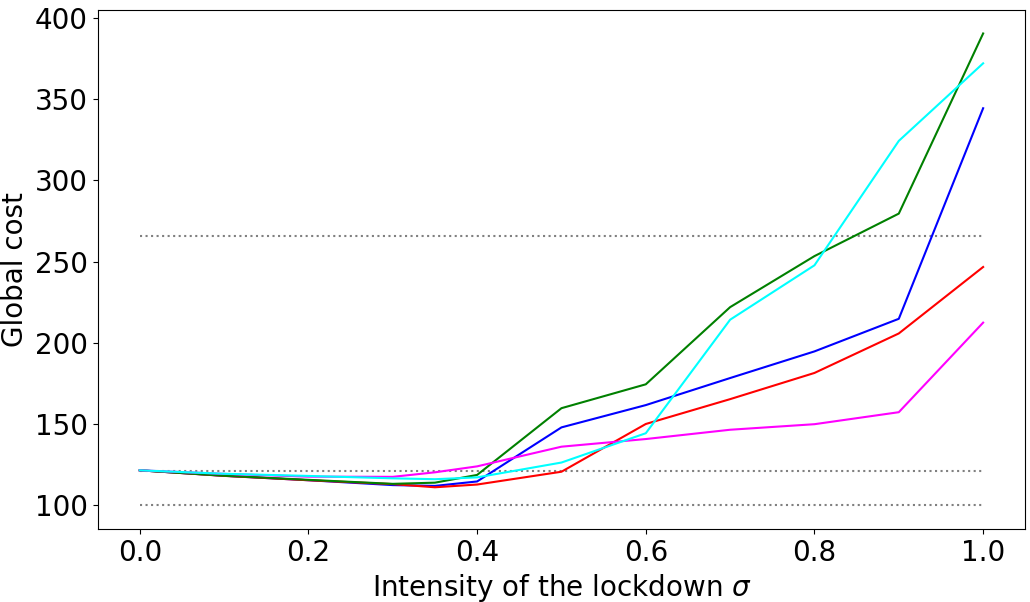}
    \caption{Comparison of global cost for different parameters of the constraints. The $x$-axis correspond to the intensity of the lockdown $\sigma$, which could vary from $0$ (no constraints) to $1$ (maximal constraints). The different curves correspond to different choices for the two threshold parameters $I_l$ and $I_d$. We choose $I_l = (0.12,0.08,0.04)$, a too low $I_l$ will clearly deteriorate the situation as it will impose a  duration of the constraints which is too long to reach collective immunity. 
    A higher $I_l$ is, on the other hand, not effective, as typically the maximum effort with the free Nash equilibrium is around $0.15$ for our choice of parameters, and thus the threshold would never be reached. For $I_d$ we took $I_d = (1.10^{-2},4.10^{-4},1.10^{-5}$). $I_d$ will have a major impact on the duration $\Delta t$ of constraints, with a log relation of the form $\Delta t \simeq - \textrm{log}(I_d)$. Increasing $I_d$ will decrease the extent of  lockdowns and conversely. A too high $I_d$ will lead to epidemic rebounds (the constraints is lifted too early), and a too low $I_d$ will impose useless extra social cost to the population. Blue curve $(I_l,I_d) = (0.08,4.10^{-4}$), red ($0.12, 4.10^{-4}$), green ($0.04, 4.10^{-4}$), magenta ($0.08, 1.10^{-2}$) and cyan ($0.08, 1.10^{-5}$). Dotted gray horizontal lines from top to bottom correspond respectively to business as usual cost, free Nash equilibrium, and societal optimum.}
    \label{fig:global_cost_constraints}
\end{figure}

We now study how the global cost for the Nash equilibrium under constraints changes with the three parameters of the constraint; results are displayed in Fig.~\ref{fig:global_cost_constraints}. The parameters used in Fig.~\ref{fig:contact_constraints} correspond to the minimum found here.

At $\sigma = 0$ we recover the free Nash equilibrium, with the same global cost, around $C_{\rm{glob}} = 120$. When the intensity $\sigma$ is increased, society carries a lower cost than in the free Nash equilibrium, because all individuals are forced to make some efforts. But at a certain intensity, a minimum is reached; the location of this minimum is mainly influenced by $r_I$, and corresponds here to the region around $\sigma = 0.3- 0.4$. In this interval, we find the optimal lockdown configuration that we presented above with $\sigma = 0.35, I_l = 0.12, I_d=4.10^{-4}$. Among the three parameters $(\sigma,I_l,I_d)$ characterizing the partial lockdown, the one which has the most impact on the global cost is $\sigma$, as there are  no significant variations between the different curves of Fig.~\ref{fig:global_cost_constraints}.  For $\sigma > 0.5$, the constraints become too strong with respect to the epidemic threat for all choices of thresholds, but especially  for low $I_l$ and  $I_d$, because this imposes long constraints which become very costly as $\sigma$ increases. 
When $\sigma$ approaches $1$ we even reach a point above the business as usual scenario (which had $C_{\rm{glob}} = 266$), as we enter a regime characterized by a succession of lockdowns followed by epidemic rebounds which are suppressed by the next lockdown before herd immunity can be reached.

\section{Optimal scenarios for dealing with an epidemic from the health authority point of view}

\label{section:strategies}

Up to this point, we have only considered dynamics with a very long end-time $T$, and a large number of agents $N$, so that the only option to terminate the epidemic is to reach herd immunity.  However there are many circumstances (expected production of a vaccine, seasonality of the virus which is expected to disappear in the summer, etc..) where the finiteness of $T$ plays a role, and others (isolated geographic configuration such as islands, strict control of borders, etc..) where the finiteness of $N$ does. This opens the way to other possible strategies, from the point of view of the centralized health authority, to control the epidemics.  We review them in this section.

\subsection{The threefold way of controlling an epidemic} 
\label{section:different_strategies}

Based on these considerations, we can identify three possible ways to deal with an epidemic: reach collective immunity (typically for $T,N$ large), contain the epidemic (for $T$ small), or eradicate the epidemic (for $N$ small). We characterize these three ways as follows.\\

\paragraph*{Strategy n°1 : reach collective immunity.}
This is the strategy that was implicitly used in the previous sections since we assumed both $T$ and $N$ very large. More formally, we consider that collective immunity has been reached at time $t$ if the proportion of infected individuals is a decreasing function of time for $t' > t$ even in the absence of efforts after $t$. For the basic SIR model Eq.~\eqref{eq:SIR} with constant $\chi$, let $R_{\rm eff}(t) = S(t) R_0$ be the effective reproduction number at time $t$, that is, the average number of secondary infected caused by a single infected agent, with $R_0 = q \chi/\xi$ the initial value of  $R_{\rm eff}$ when $S=1$. For this model we have  $\dot{I}(t) = \xi I (R_{\rm eff}(t) - 1)$. In this case, collective immunity is reached as soon as $R_{\rm eff} (t)< 1$ since $S$ is decreasing.  In a similar way, for our compartmental model we introduce
\begin{equation}
\label{eq:R_alpha}
R_\alpha(t) = \frac{\mu q}{\xi} \sum_{\beta,\gamma} n_\alpha^\gamma(t) n_\beta^\gamma(t) M^\gamma_{\alpha \beta}  S_\beta(t) \; ,
\end{equation}
the average number of secondary infected caused by a single infected agent of age class $\alpha$. We stress that $R_\alpha < 1$ does not imply  $\dot{I}_\alpha < 0$, since the number of infected in the age class $\alpha$  involves the $R_\beta$ of all classes, and some of them may be greater than 1. On the other hand, if {\em all} the $R_\alpha$ are less than one, the average proportion of infected individuals, $I \equiv \sum_\alpha K_\alpha I_\alpha $  can be easily shown to be a decreasing function.  Indeed, from Eq.~\eqref{eq:SIR-ss}, we have $\dot{I} = \sum_\alpha K_\alpha S_\alpha \lambda_\alpha - \xi I$, and
\begin{equation}
\sum_{\alpha} K_\alpha S_\alpha \lambda_\alpha = 
{\mu q } \sum_{\beta, \gamma, \alpha} K_\alpha S_\alpha n_\alpha^\gamma(t) n_\beta^\gamma(t) M^{\gamma}_{\alpha \beta} I_\beta 
= \xi \sum_\beta K_\beta I_\beta R_\beta  \; ,
\end{equation}
where we used the sum rule $M_{\alpha \beta} K_\alpha = M_{\beta \alpha} K_\beta$ enforced by the symmetric nature of contacts.  We therefore have
\begin{equation}
\label{eq:Idot}
\dot I = \xi \sum_\alpha K_\alpha I_\alpha (R_\alpha -1) \; .
\end{equation}
In the absence of effort, the rates $R_\alpha(t)$ become $R^{(0)}_\alpha(t) = \frac{\mu q}{\xi} \sum_{\beta,\gamma} M^\gamma_{\alpha \beta}  S_\beta(t)$, and Eq.~\eqref{eq:Idot} becomes
\begin{equation}
\label{eq:Idot0}
\dot I^{(0)} = \xi \sum_\alpha K_\alpha I_\alpha (R^{(0)}_\alpha -1) \; ,
\end{equation}
where the superscript denotes the absence of effort. Since the $R^{(0)}_\alpha$ are obviously decreasing functions of time, the constraint that $R^{(0)}_\alpha(t) < 1$ for all age classes $\alpha$ is a sufficient, but not necessary, condition to have reached herd immunity.  This constraint is, however, too strong, and is actually not met in our simulations, even when herd immunity is achieved. We thus find more effective to replace it by  a heuristic condition obtained by assuming the $I_\beta$ to be not very different from the average $I$ (as can be seen for example in Fig.~\ref{fig:epidemic_simulations} towards the end of the epidemics). Using Eq.~\ref{eq:Idot0}, we get  $\dot I^{(0)} \simeq  \xi I ( R^{(0)} - 1) $, with 
\begin{equation} \label{eq:Rbar}
     R^{(0)} \equiv \sum_\alpha K_\alpha R^{(0)}_\alpha \; . 
\end{equation}
$R^{(0)}$ is also a decreasing function of time, and the heuristic criterion $R^{(0)}(t) < 1$ indicates that herd immunity has been reached at $t$. This empirical condition does not guarantee mathematically the absence of an epidemic rebound once $R^{(0)}(t) < 1$ (heterogeneous $I_\alpha$ could allow $\dot I^{(0)} > 0$). Nevertheless, we will check below numerically that for the cases we considered it does actually correspond to herd immunity \footnote{Our criterion is actually better suited to describe herd immunity at the end of the epidemics than, for instance, the one which requires $S < 1/ \tilde R_0$ with $\tilde R_0 = \rho(q \mu M/ \xi)$ \cite{Inferring_social_structure,diekmann1990definition}}. This strategy, where $S$ needs to be low at the end of the epidemics, is often used for moderate epidemics and for epidemics where no other strategy is available.\\


\paragraph*{Strategy n°2 : contain the epidemic.} If an external event (e.g.~vaccine) is expected to end the epidemic within a relatively short time, another possibility to deal with an epidemic is to contain it during  the period of optimization $T$, keeping the epidemic at a low level, and end at $T$ with a number of susceptible far above the collective immunity threshold. In practice, we are in this phase if $R^{(0)}(T) > 1$. This is the strategy adopted by most countries during the Covid-19 pandemic: hold on and contain the epidemic until  a vaccine is available.  \\

\paragraph*{Strategy n°3 : eradicate the epidemic.} A final possibility is to act 
on the epidemic sufficiently early and sufficiently  intensely, that one will be able to eradicate it before it spreads to the general population. 
To implement such an idea, we need to assume a finite size $N$ of the population, and state that below a certain rate of infected, of order $1/N$, the epidemic vanishes or is at least under control so that there is no propagation anymore. Of course in practice, one would need to know precisely who is infected and insulate them from the rest of the population (by keeping them  in quarantine at hospital for instance), which would induce an extra cost of coordination which is not taken into account here. Discussing this strategy requires to add one parameter, $I_{\rm{thr}}$,  which corresponds to the threshold at witch we consider that the epidemic vanishes, with a value for $I_{\rm{thr}}$ of order $1/N$. This approach  is in practice possible only during the early stages of the epidemic, otherwise it will induce a  considerable cost. This  strategy has been used many times in China and some insular countries during Covid-19 pandemic, with strong restrictions at the early stages of the epidemic to avoid a massive spreading.

\subsection{Template strategies}
\label{section: strategies_comparison}

The above scenarios can be classified according to whether $\dot{I}^{(0)}(t)<0, \; \forall t>T$ (herd immunity), and if this is not the case, whether $I(T) > I_{\rm{thr}}$ (containment) or $I(T) < I_{\rm{thr}}$ (eradication).  Thus, any set of strategies $n(.) \equiv \{n^\gamma_\beta(.)\}$ (i.e.~defined for each age class, in each setting, and  all times $t$) belongs to one and only one of these classes.  We can, however, do a little bit more than this formal classification, and introduce for each of these scenarios what we will call a ``template strategy'', that is, a  set of strategies $n(.)$  which  provides a good approximation to the optimal one within a given scenario.  These ``templates'' can be defined as follows:




\begin{itemize}
    \item \textit{Reach collective immunity} $n_{\textrm{im}}$ : Our template for the herd immunity scenario is defined as the optimal strategy defined in Section \ref{section:societal_optimum} taken in the limit $T \to \infty$ (with $I_{\rm{thr}} \equiv 0$), namely
    \begin{equation}
        n_{\textrm{im}}(.) = \underset{n(.)}{\textrm{argmin}} \left[ \ C_{\textrm{glob}}\left(n(.),T \longrightarrow \infty\right) \right] \; .
    \end{equation}   
Indeed, we can expect that when the best approach is to use herd immunity, there is little end-time effect and the optimal strategy for a finite $T$ will be quite close to the one corresponding to  $T\to\infty$.
    As seen in Fig.~\ref{fig:cost_template_strategies}, the global cost associated with $n_{\textrm{im}}$ rises quite significantly at the beginning of the epidemic, as a significant number of agents assume the cost of infection, but once herd immunity is reached this cost flattens out since infection decreases while no effort is required anymore. It can be noted furthermore that $n_{\textrm{im}}$ does not depend much on  $r_I$, as it minimizes the cost due to social contacts (which is independent from $r_I$), while reaching collective immunity. This leads in first approximation to a constant number of agents who have been infected at the end time $T$, as the collective immunity threshold is unchanged for any value of $r_I$. Therefore, the associated final cost of this strategy $n_{\textrm{im}}$ grows with a  form  $C_{\textrm{glob}} (n_{\textrm{im}}) \simeq F_{\textrm{tot}}(n_{\textrm{im}}) + (S_0-S_\infty)r_I$, where $F_{\textrm{tot}}$ is the total amount of efforts made by agents for a strategy $n(.)$, which is (almost) independent of $r_I$, and the second term grows linearly with $r_I$.\\   
    
    \item \textit{Contain epidemic} $n_\textrm{cont}$ : We define the reproduction factor $R$ as the $R^{(0)}$ which was introduced in Eq.~\eqref{eq:Rbar}, with here arbitrary value for $n(t)$ instead of 1. One can easily claim that a sufficient condition to strictly contain the epidemic in a homogeneous infected population is to keep $R(t) = 1$. With that condition, one will enforce $I(t)$ to stay as the same level or below the initial condition $I(0)$ with a priori the lowest possible cost from the social point of view (keep $R(t) < 1$ will be more expensive).  We can therefore define the template strategy of the containment scenario  as the one coming from the  optimization 
    \begin{equation}
        n_{\textrm{cont}}(t) = \underset{n(.)}{\textrm{argmin}} \left[ F_{\rm tot}(n(.)) {\rm \; such \; that \; } R(t) = 1 \quad\forall t\right] \; ,
    \end{equation} 
where we furthermore assume that for all age classes   $S_\alpha(t) \simeq S_\alpha(0) \simeq 1$, so that $ n_{\textrm{cont}}$ is actually time independent. Since the social cost only involves current time $t$, the problem reduces to a simple, local in time, optimization problem, where $n(t)$ becomes a constant $n$ which must respect $R = 1$ and minimize $f(n)$.
The result of this  optimization, obtained numerically through a gradient descent under constraints, is illustrated in Fig.~\ref{fig:cost_template_strategies}. Note that this (constant) strategy $n_\textrm{cont}$ is independent of $r_I$, and the associated global cost  $C_{\textrm{glob}} (n_\textrm{cont}) \simeq T f(n_\textrm{cont})$  is essentially independent of $r_I$ and  grows linearly with $T$.\\   

    \item \textit{Eradicate epidemic} $n_{\rm{era}}$ : For this case, it can be shown (see appendix \ref{sec:eradication}) that, for the parameters we consider,  the optimal eradication strategy is always obtained by an application of the maximal effort  until the time $t_{\rm{thr}}$ corresponding to the eradication of the epidemics, $I(t_{\rm{thr}}) \equiv I_{\rm{thr}}$. This strategy, will be taken as our template eradication strategy. The associated final cost is therefore expected to be of the form $C_{\textrm{glob}}(n_\textrm{era}) \simeq T f_{\textrm{max}}$  if $T < t_{\rm{thr}}$ ,  the cost grows linearly with $T$ , and $C_{\textrm{glob}}(n_\textrm{era}) \simeq f_{\textrm{max}} t_{\rm{thr}}$ if $T > t_{\rm{thr}}$, where $f_{\textrm{max}}$ denotes the social cost (rate) associated with a maximum amount of efforts and $t_{\rm{thr}}$ mainly depends on $I_{\rm{thr}}$. 

\end{itemize}

\subsection{Phase transition}
\label{sec:phase_transition}
\begin{figure}[ht!]
    \centering
    \includegraphics[scale=0.45]{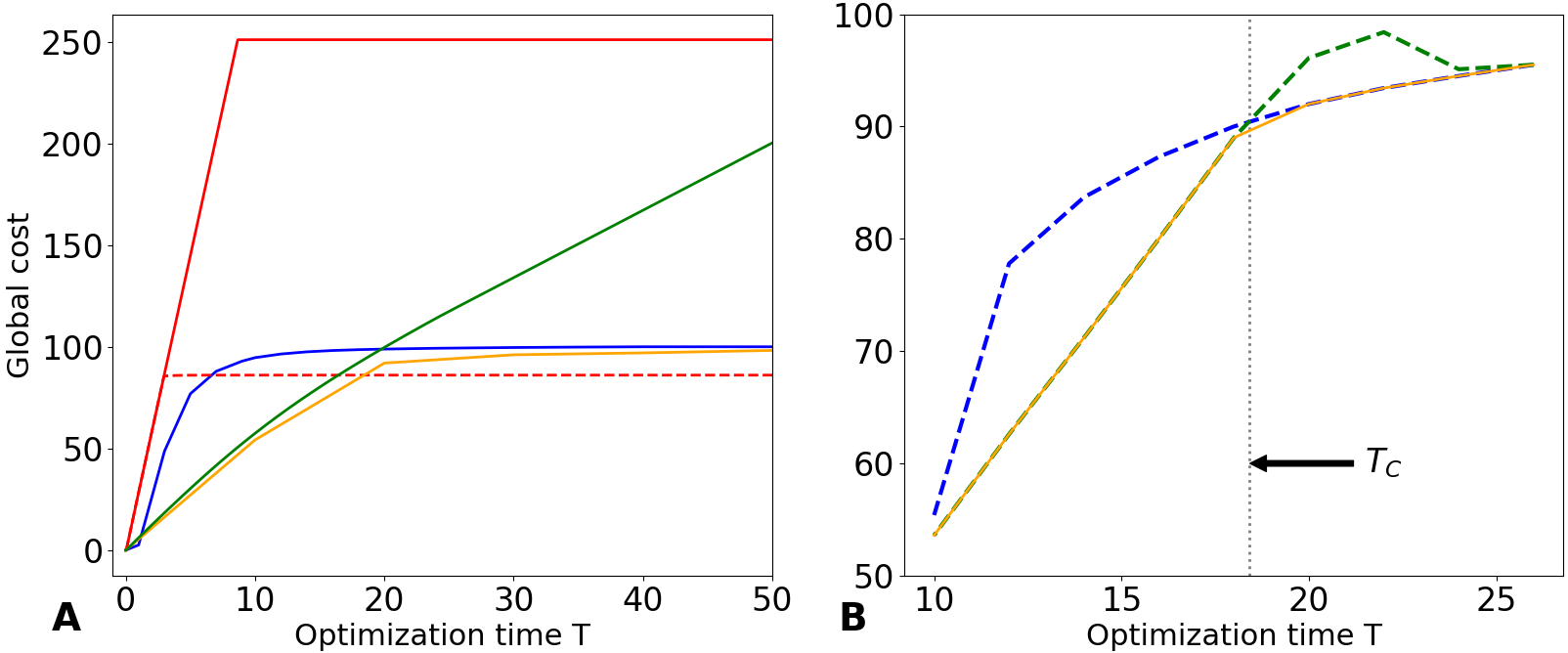}
    \caption{\textbf{A}. Comparison of the evolution of the global cost $C_{\textrm{glob}}(n,T)$ for the three template strategies $n_{\textrm{im}}$ (blue line), $n_{\textrm{era}}$ (red lines), $n_{\textrm{cont}}$ (green line) which are well defined for any value of $t$ (from $0$ to $\infty$). For the global cost associated to the eradication strategy $n_{\textrm{era}}$ (in red) we take respectively $I_{\rm{thr}} = 1.10^{-5}$ (resp. $I_{\rm{thr}} = 1.10^{-3}$) for the solid line (resp. dotted line). Regarding the strategy $n_{\textrm{im}}$, $T=\infty$ is approximated here by $T=100$. Finally in orange, we plot the true societal optimum cost at $T$ (with $I_{\rm{thr}} = 1.10^{-5}$, solid line parameters). \textbf{B}. Evolution of the global cost of the societal optimum (orange solid line) close to the transition time $T_c$ (see text). Dotted blue (resp.~green) line : evolution of the global cost with a continuous change of the strategy $n$ for the herd immunity scenario (resp.~containment scenario). Details of the computation are explained in the main text.}
    \label{fig:cost_template_strategies}
\end{figure}

\begin{figure}
   \includegraphics[scale=0.4]{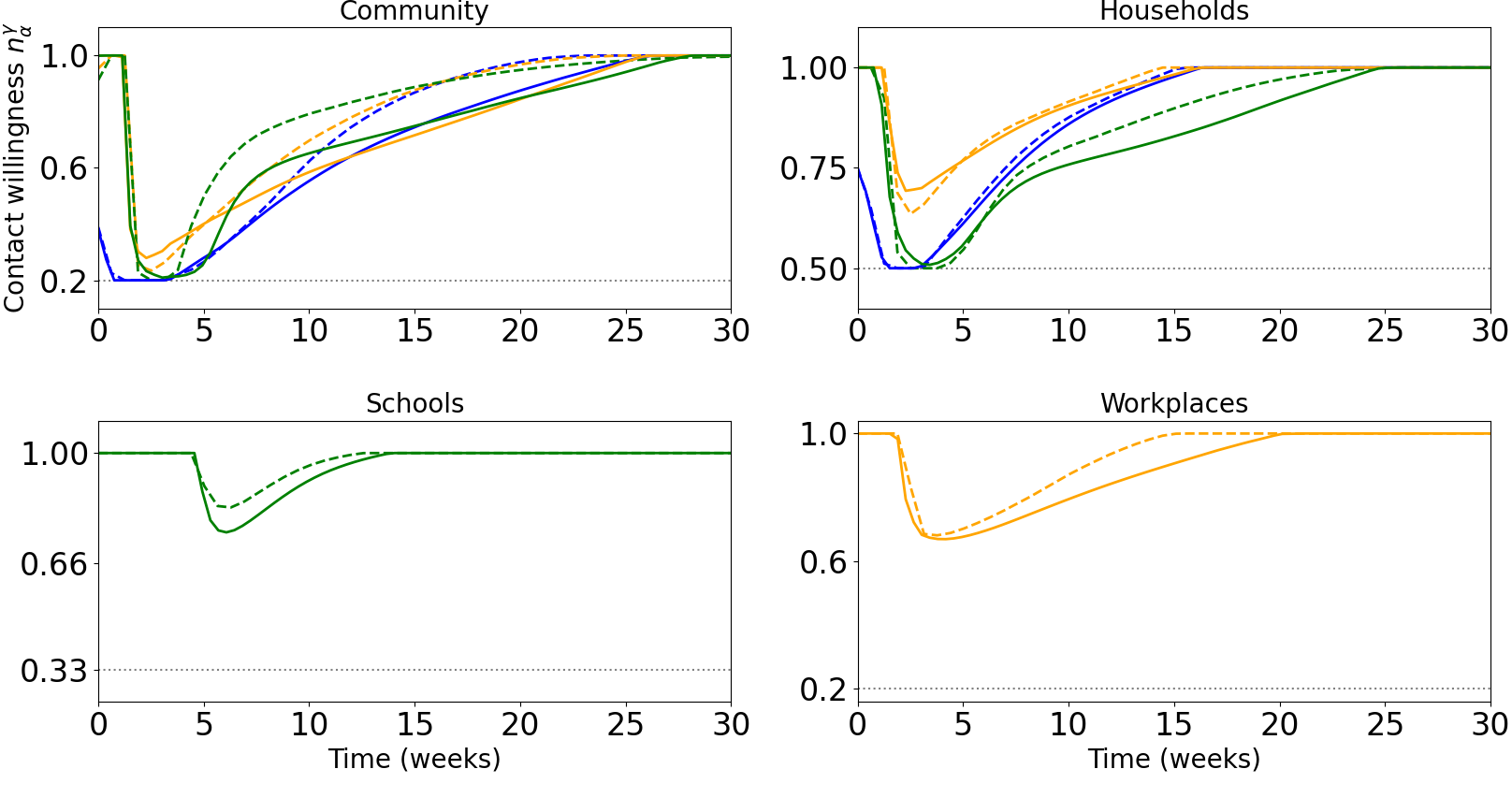}
   \includegraphics[scale=0.4]{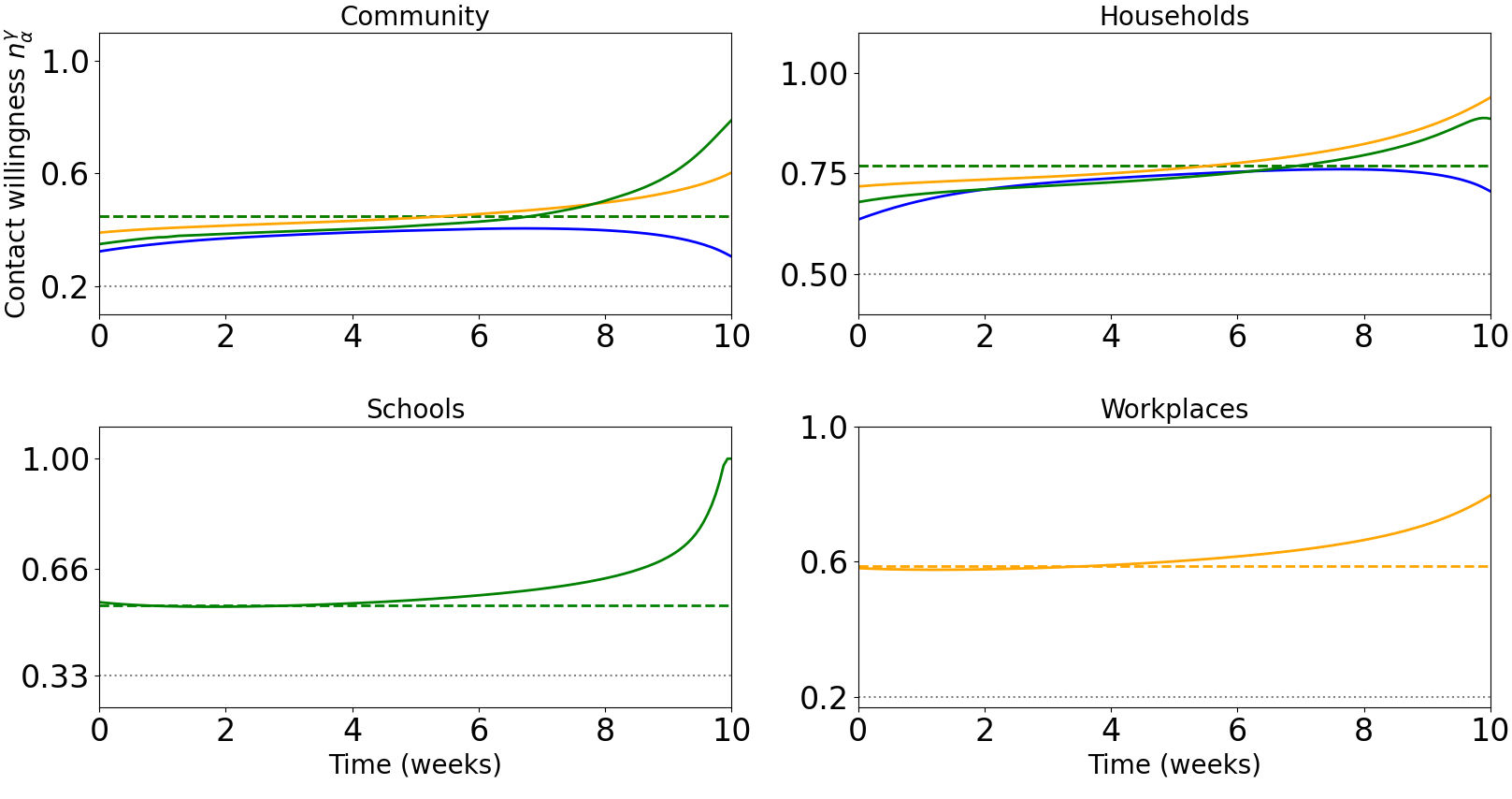}
   \includegraphics[scale=0.4]{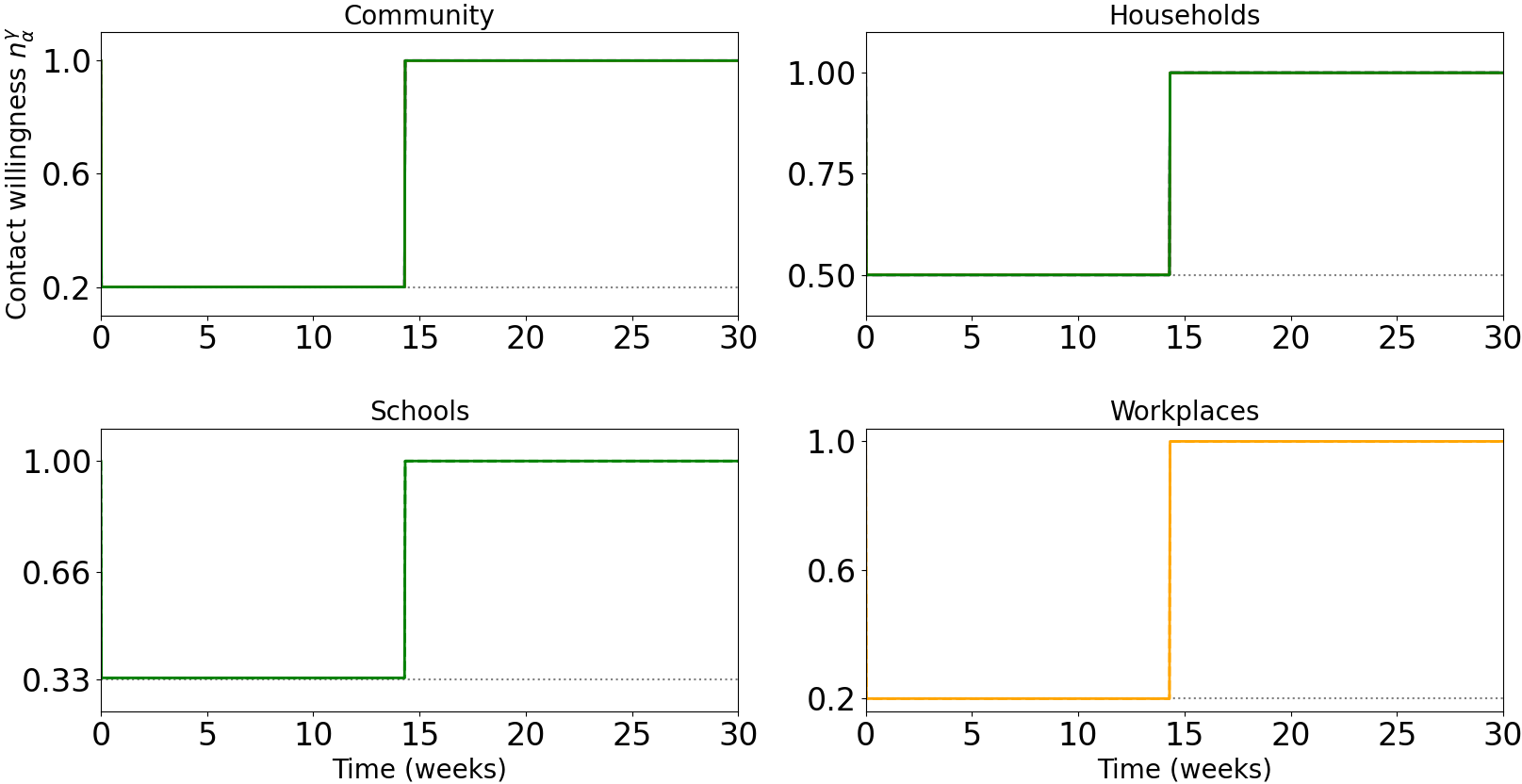}
   \caption{Contacts willingness for the three template strategies defined in Section \ref{section:strategies} (dotted lines) and the (finite-$T$) societal optimum for the corresponding parameters (solid lines). Rows 1-2: collective immunity ($T\to\infty$, computed in practice with $T=100$ and $r_I = 1$, dotted line) and societal optimum (computed with $T=30, r_I = 1, I_{\rm{thr}} = 0$, solid line). Rows 3-4: contained strategy (dotted) and societal optimum (solid) for $T=10, r_I = 1$. Rows 5-6: eradication strategy  (dotted) and societal optimum (solid) for $T=30, r_I =1, I_{\rm{thr}} = 1.10^{-5}$ -- the two strategies matches perfectly. Sub-panels and legends are the same as in Fig.~\ref{fig:contact_nash}.
\label{fig:contacts_comparison_template_reality}}
\end{figure}

For these three scenarios, we show on Fig.~\ref{fig:cost_template_strategies}A the evolution of the global cost with the optimization time $T$, for $r_I=1$ and the parameters of Table~\ref{table:M}. As expected, all costs increase with $T$, but in different ways. In blue, the collective immunity cost grows rapidly at the beginning of the epidemic, so that collective immunity is reached as soon as possible without saturating the sanitary system, after which the cost levels up. For the containment strategy $n_{\textrm{cont}}$ (green), we see that the corresponding cost increases almost perfectly linearly, as the amount of effort due to contact reduction is constant. As $S(0) = 0.99 < 1$, there is in this scenario a small spread of the infection at the beginning of  the epidemic  (and thus a small additional infection cost), before it vanishes completely.  
Finally the cost of the eradication strategy (red curve) starts with a strong linear increase (the slope of the curve here is clearly higher  than the one of the containment strategy since the maximal effort is applied), and then saturates at a level which depends on the threshold $I_{\rm thr}$.
Figure \ref{fig:cost_template_strategies}A also shows  the societal optimum cost (orange curve), which always closely follows one of the templates. At low $T$, it is a bit below the cost of the containment strategy $n_\textrm{cont}$, taking advantages of end-time effects (as illustrated in Fig.\ref{fig:contacts_comparison_template_reality}) to slightly reduce the cost. For large $T$, it follows, again from below, the collective immunity template. For the societal optimum cost, there is a transition around $20$ weeks for our choice of parameters, from a ``containment'' cost to a ``collective immunity'' cost. For $I_{\rm thr}= 10^{-3}$ (dotted line in Fig.~\ref{fig:cost_template_strategies}), the transition would go from ``containement'' to ``eradication''.  \\

This transition between different scenarios' costs strongly suggests that the associated strategies will follow the same pattern, with a transition form the neighborhood of $n_\textrm{cont}$ to the neighborhood of $n_\textrm{im}$.  
To assess this, we compare in Fig.~\ref{fig:contacts_comparison_template_reality} the optimal strategy found from the societal optimum approach with the template strategies. We observe that the small gap between template costs and societal optimum cost which was observed on Fig.~\ref{fig:cost_template_strategies}A corresponds to a small difference between the corresponding strategies. For strategy 1 (rows 1-2) we observe a finite-$T$ effect: an additional amount of efforts around $10$ to $25$ weeks appears to be profitable to limit the number of infected, even though the epidemic is almost over. The structure of the two strategies is nevertheless very similar. Regarding the ``containment'' strategy (rows 3-4), in each setting the contact willingness of each age class of agents is the same (thereby, only one constant dotted line per setting is plotted)
The societal optimum is very close to the strategy $n_{\textrm{cont}}$, but two effects make it deviate from the idealistic strategy $n_{\textrm{cont}}$. First, as $S(0)$ is not strictly equal to one (here $0.99$), there is some moderate spreading of the epidemics, which induces a small increase of effort  from retired people, as well as a small increase of infection cost. Second, there is a clear end-time effect, meaning here that individuals who are not at risk reduce their efforts just before $T$ since epidemic will not have time to propagate massively until $T$ (one can think of a vaccination campaign where individuals will start increasing their contacts before the campaign is completed). Note however that as $T$ gets close, since the epidemic begins to grow, retired individuals protect themselves and actually further limit their contacts. 
Lastly, for the eradication strategy, the societal optimum is the same as our template strategy $n_{\textrm{era}}$ (see appendix~\ref{sec:eradication} for more details).

Figures~\ref{fig:cost_template_strategies}A and \ref{fig:contacts_comparison_template_reality}  indicate that our template strategies provide an accurate approximation of the societal optimum at small and large $T$. One question we may ask now is  whether  the transition we see  at $T_c \simeq 20$ from one scenario to another can be understood as a true phase transition, or is rather of a crossover type.
To address this question, in Fig.~\ref{fig:cost_template_strategies}A we compare the societal optimum  near $T_c$,  i.e. the absolute minimum of the global societal cost, with the result of a gradient descent obtained in the following way : starting from above $T_c$ (blue) or below (green), we change $T$ by small steps $\delta T$, and use as a starting point for the gradient descent at $T+\delta T$  the result of the calculation at $T$.  What we observe is that doing this procedure, our algorithm finds, for a significant range of $T$ values around $T_c$ a local minimum which follows the herd-immunity template below $T_c$ (dotted blue) or the containment template above $T_c$ (dotted green).  This local minimum corresponds either to the true minimum when the blue or green curves match the orange one, and to a metastable state when they do not. Note that both local minima eventually fall to the global minimum (in orange) when they are sufficiently far from $T_c$, ending in a hysteresis cycle. 

There is therefore a discontinuous change of the optimal strategy at $T_c$,  which is  the  signature of a first-order phase transition. In this analogy with thermodynamics, the cost $C_{\textrm{glob}}$ represents the free energy,  and $T$ some macroscopic parameter such as temperature. The Ehrenfest classification, which defines a first-order phase transition  as a discontinuity of the first derivative of $C_{\textrm{glob}}$ with respect to $T$ at $T_c$, is clearly observed  in   Fig.~\ref{fig:cost_template_strategies}.A. We expect this phase transition to exist for a large range of parameters of our model, and we have verified its existence numerically on a number of cases. In particular, we have checked that the transition between ``containment'' phase and ``eradication'' phase is also  first-order. \\

\begin{figure}
    \centering
    \includegraphics[scale=0.55]  {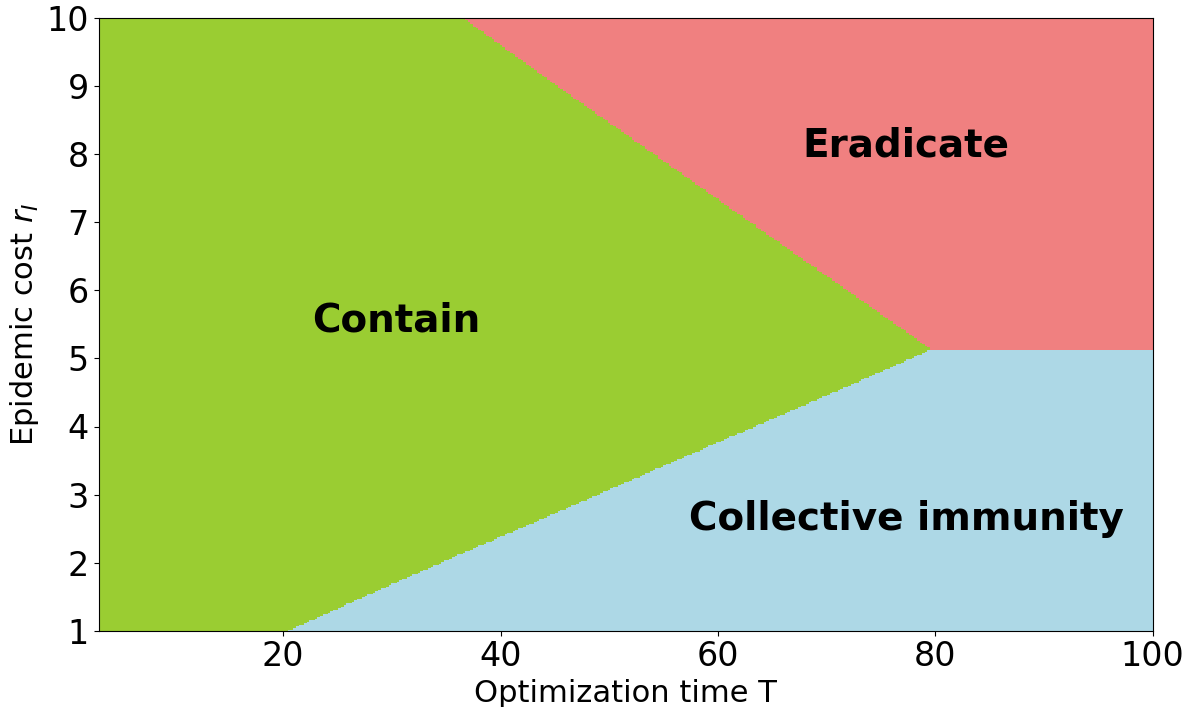}
    \caption{Phase diagram showing the best type of strategy to follow among ``reach collective immunity'' (blue), ``contain'' (green) and ``eradicate'' (red) with the parameters of Table \ref{table:M} and $I_{\rm{thr}} = 1.10^{-7}$ for the eradication strategy (it is more realistic, as it means $N \simeq 10^7$). Change $I_{\rm{thr}}$ or the initial conditions will naturally change the transition lines (between immunity and eradication areas)}
    \label{fig:color_map}
\end{figure}

We therefore end up with three distinct phases for the societal optimum, which exhibit first-order phase transitions between them, and which are well-approximated by template strategies defined above. Since these template strategies provide good approximations of the societal optimum one, we use  them in Fig.~\ref{fig:color_map} to show the ``phase diagram''  of the optimal scenarios as a function of the optimization time $T$ and the infection cost $r_I$. Of course, the optimal strategy will depend on all the parameters that we have introduced until now, but some of them (matrix of contacts $M$, capacity of the sanitary system $\nu_{\rm sat}$, proportion of agents in each age class $K_\alpha$) may be assumed to be quite similar for different epidemics affecting the same population, while $T$ and $r_I$ depend a lot on the virus under consideration and have a major impact on the best strategy. The three different scenarios appear to be optimal in distinct well-defined areas of the phase diagram. When $T$ is small (below $20$ weeks), the containment strategy is optimal whatever $r_I$. Then, there is a transient regime, where the optimal strategy can be  any of the three scenarios, collective immunity, containment,  or eradication according to $r_I$. Finally, after $T \simeq 80$ weeks, containing the epidemic is no longer an option, as the linear increase of the cost becomes prohibitive, and the best choice is either to reach collective immunity or to eradicate the epidemic. Since we use template strategies, the first-order phase transitions are represented by linear lines on the graph. 

\FloatBarrier

\section{Conclusion}
\label{section:conclusion}

In the present work we developed, following \cite{Inferring_social_structure}, an epidemic model based on the  well-known SIR compartmental model supplemented by a social structure. This social structure relies on the idea that contacts are heterogeneous in society, both because individuals socialize in different contexts, and because they react in various ways to the disease (different perception of risk). 
Therefore, one can divide society into classes of agents which differ by their behavior, by the risk that the disease represents for them, and by the settings in which socialisation takes place. 
Here we used an age differentiation, but other kinds of classification (e.g.~based on the immune status or on the presence of comorbidity) could easily be implemented within the same formalism. In the same way, one can easily add more compartments and more classes or settings to the model, without changing the global framework. 
The description of social structures obtained in this way is clearly less refined than one that would take into account the heterogeneity of social behaviors at an individual level, but it probably represents a good balance between precision and ease of application when trying to understand the dynamics of an epidemic and take appropriate, targeted action against it. 

To this compartmental epidemic model with social structure, we have, following the approach of Turinici et al. \cite{Turinici_contact_rate_SIR_simple}, added a mean field game description of the dynamics: agents may change their individual behavior depending whether they feel at risk of infection or not. 
Once the mean field game equations are derived, we computed numerically the Nash equilibrium, where each individual seeks to optimize his or her own interests. In this paradigm, individuals make a perfectly rational optimization, and are assumed to be able to performed the corresponding calculations which is  something that we cannot expect from people in practice.  The assumption here is thus rather than some central authority  will solve the system \eqref{eq:SIR_kolmogorov}-\eqref{eq:HJB2} and provide to individuals their ``best individual behavior'' $n^\gamma_\alpha$ which will be followed by agents if they sufficiently trust the institution.

The Nash equilibrium obtained within the Mean Field Game framework reduces significantly the costs associated with the epidemic when compared to the ``business as usual'' approach where social contacts are kept unchanged.  However, there is usually still a gap between the MFG cost and the one that would correspond to the societal optimal policy, which represents the minimal global cost that can be borne by the society.
To approach this optimal policy, we introduce the notion of ``constrained Nash equilibrium'', in which we assume that under some conditions, the central authority can impose some constraints, analog to the partial lockdowns that we have seen during the Covid-19 epidemic, under simple rules which are known to the agents.
In our work, we used a simple restrictive policy with three parameters ($\sigma, I_l, I_d$) and we optimized this policy (i.e.~we find the optimal set  ($\sigma^*, I_l^*, I^*_d$))  to get the lowest possible societal cost, and in this way close as much as possible the gap between the free Nash equilibrium and the societal optimum  (see Figs.~\ref{fig:contact_constraints} and \ref{fig:cost_comparison}).

In our discussion of the Nash equilibrium and of the ``constrained Nash'' approach to the societal optimum, we have implicitly limited ourself to a regime of very long optimization time $T$, and of large population $N$, for which the societal optimum policy necessary implies in some way to reach herd immunity.
In Section~\ref{section:strategies}, we go back in more details to the analysis of the societal optimum,  in particular lifting these constraints on $T$ and $N$.
Depending (mainly) on the values of $T$, $N$, and $r_I$, we can identify three {\em phases} that we  label as ``reach collective immunity" (the one implicitly assumed in the previous sections),  ``contain the epidemic'' or ``eradicate it'' (see Fig.~\ref{fig:color_map} showing which scenario is optimal depending on the  parameters $T$ and $r_I$).  The transition between any two of these phases can by understood as a  first-order phase transition, in the sense that the associated strategies present discontinuities and are different from one phase to another. An important consequence of this discontinuity is that it is primordial for an authority to clearly identify the appropriate scenario, as a wrong choice could lead to significant additional costs. 

Among these three scenarios, ``reach collective immunity'' is the one for which the time dependence of the agent strategies $\{n^\gamma_\alpha(.)\}$ are the more complex, and an authority will probably not be able to impose such exact strategy for all individuals. For this scenario, an approach through a mean field game paradigm under constraints as the one presented in this work is probably more relevant to approach the societal optimum cost, which would slightly shift the phases boundaries in Fig.~\ref{fig:color_map}.  
On the other hand, the ``containment strategy'' appears to be easier to design for an authority, as it consists in adjusting in real time the constraints, depending on whether the epidemic is growing or not, to follow $R(t) \simeq 1$.   Nevertheless, to find the best set of constraints to hold  $R(t) \simeq 1$  still involves some complexity, as one should still adapt the strategy to the response of individuals. Advantage of this scenario is that this can be performed ``on the fly'', and does not really imply any anticipation. Finally, in the ``eradication strategy'', authority has to impose the maximum admissible constraints, which is conceptually rather simple. We stress, however, that, contrarily to  the ``herd immunity'' strategy, the societal optimum obtained with strategies ``contain'' and ``eradicate'' are very far from any  Nash equilibrium, even under ``reasonable'' constraints.  The restrictions imposed with the two latters strategies lead to epidemics which stay at low levels.  In this context, the best individual strategy is to do essentially no effort, as there is almost no risk of infection. The social optimum strategy in this case  is thus extremely far from the Nash equilibrium. This emphasizes a profound difference in nature between ``herd immunity'', where individual optimization is closed to the societal optimum, and the two others where the gap is much more important. This would need to be considered by institutions when they will built collective strategies, as it is presumably very difficult to convince a population to follow on its own will  a strategy which is far from a Nash equilibrium, and the required degree of coercion would significantly vary between the two cases.   \\

The aim of this paper is to contribute to the construction of a theoretical framework on which authorities can rely to build appropriate policies against future epidemics. Our work emphasized both the challenge of this task and the extensive research which remains to be done.  Indeed, our model still involves a number of parameters. While some of them (as the matrix $M$) are known or could be relatively easily  extracted from field data, some others (as $r_I$ or the shape of $f$) are harder  to apprehend, although they are crucial if one wants to use such type of models in an appropriate way.  The model can be furthermore made more accurate with the addition of  some extra cost such as the one associated with coordination  in the case of restrictive policies. The question of evaluating quantities such as the risk induced by a possible epidemic is of course not specific to our model, and is  is actually one major task of epidemiologists.  Here however we hope to provide a more formal framework from which possible course of action can be decided  from that information. 

From a theoretical perspective, further research  could also  be performed to improve the framework. First, one may want to integrate the spatio-temporal character of the dynamics taking into account heterogeneity of populations and regions around the world. Secondly, one could include, in the impact of constraints on individuals behavior,  the feedback of the latter with respect to the imposed constraints. This is referred as Stackelberg games \cite{aurell2022optimal}, which involve a set of agents (small players) and a principal player corresponding to authorities. This sort of games should reveal the importance of getting the agreement of the population or not, depending of the choice of constraints. Thirdly, we did not incorporate explicitly in our model the possible presence of a vaccine. Vaccination campaign also involve individuals behaviors and could be studied from a mean-field game point of view \cite{Turinici_vaccination}. It can be added to the model but will rather concern another part of the epidemic, once vaccine is available, to optimize the vaccination campaign. A final active research domain is to infer accurately epidemic quantities with limited data sets, which it is almost always the case at the beginning of epidemics where limited number of tests are available. \\

Even without these improvements, the theoretical framework presented here should already be sufficiently flexible and realistic to be helpful in practice, as one could replace $f$ or the generalized infection cost $\tria$ by the precise forms that would be obtained by field data, and then pursue the same analysis. We hope that authorities and institutions in charge of design policies against epidemics could use our work to improve accuracy of epidemics prediction as well as the efficiency of non-pharmaceutical interventions. 

\acknowledgements
Louis Bremaud would like to thank the Center for Quantum Techniques and the National University of Singapore for their hospitality.
\appendix

\section{Derivation of the SIR equations}
\label{app:SIReqs}

We derive here the SIR equations \eqref{eq:SIR} in a rigorous way to make the underlying hypotheses more explicit. Let $x_k(t) \in \{ s, i, r \} $ be the state of individual $k$ at time $t$. The relative proportions of susceptible, infected and recovered in a population of size $N$ can be written as
\begin{equation}
\label{eq:SIR_stochastic}
\left\{\begin{aligned}
S(t) &= \frac{1}{N} \sum_{k = 1}^N \delta_{x_k(t),s} \\
I(t) &= \frac{1}{N} \sum_{k = 1}^N \delta_{x_k(t),i} \\
R(t) &= \frac{1}{N} \sum_{k = 1}^N \delta_{x_k(t),r} 
\end{aligned} \right.
\end{equation}
with $\delta_{a,b}$ the Kronecker symbol.

Let us consider an individual $k$ which is susceptible at time $t$ (i.e.~$\delta_{x_k(t),s}=1$).   To become infected at time $t+dt$, this individual must meet an infected individual $l$ in the time interval $[t,t+dt[$, and this encounter must lead to a transmission of the disease.  Thus the proportion of individuals which are susceptible at time $t$ and infected at time $t+dt$ is given, for a given realisation of the Markov process, by
\begin{equation}
\label{eq:average_S}
S(t+dt) - S(t) = - \frac{1}{N} \sum^N_{k=1}  \sum^N_{l=1} C_{kl}(t) \ \delta_{x_{k}(t),s}  \delta_{x_{l}(t),i} dt \; ,
\end{equation}
with $C_{kl}(t)$ the stochastic variable which take value $1$ if $k$ and $l$ met during the interval $[t,t+dt[$ with a possible infection for $k$ (if $k$ is susceptible and $l$ is infected), and $0$ otherwise. This stochastic variable has an average value (over random realizations of the Markov process) which is the product of the probability of contact during $dt$,  $\frac{1}{N} \chi_k(t) dt$ by the transmission rate $q$ since both events are independent. We then take the average over realizations assuming the independence of the two stochastic variables $\delta_{x_{k}(t),s}$, and $\delta_{x_{l}(t),i}$ which amounts to assuming that the events ``individual $k$ is susceptible at $t$'', and ``individual $l$ is infected at $t$'' are independent because $N$ is large and the population is homogeneous. We get
\begin{equation}
\begin{aligned}
\label{eq:average_S_2}
\frac{d \langle S(t) \rangle} {dt} & = - \frac{1}{N^2} \sum^N_{k=1} \sum^N_{l=1}  q \chi_k(t) \langle \delta_{x_{k}(t),s} \rangle \langle \delta_{x_{l}(t),i} \rangle \\
 &= - \frac{1}{N} \sum^N_{k=1} q \chi_k(t) \langle \delta_{x_{k}(t),s} \rangle  \langle I(t)  \rangle  \; \; .
 \end{aligned}
\end{equation}
The next simplification is to assume that the contact rate $\chi_k(t)$ only depends on the state $x_k(t)$ of individual $k$ at $t$. That means that all individuals with the same status have the same contact rate $\chi_k(t)=\chi_{x_{k}(t)}(t)$. Denoting by $\chi(t)$ the contact rate of individuals which are susceptible at $t$, Eq.~\eqref{eq:average_S_2} reduces to
\begin{equation}
\label{eq:chi_mean_field}
 \frac{d \langle S(t)\rangle  }{dt} = - q \chi(t) \langle S(t)\rangle  \langle I(t)\rangle \; .
\end{equation}
The other SIR equations in Eq.~\eqref{eq:SIR} are obtained in the same way.

\section{Numerical implementation}
\label{app:numerics}

\subsection{Numerical resolution of the Nash equilibrium}
\label{app:numerics-Nash}

We describe here two numerical methods we have implemented to reach the Nash equilibrium : an inductive sequence method and a gradient descent. Again, we omit the superscript $\gamma$ to lighten the notations. 

\subsubsection{First method : inductive sequence}
\label{method1}

The first method is the most natural one. The idea is the following. We start with an initial global strategy ${n}_\alpha^{(0)}(.)$ (the brackets $(.)$ indicate that this initial strategy is given at all times), and we compute the associated epidemic quantities $(S^{(0)}(.), I^{(0)}(.), R^{(0)}(.))$ with Eq.~\eqref{eq:SIR_kolmogorov} for these given initial conditions. Then, using Eq.~\eqref{eq:nstar_explicit}, we compute the best individual response to this epidemics dynamics, ${n_a^*}^{(0)}$. Since the latter should be followed by all individuals, we obtain a new global strategy ${n}^{(1)}_\alpha = {n_a^*}^{(0)}$. We repeat the process until we reach the Nash equilibrium condition  ${n}^{(k)}_\alpha \simeq {n^*_a}^{(k)}$  for a sufficiently large $k$.

To summarize, the global scheme of this method is the following, performed simultaneously for all age classes $\alpha$:
\begin{equation}
\label{eq:scheme_inductive_sequence}
n_\alpha^{(k)} \: \: \: \underset{\textrm{Kolmogorov}}{\longrightarrow}  \: \: \: S^{(k)},I^{(k)},R^{(k)}  \: \: \: \underset{\textrm{Bellman}}{\longrightarrow}  \: \: \:   {n_a^*}^{(k)}   \: \: \: \underset{\textrm{symmetry of agents}}{\longrightarrow}   \: \: \: n_\alpha^{(k+1)} = {n_a^*}^{(k)}  \; .
\end{equation}
Each step is quite straightforward numerically since we only deal with classical partial differential equations. Equation \eqref{eq:scheme_inductive_sequence} corresponds to an inductive sequence  $n_\alpha^{(k+1)} = F(n_\alpha^{(k)})$ where the functional $F$ is defined as $F(n_\alpha^{(k)}) = n_\alpha^{*(k)} $. However, this inductive sequence will not always converge to a fixed point of $F$, which is why we consider a second approach below. 

In practice we discretized the interval $[0,T]$ with $T=40$ weeks using $\sim 150$ time steps; typically the number of iterations to reach the fixed points is $\sim 10$.


\subsubsection{Second method : gradient descent}
\label{method2} 

To deal with cases where the inductive sequence does not converge, we use a gradient descent on the variable $n_a(.)$ of the cost $C_a$ (see \eqref{eq:final_cost}) to reach the Nash equilibrium. We use the following scheme for each age class $\alpha$ with representative individual $a$

\begin{equation}
    \label{eq:gradient_descent}
    n_a^{(k+1)}(t) =n_a^{(k)}(t) - h \cdot \nabla_1 C_a \left (n_a^{(k)}(.), \{n_\beta^{(k)}(.)\}, t \right )\Big\vert_{n_a^{(k)}(.)= n_\alpha^{(k)}(.)} \; ,
\end{equation}
where $\nabla_1$ means that the gradient is taken on $n_a^{(k)}(.)$. The dot in Eq.~\eqref{eq:gradient_descent} indicates a scalar product, $h$ and $\nabla_1$ are vectors indexed by $\gamma$.  This scheme gives $\nabla_1 C_a \left (n_a^{(k)}(.), \{n_\beta^{(k)}(.)\}, t \right ) = 0$ when we reach the equilibrium. That is, we are at a local minimum of the cost $C_a$ with respect to the first variable $n_a(.)$. We can then check numerically that we are indeed at the true Nash equilibrium, that is, at a global minimum for the variable $n_a(.)$ (for each age class $\alpha$), by checking that $F(n_{\rm Nash})=n_{\rm Nash}$ for a given Nash candidate $n_{\rm Nash}$. 

In order to make the numerical computation of the gradient $\nabla_1 C_a$ less heavy and more efficient, we first perform a few analytical steps. To avoid heavy notations, the cost at $t =0$ will be denoted as $C_a\left(n_a,n_\beta\right)$. We have
\begin{equation}
\label{eq:cost2}
  C_a\left(n_a,n_\beta\right)\equiv C_a\left(n^\gamma_a(\cdot),\{n^\gamma_\beta (.) \},0 \right) =  \int_0^{T}   \left( f_\alpha \left (n^\gamma_a(s)\right )+\lambda_a(s) \ \tria(I(s))\right)(1 - \phi_a(s)) ds \; .
\end{equation}
In order to compute the gradient of the cost with respect to the first variable, we introduce 
the functional derivative of $C_a$ with respect to its first variable $n_a$, in the direction $h$ (with $h$ a function, usually a Dirac delta). By definition, 
\begin{equation}
\label{eq:def_Gateau_derivative}
    D_{h}C_a\left(n_a,n_\beta\right)  \equiv \underset{\epsilon \longrightarrow 0}{\lim} \: \frac{1}{\epsilon}(C_a(n_a +  \epsilon h, n_\beta) \: -  \: C_a\left(n_a,n_\beta\right) )  \; \; .
\end{equation}
Using the definition of the gradient, this functional derivative can be reexpressed as
\begin{equation}
\label{eq:Functional_derivative}
    D_{h} C_a\left(n_a,n_\beta\right) =  \int_0^T h(t) \cdot \nabla_1 C_a\left(n_a,n_\beta,t \right)) dt \; .
\end{equation}
Explicitly we get  $h(t)~\cdot~ \nabla_1 C_a = \sum_\gamma h^\gamma(t) \ \frac{\delta C_a}{\delta n_a^\gamma(t)}$ with $\frac{\delta C_a}{\delta n_a^\gamma(t)}$ the functional derivative of the total cost $C_a$ with respect to $n_a^\gamma(t)$. 
Since $1-\phi_a(s)=\exp\left(-\int_0^s \lambda_a(u) du\right)$, the cost \eqref{eq:cost2} depends on $n_a$ through the terms $f_\alpha(n_a)$ and $ \lambda_a$ via \eqref{eq:lambda_a}; note that $\lambda_a$ is linear in $n_a$. Using \eqref{eq:def_Gateau_derivative} we have at first order $\lambda_a(n_a+\epsilon h)=\lambda_a(n_a)+\epsilon  h \cdot d_{n_a} \lambda_a$ with $d_{n_a} \lambda_a$ a vector indexed by $\gamma$, of components
\begin{equation}
\label{eq:lambda_a0}
d_{n_a^\gamma} \lambda_a \equiv  \mu q \sum_{\beta=1}^{n_\textrm{cl}} n^{\gamma}_{\beta}(t) M^{\gamma(0)}_{\alpha \beta} I_{\beta}(t)  \; .
\end{equation}
We then use the integral form \eqref{eq:cost2} to expand Eq.~\eqref{eq:def_Gateau_derivative} to lowest order in $\epsilon$. One of the terms involves
a double integral; in order to put $D_{h} C_a\left(n_a,n_\beta\right)$ under the form \eqref{eq:Functional_derivative}, we invert integrands and change variables, namely $\int_0^T \left [ f(t) \int_0^t g(s)ds \right ] dt = \int_0^T \left [ g(t) \int_t^T f(s)ds \right ] dt$. Once the expression is of the form \eqref{eq:Functional_derivative} we can read off the value of the gradient $\nabla_1 C_a\left(n_a,n_\beta\right)$ :  we get
\begin{multline}
\label{eq:gradient_ind}
\nabla_1 C_a(n_a, n_\beta, t) = \\
\left [ d_{n_a} f_\alpha(n_a(t)) + d_{n_a} \lambda_a (t) \tria(I(t)) \right ] (1 - \phi_a(t)) \; - \; d_{n_a} \lambda_a(t) \int_t^T \left( f_\alpha \left (n_a(s)\right )+\lambda_a(s)
      \tria(I(s))\right)(1 - \phi_a(s)) ds \;  ,
\end{multline}
with $d_{n_a^\gamma} f_\alpha(n_a(t))$ the derivative of $f_\alpha(n_a(t))$ with respect to the variable $n_a^\gamma(t)$. The gradient \eqref{eq:gradient_ind} is then computed numerically in order to follow the scheme \eqref{eq:gradient_descent}.

\subsection{Numerical resolution of the constrained Nash equilibrium}
\label{app:numerics-constr}

For the constrained Nash equilibrium, the strategies $n_a^{k}(t)$ in Eq.~\eqref{eq:gradient_descent} additionally must fulfill constraints such as Eq.~\eqref{eq:lockdown}. Since these contraints are active or not depending on the value of $I(t)$, at each step $k$ one must check that the strategies
respect the constraints defined by the values of the epidemic rate at step $k$. Each step of the gradient descent therefore comprises two parts. In the first part, we perform the same gradient descent as the one described for the Nash equilibrium \ref{method2}, but now we check that the new strategies $\{n_a^{k+1}(.) \}$ respect the constraints defined by the $I(.)$ from step $k$; if they do not, we enforce them by correcting accordingly the $\{n_a^{k+1}(.)\}$. In the second part, we  compute the new epidemic rates and find the corresponding new constraints. 

An issue appears when we approach the Nash equilibrium. The variation of the constraints and of the strategy $\{n_a^{k+1}(.)\}$ can form some cycles which impede convergence. To bypass this difficulty, we choose to freeze the constraints at some step $k$ and continue the gradient descent process as in the method \ref{method2}; after some steps, we recompute the constraints and we continue the process until the convergence.

\subsection{Numerical resolution of the societal optimum}
\label{app:numerics-optim}

We can reach this optimal strategy through different ways. Here we choose to make a gradient descent on the cost  $C_{\rm glob}$, but one can also use the Pontryagin maximum principle \cite{tchuenche2011optimal}. We optimize the behavior of individuals to minimize the total cost paid by the population 
\begin{equation}
    C_{\rm glob}(n_\beta) = \sum_\alpha K_\alpha C_\alpha(\{n_\beta\}) \; \; ,
\end{equation}
where we omit the bracket to lighten the notations. In order to do this minimization, we will follow the same scheme as described in Eq.~\eqref{eq:gradient_descent}. We thus have to compute $\nabla C_{\rm glob}(n_\beta,t)$ which only involves all the collective strategies $n_\beta$ (there is no more individual behavior in this cost) and the instant $t$ where we compute the gradient. We will compute this gradient for each age class $\alpha$,
\begin{equation}
\label{eq:functional_derivative_cost}
     D_{h} C_\alpha(n_\beta) \equiv  \int_0^T h(t) \cdot \nabla_1 C_\alpha(n_\beta,t) dt \;  , 
\end{equation}
to identify $\nabla_1 C_\alpha(n_\beta,t)$  as in Section \ref{method2}, where  $\nabla_1$ is now on the global behavior $n_\beta$ and has components along $\gamma$ and $\beta$ (as does $h$). New terms appear because quantities such as the proportion of infected individuals $I(.)$ now depend on all $n_\beta$. We now sketch the main elements of the computation. First, we compute the expression of the functional derivative of the gradient $ D_h C_\alpha(n_\beta,t)$. Starting from the expression of $C_\alpha$  \eqref{eq:final_cost} we get
\begin{equation}
\label{eq:functional_derivative_cost_SO}
    D_{h} C_\alpha(n_\beta,t) =  D_{h} \left [ \int_t^{T}   \left( f_\alpha \left (n_\alpha(s)\right )+\lambda_\alpha(s) \ \tria(I(s))\right)(1 - \phi_\alpha(s)) ds \right] \; . 
\end{equation}
Thus, we need to compute each functional derivative of the terms appearing in \eqref{eq:functional_derivative_cost_SO}, which gives
\begin{align}
\label{eq:D_h_lambda}
    & D_{h} \lambda_\alpha(t) = \underset{\epsilon \longrightarrow 0}{\textrm{lim}} \; \frac{1}{\epsilon} \left [ \sum_\gamma \sum_\beta q M^\gamma_{\alpha \beta} (n^\gamma_\alpha(t) + \epsilon h^\gamma_\alpha(t)) (n^\gamma_\beta(t) + \epsilon h^\gamma_\beta(t)) \left(I_\beta(t) + \epsilon D_{h} I_\beta(t) \right ) \right] \\
\label{eq:D_h_phi}
    &D_{h} \phi_\alpha(t) = (1 - \phi_\alpha(t)) \int_0^t D_{h} \lambda_\alpha(s) ds \\
\label{eq:D_h_I}
    &D_{h} I_\beta(t) = \int_0^t \frac{\delta I_\beta(t)}{\delta n(s)} \cdot h(s) ds \\
\label{eq:D_h_f}
    &D_{h} f_\alpha(n_\alpha(t)) = d_{n} f_\alpha(n_\alpha(t)) \cdot h(t) \\
\label{eq:D_h_g}
    &D_{h} \tria(I(t)) =  
    \frac{\kappa_\alpha r_I \nu_{\rm sat}}{I_{\rm sat}} D_{h}I(t) \exp \left[\nu_{\rm sat} \frac{I(t)-I_{\rm sat}}{I_{\rm sat}}\right] \; \; ,
\end{align}
where the dots in Eqs.~\eqref{eq:D_h_lambda}-\eqref{eq:D_h_I}-\eqref{eq:D_h_f} indicate that $h$ and $n$ are indexed by $\beta$ and $\gamma$ and indices are summed over. In \eqref{eq:D_h_I}, $\delta I_\beta(t) / \delta n(s)$ indicates the functional derivative of $I_\beta(t)$  with respect to the collective behavior $n(s)$. This ``time delayed'' derivative is the crucial term of the gradient for the societal optimum, one can perform a linearization of Eqs.~\eqref{eq:SIR-ss} to propagate linearly the elementary deformation of $I_\beta$ from time $s$ to time $t$ to avoid several numerical computation of the whole epidemic. 
As in Section \ref{app:numerics-Nash} above, we use these expressions to compute explicitly Eq.~\eqref{eq:functional_derivative_cost_SO} and put it under the form Eq.~\eqref{eq:functional_derivative_cost}, which gives the expression of $\nabla_1 C_\alpha(n_\beta,t)$. We can then perform the gradient descent scheme Eq.~\eqref{eq:gradient_descent} numerically and efficiently without several computations of the whole epidemic.  

\section{Eradication strategy}
\label{sec:eradication}
In this section, we show that the optimal eradication strategy is to hold the maximum amount  of efforts in the interval $[0, t_{\textrm{thr}}]$ until the eradication of the epidemic when $I(t_{\textrm{thr}})=I_{\textrm{thr}}$, and then completely release the efforts. This strategy is sometimes referred in the literature as a bang-bang strategy \cite{roy2023recent}.  To show that this strategy is optimal, we have to show that any small reduction of efforts $\delta n$ made during $\delta t$ in the interval $[0, t_{\textrm{thr}}]$ will increase $t_{\textrm{thr}}$ so that the total cost paid by individuals will be higher. Without loss of generality, we consider that time $0$ corresponds to the time at which we start the efforts. We refer to this slightly different strategy as the deviating strategy, and the associated epidemic is denoted $\tilde I$. On the other hand, $t_{\textrm{thr}}$ will increase by a time $\delta \tau$, as the time at which epidemic vanish will be greater. We are left with a competition between two costs : $d_nf(n_{\textrm{min}}) \delta t \delta n$ which is the (negative) cost caused by the reduction of efforts (this is a gain from the individual point of view), and $\delta \tau f(n_{\textrm{min}})$ which is the extra (positive) cost that individuals will pay to eradicate the epidemic. To compare these costs, we need to evaluate $\delta \tau$ in terms of $\delta t$ and $\delta n$. \\

At $t_{\textrm{thr}}$, one has $I(t_{\textrm{thr}}) = 0$. For the deviating strategy, one has $\tilde I (t_{\textrm{thr}} + \delta \tau) = 0$, where $\tilde I(t) \equiv I(t) + \delta I(t)$, with $\delta I(t)$ the small difference amount of infected between the two strategies. We get
\begin{equation}
\begin{aligned}
& \left(I + \delta I \right) (t_{\textrm{thr}} + \delta \tau) = I(t_{\textrm{thr}}) \\
& \dot{I}(t_{\textrm{thr}}) \delta \tau + \delta I(t_{\textrm{thr}}) = 0 \\
& \delta \tau = - \frac{\delta I(t_{\textrm{thr}})}{\dot{I}(t_{\textrm{thr}})} \; \; ,\\
\end{aligned}
\end{equation}
which allows us to evaluate $\delta \tau$. Indeed, at time $t_{\textrm{thr}}$ we have $\dot{I}(t_{\textrm{thr}}) \simeq - \xi I_{\textrm{thr}}$, as the number of new infected is completely negligible at this point. A priori, since there is a little spread of the epidemic in the population we will have $\delta I(t_{\textrm{thr}}) > \delta I(0) \exp(- \xi t_{\textrm{thr}})$, and close to this value if $I(0)$ is small enough. Therefore, we get $\delta \tau > \frac{\delta I(0)}{\xi I_{\textrm{thr}}} \textrm{exp}\left(- \xi t_{\textrm{thr}}\right)$. At this stage, we need to give an order of magnitude for $t_{\textrm{thr}}$. We use that $I(t_{\textrm{thr}}) \simeq I(0) \textrm{exp}\left(- \xi t_{\textrm{thr}}\right) = I_{\textrm{thr}}$ and thus  $\delta \tau > \frac{\delta I(0)}{\xi I(0)}$. One can then easily show that $\delta I(0) \propto \delta n \delta t$ where the proportionality coefficient can by written in a formal way as $\frac{\partial \lambda}{\partial n}(n_{\textrm{min}}) S(0)$ where we omit age class notations (generalization is straightforward). Finally, we get the extra cost $\delta C$ paid by individuals 

\begin{equation}
\delta C = d_nf(n_{\textrm{min}}) \delta t \delta n  +\delta \tau f(n_{\textrm{min}}) > \delta t \delta n \left [ d_nf(n_{\textrm{min}}) + f(n_{\textrm{min}}) \frac{\partial \lambda}{\partial n}(n_{\textrm{min}}) S(0)\right ] > 0 \; .
\end{equation}
For any positive $\delta t, \delta n$, one can check that $\left [ d_nf(n_{\textrm{min}}) + f(n_{\textrm{min}}) \frac{\partial \lambda}{\partial n}(n_{\textrm{min}}) S(0)\right ] > 0$, where $ \frac{\partial \lambda}{\partial n} \propto I(0)$ with $I(0) \ge I_{\textrm{thr}}$. The extra cost paid by individuals for the deviating strategy is always positive, it is therefore worse than the initial one. The initial strategy presented at the beginning of this section is the optimal one in this sense. One can also argue that this local minimum is the true minimum among all eradicating strategies, as the above reasoning will be a priori true for higher values of $n$, considering the shape of $f$.

\section{Exploration of the parameter space}
\label{sec:exploration}

\begin{figure}[ht!]
    \centering
    \includegraphics[scale=0.38]{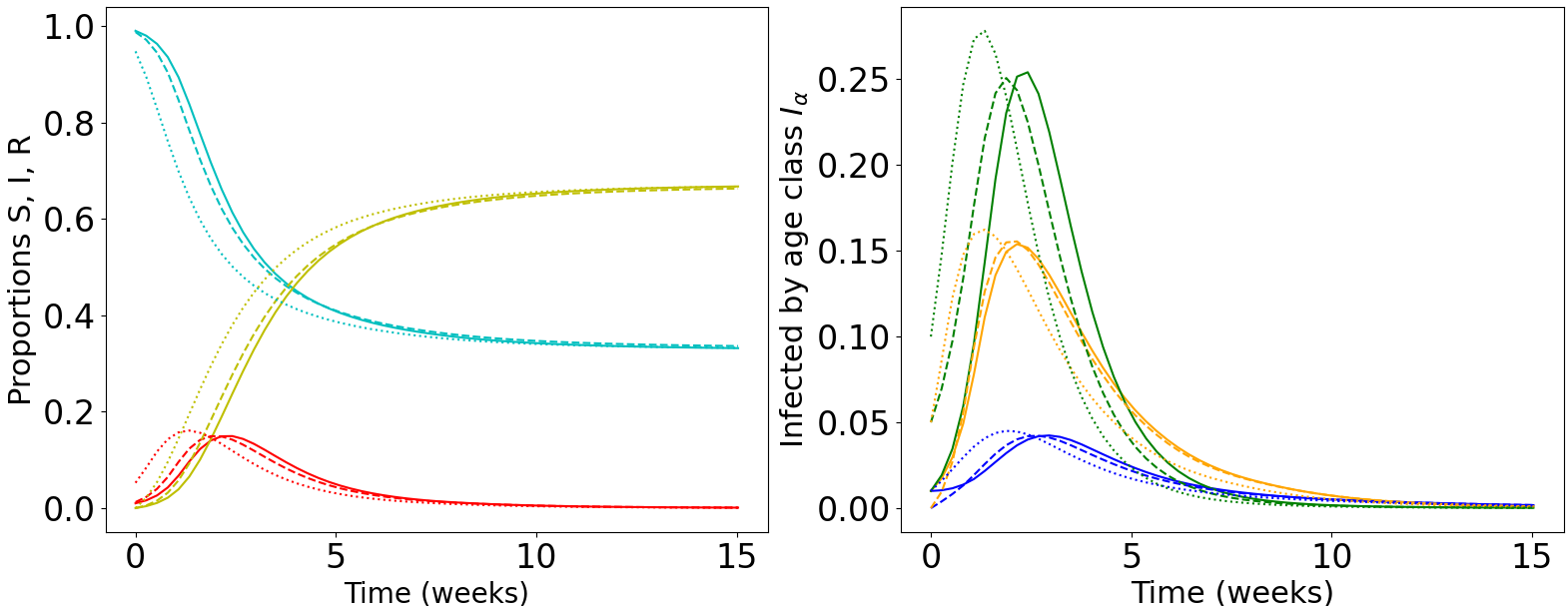}
    \includegraphics[scale=0.38]{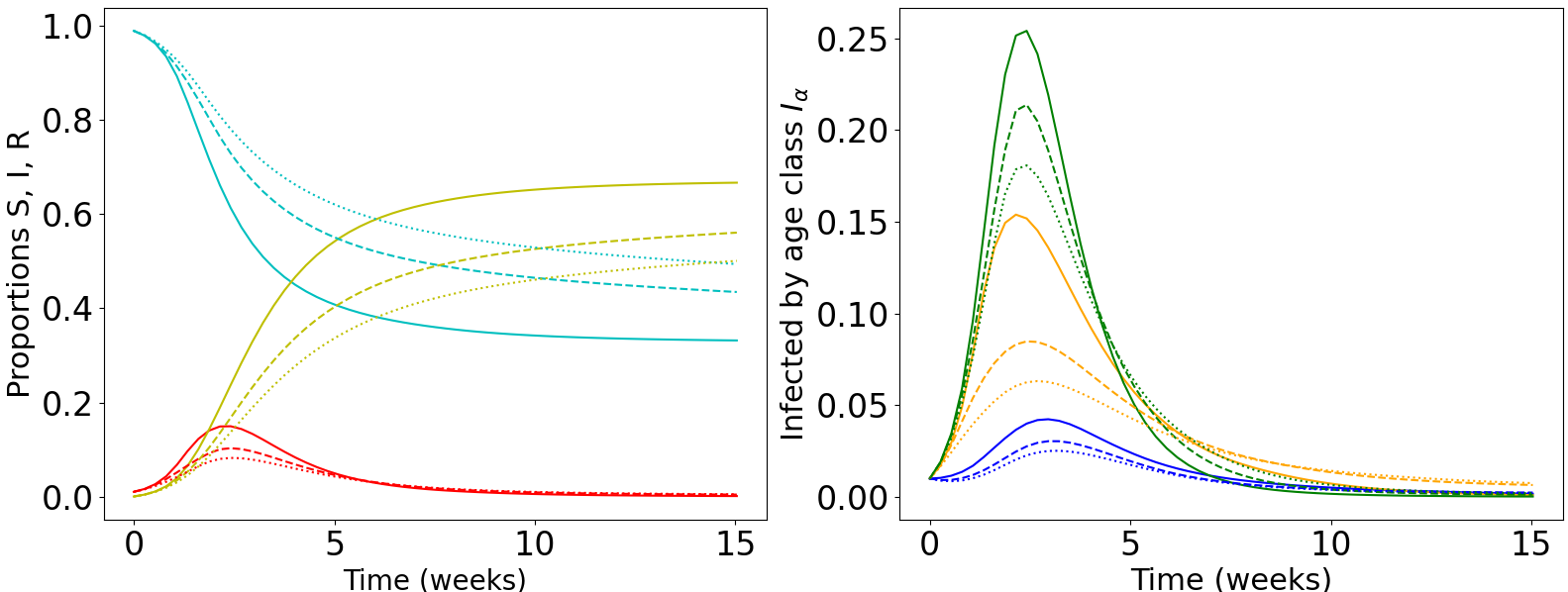}
    \includegraphics[scale=0.38]{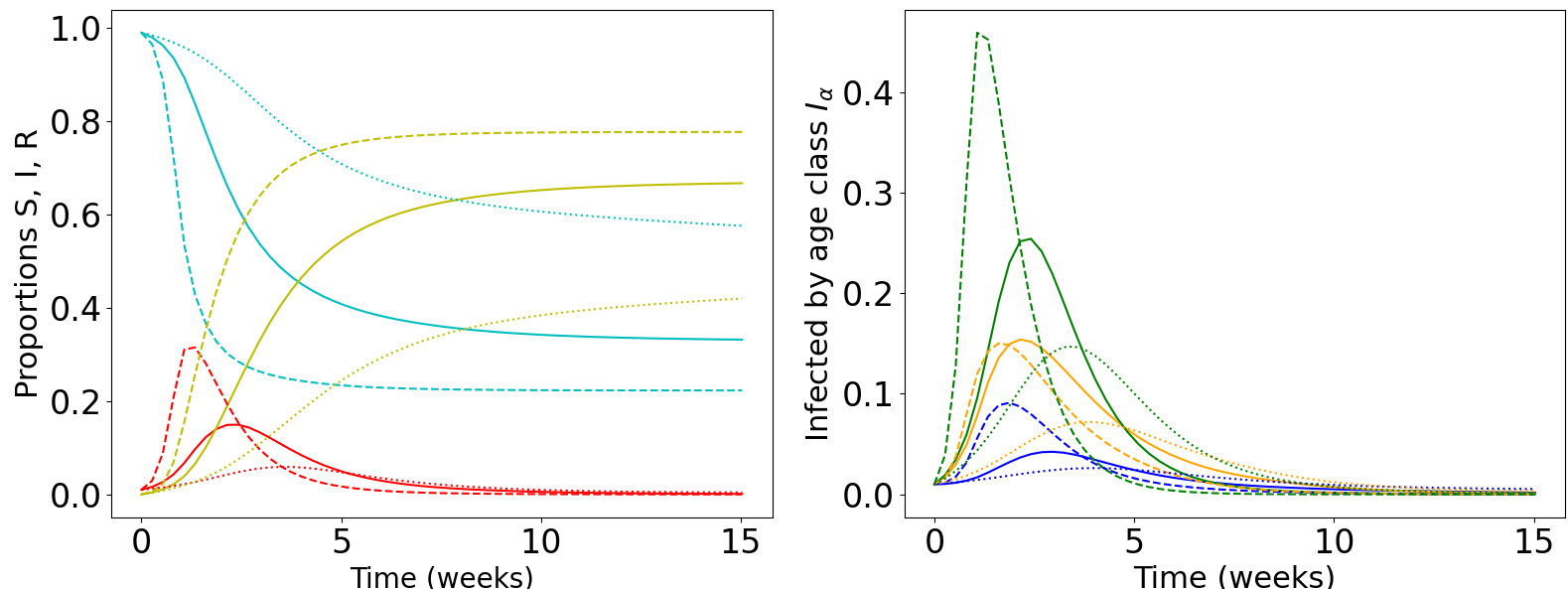}
    \includegraphics[scale=0.38]{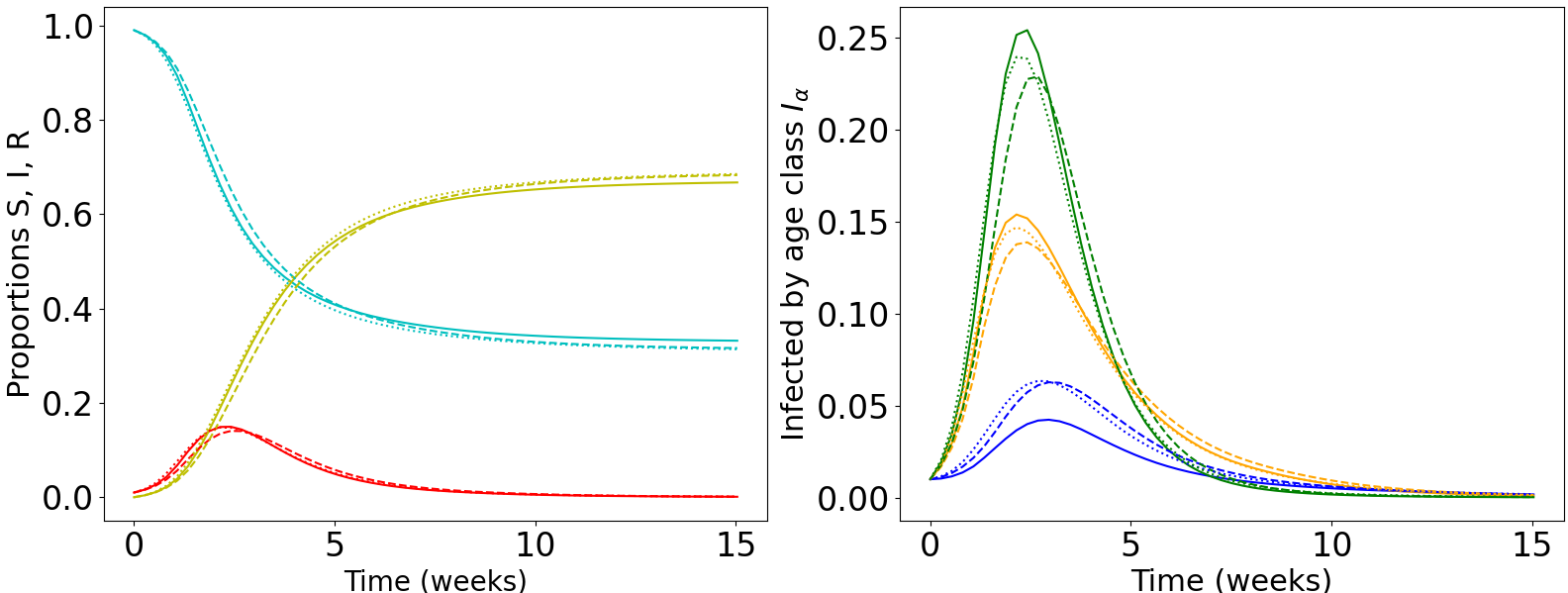}
    \caption{Comparison of Nash equilibrium epidemics for different set of parameters. We use the set of parameters of Table~\ref{table:M} with one (arbitrary but realistic) parameter change for each row (solid lines correspond always to baseline parameters). Color legend is the same as in Fig.~\ref{fig:epidemic_simulations}. First row : initial conditions change with $(S_0(0),S_1(0),S_2(0)) = (0.99,0.99,0.99) $ for solid line, dashed $(0.95,1,1)$ and dotted $(0.9,0.95,0.99)$. In each case, $I_\alpha(0) = 1 - S_\alpha(0)$ and $R_\alpha(0)=0$. Second row :  three different $r_I$ with $r_I = 1$ (solid), $r_I = 3$ (dashed) and $r_I = 5$ (dotted). Third row : three different proportions in the population, $(K_0,K_1,K_2) = (0.25,0.5,0.25)$ for solid line, $(0.6,0.2,0.2)$ for dashed lines, and $(0.2,0.2,0.6)$ for dotted lines. Fourth row : three different matrices $M_1$(solid), $M_2$(dashed) and $M_3$(dotted) defined in Table~\ref{table:Matrix_appendix}.} 
    \label{fig:epidemic_comparison_appendix}
\end{figure}

\begin{figure}[ht!]
    \centering
    \includegraphics[scale=0.32]{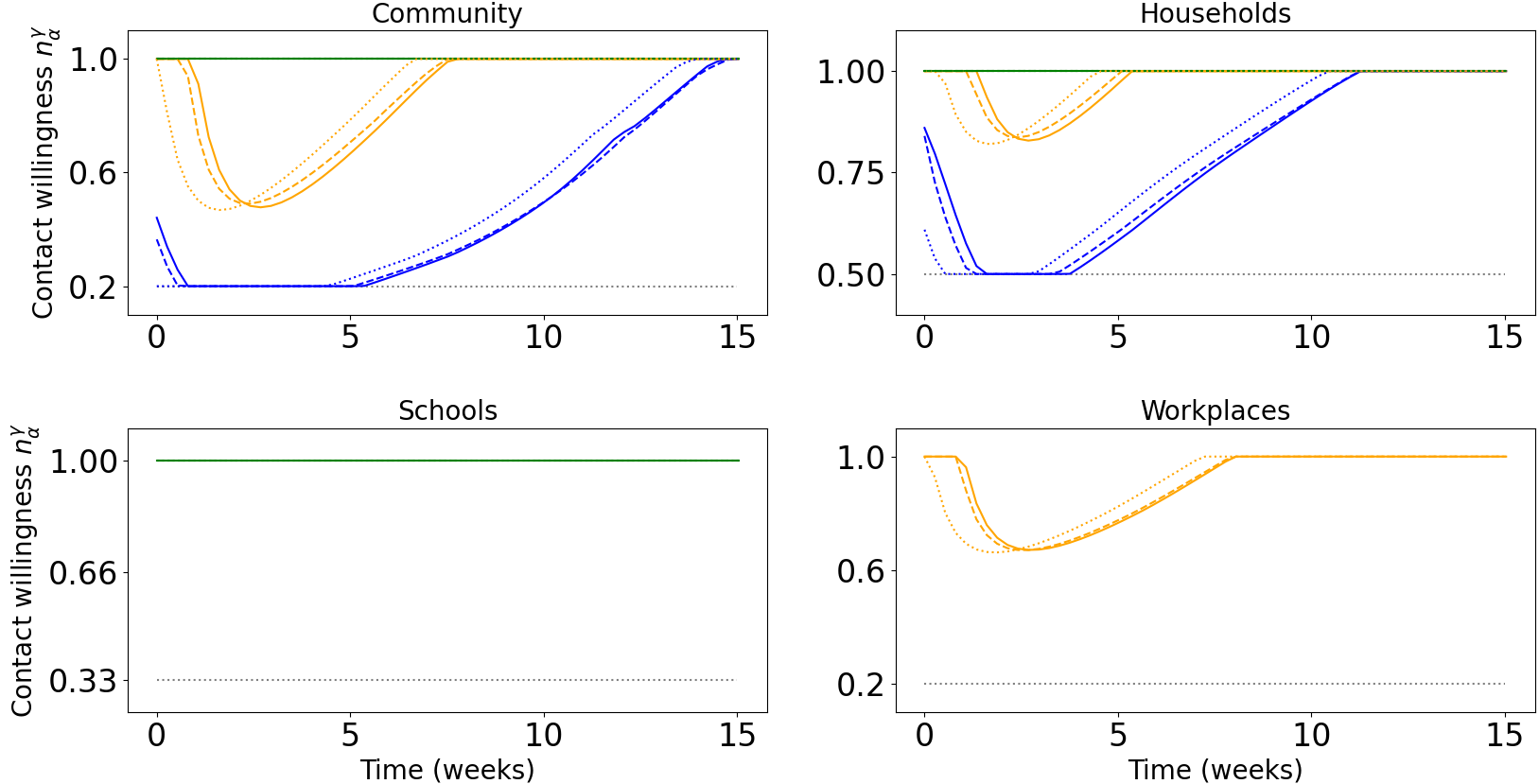}
    \par\vspace{0.2cm}
    \includegraphics[scale=0.32]{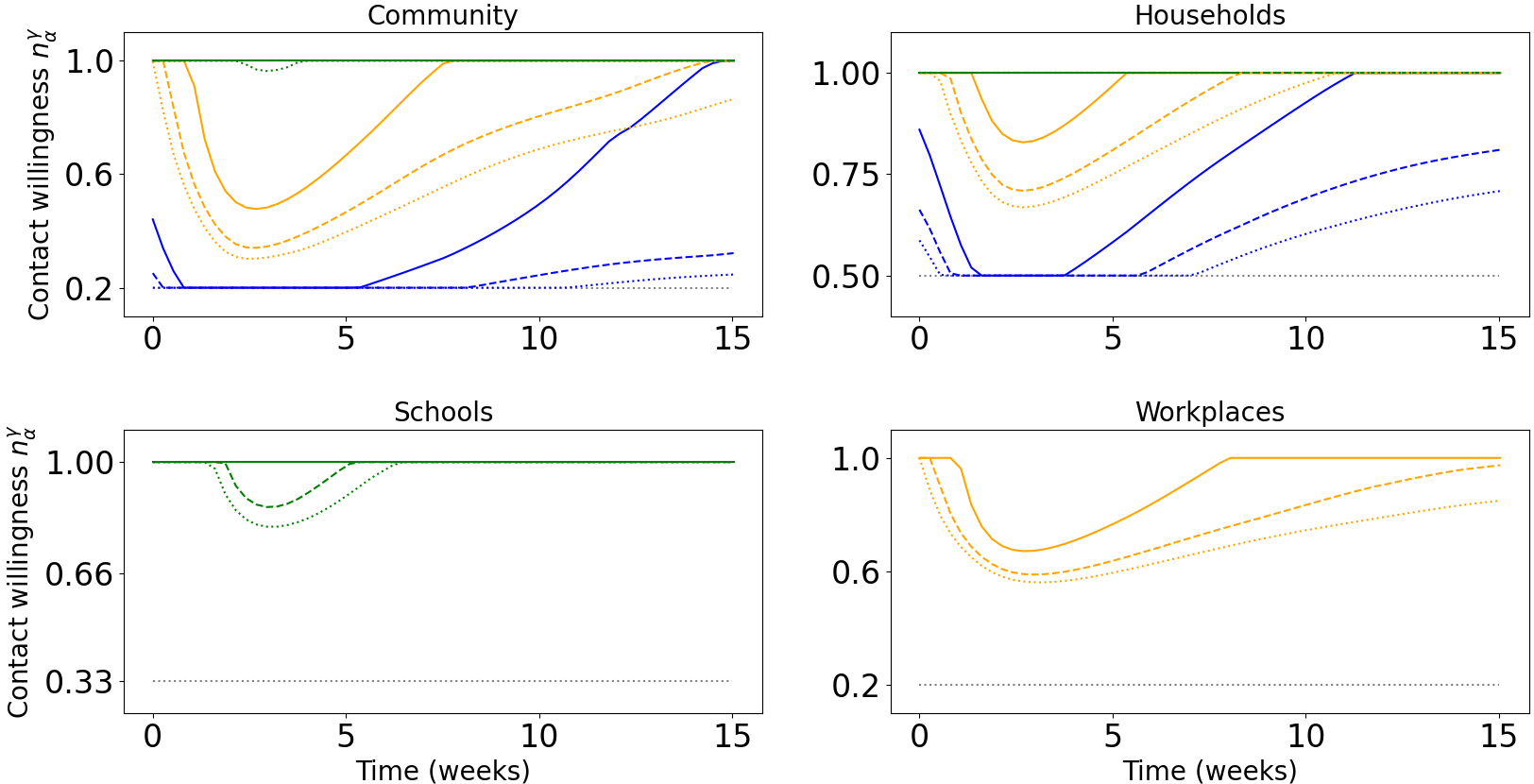}
     \par\vspace{0.2cm}
    \includegraphics[scale=0.32]{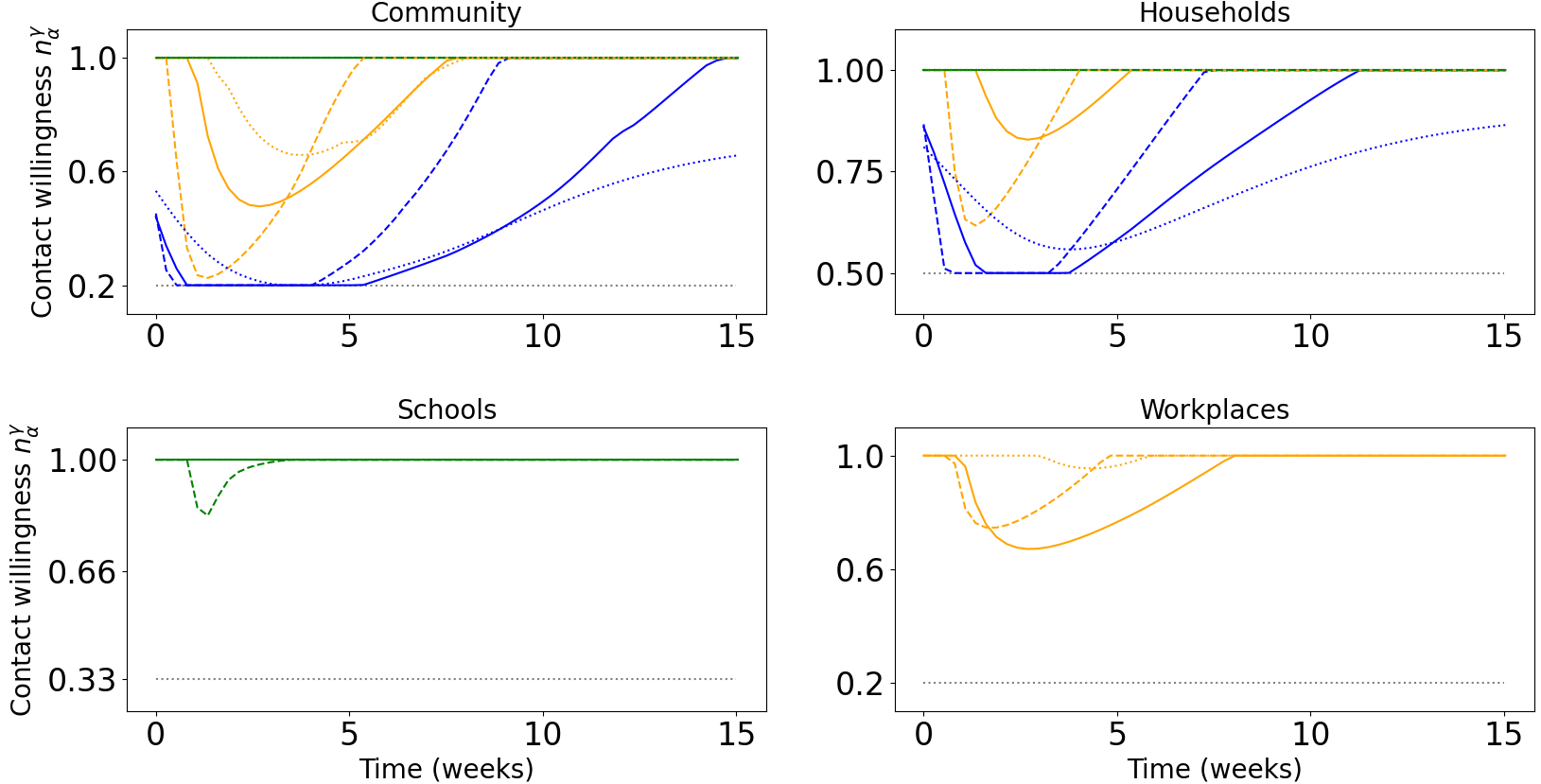}
     \par\vspace{0.2cm}
    \includegraphics[scale=0.32]{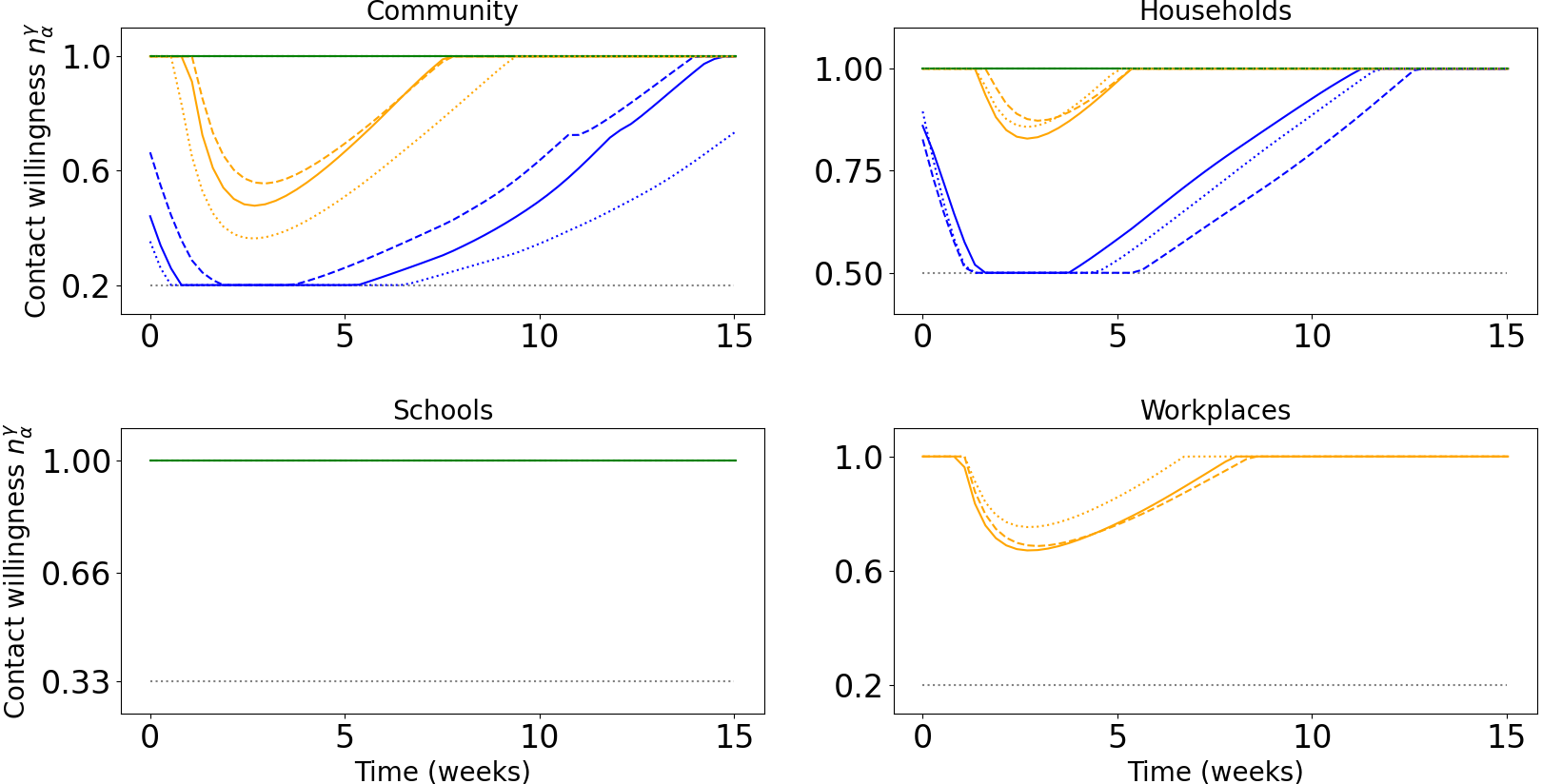}
    \caption{Comparison of Nash equilibrium contact willingness for the different set of parameters used in Fig.~\ref{fig:epidemic_comparison_appendix} and the same legend for solid, dashed and dotted lines. We keep the legend of Fig.~\ref{fig:contact_nash} regarding colors.}
    \label{fig:contact_comparison_appendix}
\end{figure}

We present below the Nash equilibrium results (first for epidemic quantities in Fig.~\ref{fig:epidemic_comparison_appendix} and then for contact willingness in Fig~\ref{fig:contact_comparison_appendix}) where we change at each time one of the parameters presented in Table \ref{table:M}. We see in Fig.~\ref{fig:epidemic_comparison_appendix} that the general behaviors observed with the original set of parameters  (unicity of the peak, reach collective immunity) are quite robust to many different changes. As expected, contacts between classes allow an epidemic spreading even in classes where no one is infected at $t=0$ (first row). Then, in second row regarding different $r_I$, we see that epidemic peak occurs at a lower level as $r_I$ increases, since individuals do more efforts to protect themselves. In third row, we see that the different proportion of age classes in the population will have a huge impact on the epidemic. Indeed, it will affect both the matrix of effective contacts (which are higher between young people) and the risk due to infection (which is lower for young). Hence, the observed behavior results in a high and quick epidemic for a young population, while it is significantly lower and slower for an old population. Finally in fourth row, the precise matrix of contacts $M$ affects the epidemic in each class, but in a relatively moderate way regarding the global evolution of infected proportion in the population.

\begin{table}
    \centering
    \begin{tabular}{ c c c c}
     \hline 
     $M_1^S$ & $M_1^W$ & $M_1^C$ & $M_1^H$ \\ 
     \hline
    $\begin{pmatrix}
    100 & 0 & 0 \\
    0 & 0 & 0 \\
    0 & 0 & 0
    \end{pmatrix}$  & 
    $\begin{pmatrix}
    0 & 0 & 0 \\
    0 & 75 & 0 \\
    0 & 0 & 0
    \end{pmatrix}$ & $\begin{pmatrix}
    12.5 & 25 & 12.5 \\
    12.5 & 25 & 12.5 \\
    12.5 & 25 & 12.5
    \end{pmatrix}$ & $\begin{pmatrix}
    15 & 25 & 10 \\
    12.5 & 32.5 & 5 \\
    10 & 10 & 30
    \end{pmatrix}$  \\ 
    \hline 
    $M_2^S$ & $M_2^W$ & $M_2^C$ & $M_2^H$ \\ 
    \hline
    $\begin{pmatrix}
    100 & 0 & 0\\
    0 & 0 & 0 \\
    0 & 0 & 0 \\
    \end{pmatrix}$ & $\begin{pmatrix}
    0 & 0 & 0\\
    0 & 75 & 0 \\
    0 & 0 & 0 \\
    \end{pmatrix}$ & $\begin{pmatrix}
    12.5 & 15 & 5\\
    7.5 & 25 & 5 \\
    5 & 10 & 12.5 \\
    \end{pmatrix}$ & $\begin{pmatrix}
    12.5 & 15 & 20\\
    7.5 & 30 & 17.5\\
    20 & 35 & 12.5\\
    \end{pmatrix}$ \\
    \hline 
    $M_3^S$ & $M_3^W$ & $M_3^C$ & $M_3^H$ \\ 
    \hline
    $\begin{pmatrix}
    75 & 0 & 0\\
    0 & 0 & 0 \\
    0 & 0 & 0 \\
    \end{pmatrix}$ & $\begin{pmatrix}
    0 & 0 & 0\\
    0 & 50 & 0 \\
    0 & 0 & 0 \\
    \end{pmatrix}$ & $\begin{pmatrix}
    25 & 50 & 25\\
    25 & 50 & 25 \\
    25 & 50 & 25 \\
    \end{pmatrix}$ & $\begin{pmatrix}
    12.5 & 25 & 12.5\\
    12.5 & 25 & 12.5\\
    12.5 & 25 & 12.5\\
    \end{pmatrix}$ \\
    \hline
    \end{tabular}
    \caption{Table of matrices $M_1$, $M_2$ and $M_3$ (given with the form $M^\gamma$) used for the fourth row of Fig.~\ref{fig:epidemic_comparison_appendix}. The first one corresponds to the one we took in our previous simulations (Table~\ref{table:M}), while the two others are chosen to explore two behaviors : matrix $M_2$ corresponds to a society with important heterogeneous contacts, especially in households; while matrix $M_3$ is a society which is more homogeneous with a lot of contacts in community. Matrix elements are contact rates (per week) in our model.}
    \label{table:Matrix_appendix}  
\end{table}

\FloatBarrier

\bibliography{biblio}

\end{document}